\documentclass[%
	a4paper,		
	twoside,		
	11pt,			
	toc=listof,		
	toc=bibliography, 
 	titlepage		
	BCOR=10mm,		
    captions=tableheading,
    numbers=noenddot,
	DIV=12,			
]{scrbook}

\usepackage[utf8]{inputenc}
\usepackage[british]{babel}
\usepackage{blindtext}
\usepackage[%
	autostyle,			
	german=quotes		
]{csquotes}				
\MakeOuterQuote{"}

\usepackage{marvosym}

\usepackage{pifont}

\usepackage{setspace}
\usepackage{scrhack}

\usepackage{array}
\usepackage{booktabs}			
\usepackage{tabularx}			
\usepackage[dvipsnames]{xcolor}	

\usepackage{graphicx}	
\usepackage{caption}	
\usepackage[format=plain,
			 font={footnotesize,sf},
			 labelfont={bf},
			 justification=justified,
			 margin=1cm,
			 aboveskip=10pt,
			 belowskip=10pt,
			 position=bottom
]{caption}		

\usepackage{subfigure}
\usepackage{wrapfig}
\usepackage{float}		
\usepackage{placeins}	
\usepackage{tabularx}
\usepackage{eso-pic}
\usepackage{chngpage}

\usepackage{amsmath}
\usepackage{amssymb}
\usepackage{empheq}
\usepackage{dsfont}
\usepackage{amstext}
\usepackage{amsfonts}
\usepackage{amsthm}
\usepackage{wasysym}

\usepackage{color}
\usepackage{tikz}

\usepackage{listings}

\usepackage{algorithm}
\usepackage{algpseudocode}

\usepackage{nicefrac} 
\usepackage{units}

\usepackage{listings}
\usepackage{parskip}
\setparsizes{0}{0.25\baselineskip plus .25\baselineskip}{1em plus 1fil}

\usepackage{fancyhdr}

\usepackage{ifthen}

\usepackage[
	bookmarks,		
	raiselinks,		
	pageanchor,		
    colorlinks = true,		
	citecolor=black,
	linkcolor=black,	
	urlcolor=magenta,
	filecolor=cyan,	
	menucolor=black,	
	pagebackref,	
	breaklinks = true
]{hyperref}

\usepackage{breakcites}

\usepackage{cleveref} 
\crefname{equation}{equation}{equations}
\crefname{figure}{figure}{figures}
\crefname{section}{section}{sections}
\creflabelformat{equation}{#2\textup{#1}#3}

\usepackage[
	left=3cm,right=2.5cm,top=1.5cm,bottom=1.5cm,	
	headheight=7mm,headsep=0.9cm,					
	footskip=1.5cm,includeheadfoot					
]{geometry}

\theoremstyle{plain} 
\theoremstyle{remark} 
\theoremstyle{plain} 
\theoremstyle{plain} 
\theoremstyle{plain} 

\usepackage{suffix} 

\newcommand{\Vth}{V\textsubscript{th}}

\newcommand{\Glass}{Glass(t,T\textsubscript{hist})}
\newcommand{\Thist}{T\textsubscript{hist}}
\newcommand{\GST}{Ge\textsubscript{2}Sb\textsubscript{2}Te\textsubscript{5}}
\newcommand{\SbTe}{Sb\textsubscript{2}Te\textsubscript{3}}
\newcommand{\AIST}{Ag\textsubscript{4}In\textsubscript{3}Sb\textsubscript{67}Te\textsubscript{26}}

\newcommand{\Rsc}{R\textsubscript{s,cryst}}
\newcommand{\Rsa}{R\textsubscript{s,amo}}
\newcommand{\Rsp}{R\textsubscript{s,proj}}
\newcommand{\RePCM}{R\textsubscript{ele-PCM}}
\newcommand{\RePROJ}{R\textsubscript{ele-proj}}
\newcommand{\Rc}{R\textsubscript{cryst}}
\newcommand{\Ra}{R\textsubscript{amo}}
\newcommand{\Rint}{R\textsubscript{int}}
\newcommand{\Rprojc}{R\textsubscript{proj,c}}
\newcommand{\Rproja}{R\textsubscript{proj,a}}

\newcommand{\kb}{k\textsubscript{b}}
\newcommand{\Es}{E\textsubscript{s}}
\newcommand{\Ea}{E\textsubscript{a}}
\newcommand{\TMN}{T\textsubscript{MN}}
\newcommand{\nuo}{$\nu$\textsubscript{0}}
\newcommand{\nuoo}{$\nu$\textsubscript{00}}

\newcommand{\Ed}{E\textsubscript{d}}

\newcommand{\figref}[1]{Figure~\ref{#1}}					
\WithSuffix\newcommand\figref*[1]{Figure~\subref{#1}}	
\newcommand{\tableref}[1]{Table~\ref{#1}}					
\WithSuffix\newcommand\tableref*[1]{Table~\subref{#1}}	
\WithSuffix\newcommand\eqref*[1]{~\eqref{#1}}		
\newcommand{\equref}[1]{Equation~\cref{#1}}
\WithSuffix\newcommand\equref*[1]{Equation~\subref{#1}}

\usepackage[textwidth=2cm,			
            textsize=footnotesize,	
            english,					
            shadow,					
            colorinlistoftodos]{todonotes}

\ifthenelse{\boolean{@twoside}}
{														
	\fancyhf{}											
	\fancyhead[RO,LE]{\small\sffamily\bfseries\thepage}	
	\fancyhead[LO,RE]{\small\sffamily\nouppercase{\leftmark}}
	\fancypagestyle{plain}{								
		\fancyhf{}										
		\fancyhead[RO,LE]{\small\sffamily\bfseries\thepage}
		\fancyhead[LO,RE]{\small\sffamily\nouppercase{\leftmark}}
	}
}
{														
	\fancyhf{}											
	\fancyhead[R]{\small\sffamily\bfseries\thepage}		
	\fancyhead[L]{\small\sffamily\nouppercase{\leftmark}}
	\fancypagestyle{plain}{								
		\fancyhf{}
		\fancyhead[R]{\small\sffamily\bfseries\thepage}	
		\fancyhead[L]{\small\sffamily\nouppercase{\leftmark}}
	}
}

\hypersetup{	
	pdftitle    = {Titel der PDF Datei},
	pdfsubject  = {thesis},
	pdfauthor   = {Benedikt Kersting}
	}

\begin{document}	

\begin{titlepage}
\thispagestyle{empty}
\begin{center}
 \par
 \vspace*{20ex}
 \huge
 \textsc{Benedikt Johannes Kersting}
 \par
 \vspace*{3ex}
 \textsc{Quest for a solution to drift \\ in phase change memory devices}
 \par
 \vspace*{3ex}
 \textsc{2022}
 
\end{center}
\clearpage

\end{titlepage}

	\cleardoublepage
	\begin{center}
	\thispagestyle{empty}
	\par
	\vspace*{1ex}
	\Large
	Materialphysik
	\par
	\vspace*{10ex}
	Quest for a solution to drift in phase change memory devices
	\par
	\vspace*{25ex}
	Inaugural-Dissertation \\
	zur Erlangung des Doktorgrades (Dr. rer. nat./Dr. phil.) \\
	der Naturwissenschaften im Fachbereich Physik \\
	der Mathematischen-Naturwissenschaftlichen Fakultät \\ 
	der Westfälischen Wilhelms-Universität Münster \\ 
	
	\par
	\vspace*{22ex}
	
	vorgelegt von \\
	Benedikt Johannes Kersting \\
	aus München \\
	- 2022 - 
	 
\end{center}
\clearpage
\thispagestyle{empty}
\par 
\vspace*{110ex}
\large

\hspace*{5mm} Dekan: \hspace*{55.1mm} Professor Michael Rohlfing

\hspace*{5mm} Erster Gutachter: \hspace*{35.78mm} Professor Martin Salinga

\hspace*{5mm} Zweiter Gutachter: \hspace*{33.4mm} Professor Gerhard Wilde

\hspace*{5mm} Tag der mündlichen Prüfung: \hspace*{14.35mm} ........................................

\hspace*{5mm} Tag der Promotion: \hspace*{32.05mm} ........................................

\thispagestyle{empty}

\vspace*{50ex}

\noindent\hspace*{55mm}
All models are wrong, \\
\noindent\hspace*{55mm}
but some are useful. \\
\noindent\hspace*{70mm}
		\textit{- George Box}
	\cleardoublepage
	
	\pagenumbering{roman}

\chapter*{Abstract}
\addcontentsline{toc}{chapter}{Abstract}

Phase change memory (PCM) is an electronic memory technology. In a phase change memory device, switching between a crystalline and amorphous material state occurs in nanoseconds and is thermally induced by Joule heating. The resistivity contrast between the two material states allows information to be encoded. Dependent on the amount of low resistive crystalline and high resistive amorphous material, a quasi-analog resistance can be programmed into the device. This multi-level storage capability could give PCM the edge compared to other more established technologies. Additionally, it makes PCM suitable for in-memory computing. By executing certain computational tasks in the memory, this non-von Neumann computing paradigm aims to reduce the amount of data that needs to be shuttled between the processing (CPU) and memory unit (DRAM). Multi-level storage, however, is impeded by a continuous increase of the amorphous state's resistivity. This phenomenon, called drift, remains one of the most pressing issues of PCM technology. To date, only binary PCM chips have been productized. 

The goal of this thesis is to gain new insights into the drift phenomenon and identify strategies to mitigate it. An extensive experimental characterization of PCM devices and in particular drift forms the foundation of each chapter. With respect to time-scales, ambient temperatures, device dimensions, and combinations thereof, drift is studied under unprecedented conditions. In three studies, different aspects of drift are examined.

Study 1 - The origin of structural relaxation: Drift has been attributed to structural relaxation of the non-equilibrium amorphous state created by melt-quenching. Models predict that it takes a certain amount of time before the thermally activated structural relaxation processes begin. Drift measurements at ambient temperatures between \unit[100]{K} and \unit[300]{K} over 9 orders of magnitude in time reveal the onset of relaxation in a melt-quenched state. The data is used to appraise two phenomenological models for structural relaxation, specifically the Gibbs relaxation model and the collective relaxation model. Additionally, a refined version of the collective relaxation model is introduced and the consequences of a limited number of structural defects are discussed.  

Study 2 - Exploiting nanoscale effects in phase change memories: Scaling devices to ever-smaller dimensions is incentivized by the requirement to achieve higher storage densities and less power consumption. Eventually, confinement and interfacial effects will govern the device characteristics, not the bulk properties of the phase change material. Anticipating these consequences, the feasibility to use a single element is assessed for the first time. In the devices, pure antimony is confined one-dimensionally in the range from $\sim$9 atomic layers (\unit[3]{nm}) to $\sim$30 atomic layers (\unit[10]{nm}). The power efficiency, stability against crystallization, and drift are characterized under different degrees of confinement. 

Study 3 - State-dependent drift in a projected memory cell: New device concepts are aiming to reduce drift by decoupling the programmed device resistance from the electronic properties of the amorphous phase. To this end, a shunt resistor that scales with the amount of amorphous material is added to the projected memory cell. Simulations with an equivalent circuit model and the drift characteristics of a projected device based on antimony put the idealized concept to the test. Compared to an unprojected device the resistance evolution with time changes notably. The contact resistance between the phase change material and the shunt resistor is identified as a decisive parameter to realize a device with the desired properties. 

In the conclusion, strategies to mitigate drift by material optimization, confinement, and new device concepts are revisited. The findings of this work are used to assess the prospect of different ideas discussed in the literature. Additionally, they inspire alternative explanations for the remarkably small drift reported in some studies and stimulate two novel ideas on how drift can be reduced in future PCM devices.

	\chapter*{Zusammenfassung}
\addcontentsline{toc}{chapter}{Zusammenfassung}

Phasenwechselspeicher sind ein elektronisches, resistives Speichermedium. Das Phasenwechselmaterial in diesem resistiven Speicher (Zelle) kann innerhalb von Nanosekunden in einen kristallinen oder amorphen Zustand versetzt werden. Um einen Phasenwechsel zu erreichen, wird das Material mittels Joulscher W{\"a}rme erhitzt. Die amorphe Phase hat einen deutlich gr{\"o}\ss eren spezifischen elektrischen Widerstand als die kristalline Phase. Indem man den Anteil des amorphen und kristallinen Materials in der Zelle {\"a}ndert, kann ein quasi analoger Widerstand programmiert werden. Das Potenzial, als Multi-Level-Speicher genutzt zu werden, k{\"o}nnte Phasenwechselspeichern im Vergleich zu anderen Speichermedien einen Wettbewerbsvorteil verschaffen. Dar{\"u}ber hinaus er{\"o}ffnet dies die M{\"o}glichkeit Phasenwechselspeicher im Bereich des In-Memory-Computing zu verwenden. Mit diesem Konzept versucht man, einige Schw{\"a}chen der klassischen von-Neuman-Computer-Architektur zu umgehen. Indem gewisse Rechenoperationen mit dem analogen Speicher ausgef{\"u}hrt werden, kann die Menge der Daten, die zwischen Speicher und Prozessor transferiert werden muss, reduziert werden. Bislang k{\"o}nnen Phasenwechselspeicher jedoch nur begrenzt als Multi-Level-Speicher genutzt werden, da der spezifische Widerstand des amorphen Zustands kontinuierlich zunimmt. Dieses Ph{\"a}nomen wird als Drift bezeichnet. Drift stellt eine der gr{\"o\ss}ten Herausforderungen f{\"u}r diese Technologie dar. Bislang wurden nur bin{\"a}re Phasenwechselspeicher zur Produktreife entwickelt. 

Das Ziel dieser Arbeit ist es, das Drift Ph{\"a}nomen besser zu verstehen und neue Konzepte, mit denen Drift unterdr{\"u}ckt oder reduziert werden kann, zu untersuchen. Die Arbeit ist in drei Studien unterteilt, die in je einem Kapitel dargestellt werden. Eine sorgf{\"a}ltige experimentelle Charakterisierung von Phasenwechselspeichern und insbesondere ihres Drift bildet die Grundlage eines jeden Kapitels. In vielerlei Hinsicht wird Drift von Phasenwechselspeichern unter neuen und extremen Bedingungen vermessen. Dies betrifft die Zeitskalen, die Umgebungstemperatur und die Art der Speicherzellen. 

Studie 1 - Der Beginn struktureller Relaxation: Die meisten Theorien gehen davon aus, dass Drift durch die strukturelle Relaxation des amorphen Zustandes hervorgerufen wird. Relaxationsmodelle sagen voraus, dass es eine gewisse Zeit dauert, bevor thermisch-aktivierte Relaxationen stattfinden. In den hier vorgestellten Messungen, {\"u}ber 9 Gr{\"o}\ss enordnungen in der Zeit und bei Umgebungstemperaturen zwischen \unit[100]{K} und \unit[300]{K}, kann der Beginn der Relaxation beobachtet werden. Die Daten werden genutzt, um zwei ph{\"a}nomenologische Relaxationsmodelle, das Gibbs-Modell und das Collective-Relaxation-Modell, zu vergleichen und zu pr{\"u}fen. Desweiteren wird eine erweiterte Version des Collective-Relaxation-Modell entwickelt und diskutiert, welche Folgen eine begrenzte Anzahl struktureller Defekte f{\"u}r das Gibbs-Modell h{\"a}tte.

Studie 2 - Wie sich Effekte im Nanobereich nutzen lassen: Um h{\"o}here Speicherdichten und eine bessere Energieeffizienz zu erreichen, ist es erforderlich, elektronische Speicher in immer kleineren Dimensionen zu bauen. Entsprechend wird das Verhalten des Speichers irgendwann durch Grenzflächeneffekte dominiert werden und nicht mehr nur von den Eigenschaften des Phasenwechselmaterials abh{\"a}ngen. In Anbetracht der zu erwartenden Folgen stellt sich die Frage, ob auch andere Materialien als die etablierten Legierungen f{\"u}r Phasenwechselspeichern genutzt werden k{\"o}nnten. In den hier untersuchten Zellen wird erstmals pures Antimon als Phasenwechselmaterial eingesetzt. Die Energieeffizienz, Stabilit{\"a}t gegen Rekristallisation und Drift von Zellen mit unterschiedlichen Antimon Schichtdicken (\unit[3]{nm} bis \unit[10]{nm}) werden verglichen.

Studie 3 - Drift in Projizierten-Zellen: In der Projizierten-Zelle soll der programmierte Widerstand von den elektronischen Eigenschaften der amorphen Phase entkoppelt werden, um Drift zu reduzieren. Zu diesem Zweck wird das Phasenwechselmaterial auf einem elektrisch leitf{\"a}higen Material deponiert (Projektions-Schicht). Es fungiert als paralleler Widerstand, der mit der Gr{\"o}\ss e des amorphen Volumens skaliert. Die Analyse eines Ersatzschaltbildes der Zelle und Experimente mit Projizierten-Zellen zeigen, dass diese Zellen ein deutlich anderes Driftverhalten aufweisen. Der Kontaktwiderstand zwischen Phasenwechselmaterial und Projektions-Schicht muss minimiert werden, um eine Zelle mit den gew{\"u}nschten Eigenschaften herzustellen. 

Im Fazit wird noch einmal diskutiert, wie sich Drift reduzieren l{\"a}sst. Anhand der in dieser Arbeit gewonnen Erkenntnisse werden unterschiedliche in der Literatur vorgeschlagene Konzepte {\"u}berdacht. Au\ss erdem ergeben sich aus den Ergebnissen neue Erkl{\"a}rungen, warum in einigen Ver{\"o}ffentlichungen ein sehr geringer Drift gemessen wurde. Daraus folgen auch zwei neue Ans{\"a}tze, wie Drift in Phasenwechselspeichern reduziert werden k{\"o}nnte.

	\cleardoublepage
	
	\tableofcontents
	\cleardoublepage
	
	\newpage
	\pagenumbering{arabic}

		\chapter{Introduction}
\label{Chapt:Intro}

In this chapter, we will recap the basic operation principle of a phase change memory (PCM) device and introduce two original device concepts. Next, it is discussed how multi-level storage and programming are achieved. A brief presentation of the two most promising emerging applications of phase change memory technology sets the stage for the key problem addressed in this thesis: temporal changes of the amorphous state. The ensuing section is a review of how device characteristics drift, the theories proposed to explain this phenomenon, and potential approaches to mitigate drift. Having defined the problem at hand, the chapter concludes with the scope and structure of this thesis. 

\section{Operation principle of phase change memory}

The key component of a PCM device is the phase change material. Dependent on its atomic configuration, the device stores a high resistance value (0;~RESET) or a low resistance value (1; SET). Phase change materials are typically chalcogenide compounds such as \GST. They exhibit a pronounced resistivity contrast between the disordered amorphous phase (high resistivity) and the periodically ordered crystalline phase (low resistivity). Additionally, their atomic mobility is highly temperature-dependent. Below the temperature regime of fast crystallization, both phases can be retained for years, whereas the phase transition happens within nano-seconds at elevated temperatures. 

\begin{figure}[bth!]
	\centering
	\includegraphics[width=0.8\linewidth]{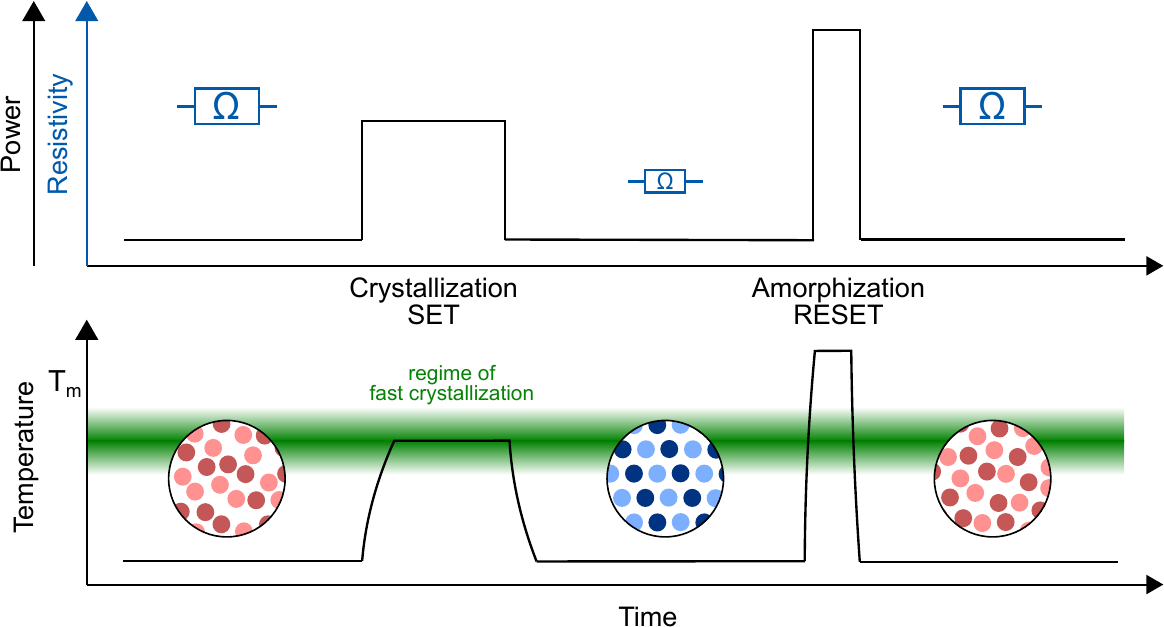}
	\caption[Phase change memory - Basic write operations]{\textbf{Phase change memory - Basic write operations.} A PCM device can be reversibly switched between a highly resistive amorphous state and a low resistive crystalline state. The amplitude and duration of electronic pulses determine the temperature profile created in the device by joule heating. The SET operation erases the amorphous state by heating the material to a temperature range that enables recrystallization on the order of nanoseconds. To RESET the device the phase change material is molten and melt-quenched with a sharp pulse.}
	\label{fig:Intro_OperationPrinciple}
\end{figure}

To switch a device from the amorphous phase to the crystalline phase, the material is heated to the temperature regime of fast crystallization. The amorphous phase is a thermodynamic non-equilibrium state. Thus, heating the material, i.e., increasing the atomic mobility, results in a transition to the thermodynamic equilibrium state, the crystalline phase (\Cref{fig:Intro_OperationPrinciple}). 

To create an amorphous phase again, the material is heated to even higher temperatures and quenched rapidly. Above the melting temperature (typically T\textsubscript{m}\textless  \unit[850]{K} \cite{Madelung2005,Kalb2003,Crawley1972}), the ordered structure of the crystal breaks down and the material assumes a molten state. Melt-quenching allows freezing in the disordered structure of the molten state. The melt is cooled so rapidly below the temperature regime of fast crystallization that the atomic rearrangement to the ordered crystalline structure is suppressed. 

\begin{figure}[bth!]
	\centering
	\includegraphics[width=0.6\linewidth]{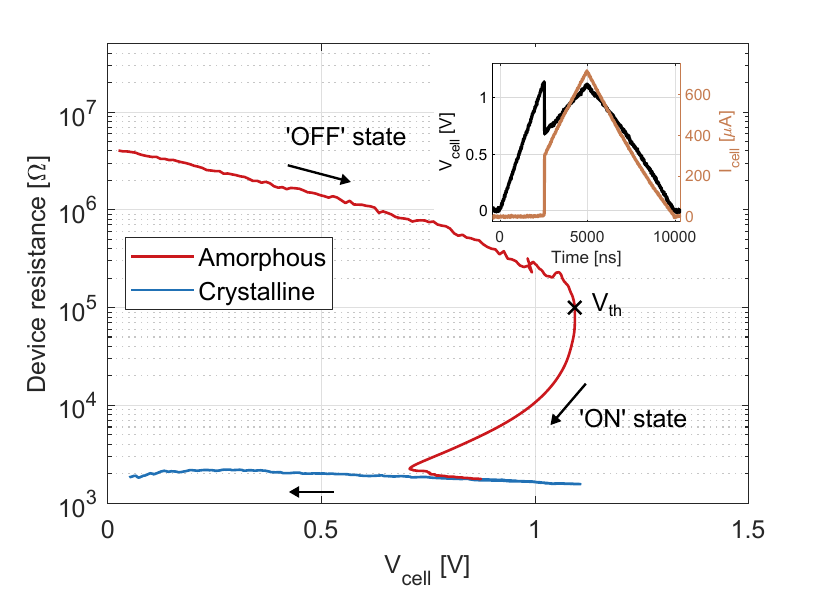}
	\caption[Exemplary threshold switching characteristic]{\textbf{Exemplary threshold switching characteristic.} At elevated fields, the resistance of the amorphous state is exponentially field-dependent. Beyond a threshold value (V\textsubscript{th}) the material switches to a highly conductive amorphous on-state. After threshold switching, Joule heating becomes sufficiently large to recrystallize the amorphous state. Since the PCM device is operated in series with a resistor, the voltage drop over the cell reduces in the moment of rapid switching. The inset shows the transient device voltage and current traces. In the subthreshold regime, the amorphous phase is characterized with a source meter unit (cell voltage\textless \unit[1]{V}). Threshold switching is captured with an oscilloscope.}
	\label{fig:Intro_ThresholdSwitching}
\end{figure}

The temperature changes, required to switch the device, are induced by Joule heating. The power, dissipated in the phase change material by applying electrical pulses to the device, locally heats the material to T\textsubscript{m}. Efficient reversible switching between the high and low resistance device states is only feasible due to the highly field-dependent resistance of the amorphous phase (\Cref{fig:Intro_ThresholdSwitching}). With increasing field strength it transitions first from ohmic to exponential to super-exponential before, finally, threshold-switching to orders of magnitude lower resistivity occurs \cite{Krebs2010,Ovshinsky1968,Ovshinsky1973a}. Thus,  at low voltages (read), the amorphous phase exhibits several orders of magnitude larger resistance, but, with moderate voltages (write), sufficient Joule heat to crystallize it can still be created.

\Cref{fig:Intro_ThresholdSwitching} shows the resistive switching of a device that was programmed to a resistance of \unit[28]{M$\Omega$} with a programming power of \unit[900]{$\mu$W}. Threshold switching occurs at \unit[1.1]{V}. To get a better feeling for the criticality of the field-dependent transport, consider a scenario where the amorphous state is purely ohmic. Even if just \unit[5]{\%} of the power required to melt the device was sufficient to reach the temperature regime of fast crystallization, it would require a voltage of $\sqrt{P\cdot R} = \sqrt{\unit[45]{\mu W}\cdot \unit[28]{M\Omega}} = \unit[35]{V}$ to crystallize the device.

To summarize, phase change materials combine three key characteristics enabling their usage for a non-volatile memory device. First, the resistivity contrast between crystalline and amorphous phase encodes information. Second, due to the highly temperature-dependent atomic mobility, information can be stored for years (\unit[3$\cdot$10\textsuperscript{7}]{s}) and written in nano-seconds (\unit[10\textsuperscript{-9}]{s}). Third, as a result of its highly field-dependent electrical transport, culminating in threshold-switching, the amorphous state can be recrystallized using moderate write voltages.  

\section{Device concepts}

\begin{figure}[bth!]
	\centering
	\includegraphics[width=1\linewidth]{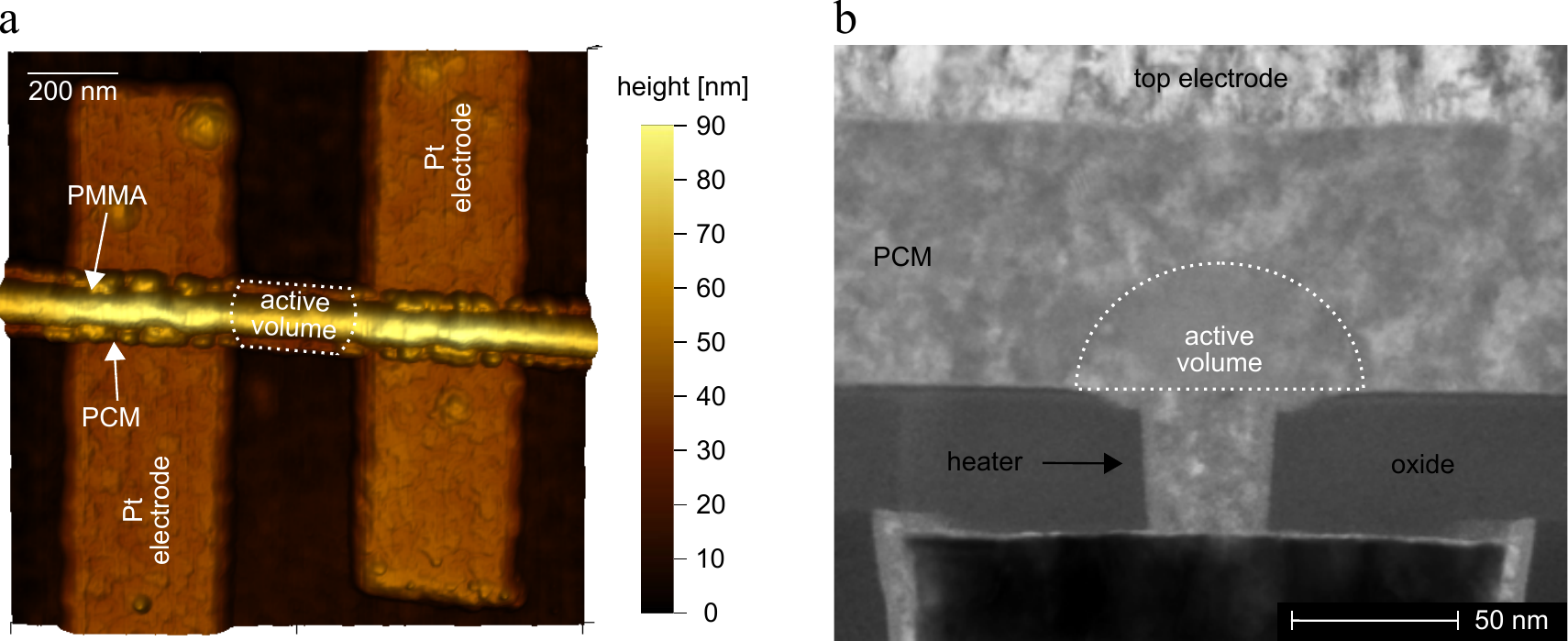}
	\caption[Two basic device concepts]{\textbf{Two basic device concepts.} \textbf{a}, Atomic force microscopy image of a bridge cell. The phase change material is patterned to a narrow stripe connected to two metal electrodes. The material volume between the electrodes can be molten and amorphized. \textbf{b}, Transmission electron micrograph of a mushroom cell. The bottom electrode (heater) is patterned to a small cylinder. The highest current densities are reached in the vicinity of the heater. Above, the heater forms an amorphous dome in a half-spherical shape.   
	}
	\label{fig:Intro_DeviceConcepts}
\end{figure}

A PCM device is programmed to a high resistance state when the current path through the device is blocked by amorphous phase change material. The power required to write a RESET state scales with the material volume that must be melt-quenched to achieve this \cite{Chen2006, Raoux2014, Xiong2017}. Accordingly, PCM devices are designed such that the amorphization and crystallization of a small, localized volume allow to switch the memory cell. This is achieved either by confining the phase change material itself or by confining one of the contact electrodes to the phase change material. The bridge cell is an example of the former and the mushroom cell of the latter (\Cref{fig:Intro_DeviceConcepts}). 

In a bridge cell, the phase change material is patterned to a narrow line connected to two electrodes \cite{Lankhorst2005}. The active material volume, switching between crystalline and amorphous state, is located in the proximity of the center between the electrodes~\footnote{The amorphous volume can shift away from the center of the bridge due to thermoelectric effects. This shift has been experimentally observed by electron microscopy \cite{Castro2007, Oosthoek2015a}.}. In this layout, the volume that needs to be amorphized to program the cell is defined by the cross-section of the bridge. It is limited by the smallest width to which the material can be patterned and the material thickness that can be deposited uniformly and still switched. Device scaling to \unit[20]{nm} wide and \unit[3]{nm} thick bridges has been demonstrated, proving that PCM is a scalable technology \cite{Chen2006, Raoux2008a}.

The mushroom cell is a vertical device structure, in which the phase change material film is sandwiched between a bottom electrode, confined to a narrow cylinder (heater), and a top electrode. During programming, the highest current density and thus largest heat dissipation in the phase change material is at the bottom electrode. The active volume has the shape of a half-sphere. In the high resistance state, an amorphous dome covers the bottom electrode contact. Compared to the bridge cell, the separation into two functional units (heater and memory element; phase change material) may provide more degrees of freedom for device optimization. 

Besides these two original cell geometries, other concepts of material and/or electrode confinement have been realized in numerous alternative designs. Most prominently, the bridge cell has also been fabricated in a vertical structure to reduce the device footprint. A phase change material pillar is sandwiched between a top and bottom electrode. This confined or dashed cell design, promising the highest memory densities, is being explored by key industry players \cite{Kim2010a, Kim2017a, Kim2019, Choe2017}. Other devices, aiming to reduce the electrode size, are for example the $\mu$-trench cell \cite{Pellizzer2004} and the edge contact type cell \cite{Ha2003}.

\section{Multi-level programming}
	
Even though the phase change material is switched only between two phases (amorphous and crystalline), it is in principle possible to use PCM as a quasi-analog device. This is achieved by altering the phase configuration, i.e., the volume ratio of amorphous and crystalline material in the device. Within the range, defined by the fully crystalline device and the largest amorphous volume, any resistance can be programmed. The programming accuracy is ultimately limited by the read-noise \cite{Nandakumar2021}. Traditional memory applications, however, require the storage of reliably distinguishable and thus well-separated states. The feasibility of storing two bits per PCM device has been demonstrated in array-level studies (256-Mcell) \cite{Close2013} and encoding schemes with up to \unit[4]{bit} per cell have been shown on small arrays (\textless \unit[1000]{devices}) \cite{Nirschl2007,Papandreou2011b}. It should be noted that closed-loop iterative programming schemes are required to achieve a target resistance. In the following, the change of the fraction of amorphous material, and thus the programmed resistance, dependent on the profile of the electrical pulses applied to the device will be discussed. The SET operation (recrystallization of the amorphous material) is determined by the time spent in the temperature regime of fast crystallization. The RESET operation (melt-quench process) is determined by the volume of molten material and the quench rate. 

\begin{figure}[bth!]
	\centering
	\includegraphics[width=1\linewidth]{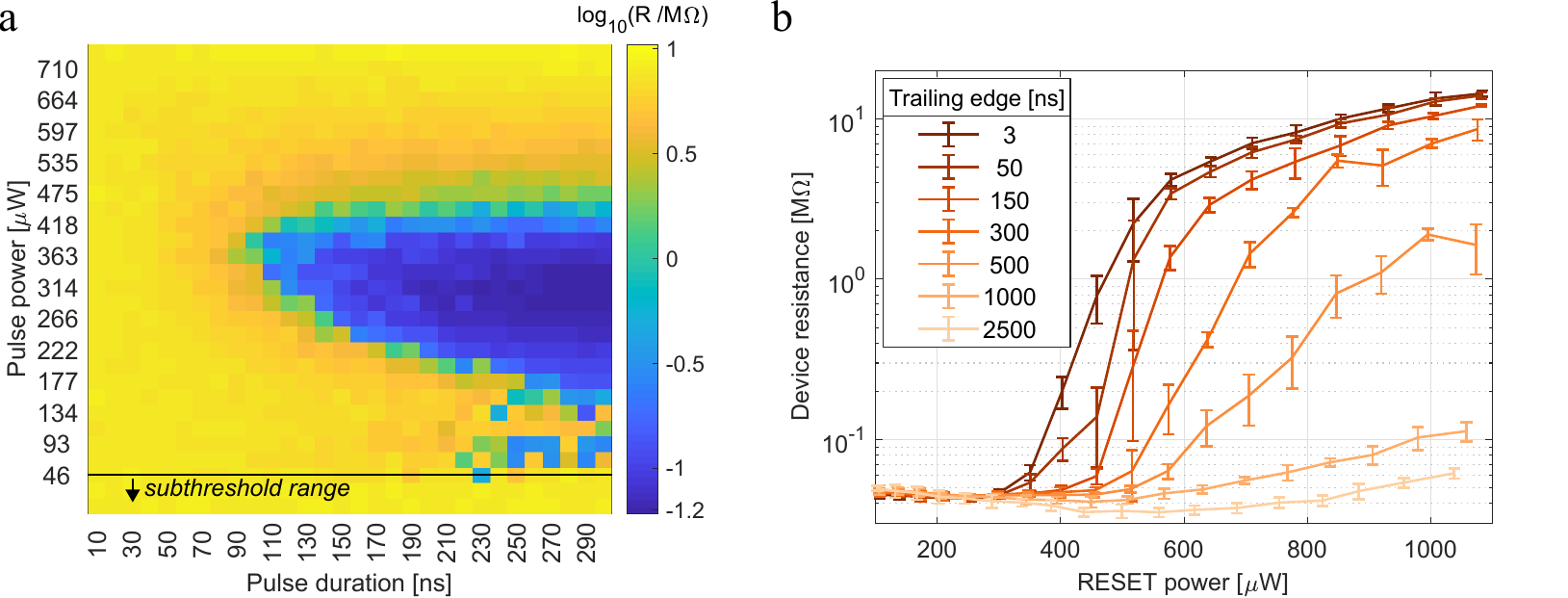}
	\caption[Multi-level programming]{\textbf{Multi-level programming. a}, SET window. Square-shaped programming pulses with varying amplitude and duration were applied to a doped \GST{} mushroom cell in an amorphous state (RESET). The color map shows the device resistance as a function of pulse power and duration. The peak of the temperature regime of fast crystallization and thus the fastest SET is achieved with a pulse power of around \unit[314]{$\mu$W}. With pulse powers below \unit[46]{$\mu$W} the amorphous 'ON' state (\Cref{fig:Intro_ThresholdSwitching})  cannot be sustained (subthreshold range).  
		\textbf{b}, RESET programming. Programming pulses with varying amplitude and trailing edge were applied to a doped \GST{} mushroom cell in the crystalline state (SET). The device resistance measured after programming is plotted as a function of pulse power. With increasing power higher device resistances are achieved. Longer trailing edges result in a smaller device resistance. Compared to the box pulses studied in \textbf{a}, an even better crystallization can be achieved with long pulse trailing edges. Error bars denote the standard deviation of five measurements.}
	\label{fig:Intro_Programming}
\end{figure}

The impact of the amount of time spent at a certain temperature during the SET operation is visualized in \Cref{fig:Intro_Programming}a. To recrystallize the amorphous device, the pulse power must be large enough to heat the phase change material to the temperature regime of fast crystallization. This is achieved through powers in the range of \unit[150]{$\mu$W} to \unit[380]{$\mu$W}. Smaller pulse powers create insufficient heat to recrystallize the material and too large pulse powers either heat the device to temperatures where recrystallization is slow again or could even melt the phase change material. The longer the device is heated to the temperature regime of fast crystallization the more amorphous material recrystallizes. 

The RESET operation can be controlled by the pulse power and the pulse trailing edge. \Cref{fig:Intro_Programming}b shows an exemplary measurement of the programmed device resistance dependent on both parameters. 

With increasing pulse power the device is heated to higher temperatures. Once the phase change material is heated locally to its melting point (here with a power \textgreater \unit[350]{$\mu$W}), an amorphous state can be created, resulting in an increase of the device resistance. This power is defined as programming onset. It should be noted that the temperature inside the device is spatially non-uniform. Initially, at the programming onset, only a small fraction of the phase change material is molten, but with increasing power an increasingly large volume is molten, and the device can be programmed to higher resistances.

The RESET pulse trailing edge allows to alter the cooling profile, i.e., how much time is spent at a certain temperature. Importantly, the time spent in the temperature regime of fast crystallization changes. Thus, with increasing pulse trailing edge, a larger fraction of the molten volume recrystallizes during the melt-quenching process. Accordingly, at a constant programming power the device resistance decreases, and the programming onset shifts towards larger powers.
	   
\section{Emerging applications of phase change memory}

\begin{figure}[bth!]
	\centering
	\includegraphics[width=1\linewidth]{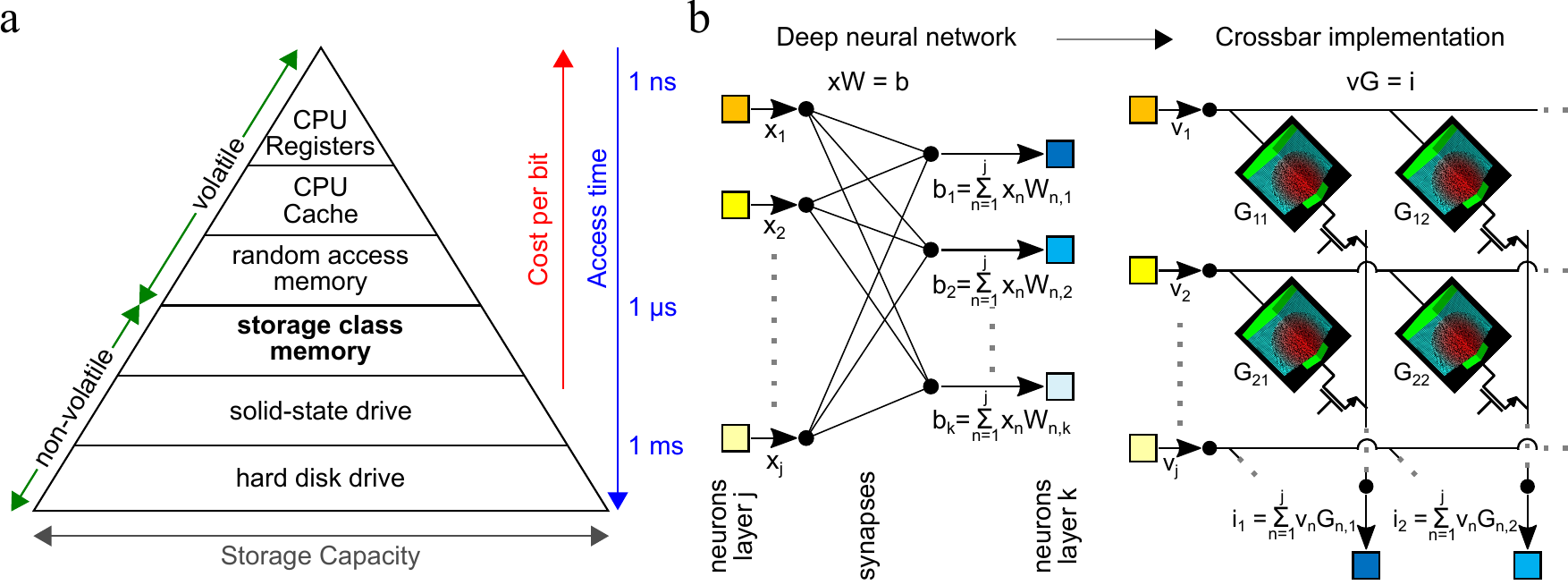}
	\caption[Phase change memory applications]{\textbf{Phase change memory applications. a}, Memory hierarchy. To achieve high compute performance and large memory capacity multiple memory technologies are combined.  PCM devices have been proposed as a potential technology for storage class memory. This new element in the memory hierarchy is supposed to bridge the gap in access time of almost three orders of magnitude between solid-state drives and random-access memory \cite{Burr2010,Fong2017}. \textbf{b}, Crossbar implementation of deep neural network computational workloads. The information propagation from one layer of a neural network to the next through the synapses is computed by a vector multiply operation. This computation can be executed on a resistive crossbar array. The synaptic weights (W) are mapped to the device conductance (G) and the inputs from layer j (x) are mapped to a read voltage (v). The column-wise summed crossbar currents (i) correspond to the product vector (b).}
	\label{fig:Intro_MemoryHierarchy}
\end{figure}

The ideal memory technology should combine high storage density (small feature size \& multi-level storage), low latency, i.e., fast read and write access, energy efficiency (non-volatile, small read \& write energy), high endurance, and low cost per bit. The perfect hardware, fulfilling all requirements at once, is yet to be invented. Thus, in today's computer architectures multiple different technologies are combined in the memory hierarchy (\Cref{fig:Intro_MemoryHierarchy}a). While tera-bytes of information are stored on hard drives, the capacity of random-access memories is on the order of tens of giga-bytes. Dependent on the timescales on which data must be accessible to the processor it is stored on different levels of the hierarchy and temporarily moved to higher levels. 

Introducing a new technology to the hierarchy is a particularly challenging endeavor. First, the new technology must match the performance of established hardware that has been extensively studied and optimized over years. Furthermore, it must be scalable and manufacturable with little infrastructural changes in the semiconductor fabrication plant to guarantee future competitiveness. 

Regardless of these obstacles Intel and Micron Technology introduced a PCM-based product to the market in 2015. In the 3D-XPoint\textsuperscript{TM} technology, arrays of binary PCM cells are stacked vertically \cite{Choe2017, Forbes_3DX, Intel2017}. The technology is designed to bridge the performance gap between solid-state drives and random-access memory. Despite the first productization with Intel, Micron announced to cease the development of 3D-XPoint\textsuperscript{TM} in 2021 due to an "insufficient market validation to justify the ongoing high levels of investments required to successfully commercialize 3D-XPoint\textsuperscript{TM} at scale" \cite{Micron_3DX_stop}. 

A second product, also making use of PCM, was presented by ST microelectronic and CEA-Leti at IEDM in 2018. In this case, a \unit[16]{MB} PCM array is integrated as embedded memory on automotive microcontrollers. Again, the PCM is presented as a binary memory \cite{Arnaud2019,STmicroelectronics2018}.     

Besides the continuous scaling of hardware components to smaller dimensions, the design of application-specific integrated circuits (ASICs) has become an important approach to improve compute performance. In particular, hardware dedicated to the training and inference of deep neural networks has been developed. Examples include the IBM Telum processor, the Google Tensor Processing Unit (TPU), and the AWS Inferentia \cite{Telum,GoogleTPU,Inferentia}.  

In this context, in-memory computing with quasi-analog PCM crossbar arrays is also being explored \cite{Burr2015,Nandakumar2018,Joshi2020,Khaddam-Aljameh2021,Narayanan2021,Nandakumar2021}. The key idea is to avoid shuttling the enormous number of network parameters between CPU and memory, by executing certain computations inside the memory. Specifically, the vector-matrix multiplication, a key operation required to process a deep neural network, is executed in-memory. To this end, the matrix of synaptic weights is mapped to the conductance of the devices and the activations are translated to the duration or amplitude of the voltage signal applied to the crossbar rows (\Cref{fig:Intro_MemoryHierarchy}b). Each PCM device executes the multiply operation (i\textsubscript{j,k} = G\textsubscript{j,k} $\cdot$ v\textsubscript{j}) and the currents of one crossbar column are summed (i\textsubscript{k} = $\sum_{n=1}^{j}$ i\textsubscript{n,k}). Here, G denotes the device conductance, v is the read bias signal, i is the current, j is the row index and k is the column index. 

Apart from these fairly matured applications for PCM, a multitude of alternative ideas have been discussed. For example, input signals can be accumulated and integrated. This is achieved by applying pulses that crystallize an amorphous device only gradually (c.f. \Cref{fig:Intro_Programming}a). Here, the device resistance encodes the information of how many signals and/or what signal amplitude have been applied. This approach has been used to emulate synaptic behavior \cite{Tuma2016} and find temporal correlations \cite{Sebastian2017b}. Additionally, concepts to implement bit-wise logical operations on resistive crossbar arrays have been proposed \cite{Siemon2015,Talati2016,Cheng2019,Giannopoulos2020}. This way, for example, database queries could be executed efficiently on memory crossbars.   

At present, PCM is a fairly matured technology and at least two promising applications, namely storage class memory and in-memory computing for deep neural network inference, have been identified. In the memory space, first products have entered the market, and inference of deep neural networks with millions of synaptic weights on fully integrated chips is actively explored. Yet, device non-idealities inherent to the amorphous state remain to be addressed. Most importantly, the multi-level storage capability is deterred by temporal changes of the amorphous state.

\section{Temporal changes of the amorphous state}
	
Key metrics of a PCM device are not stable but change systematically with the time elapsed after creating an amorphous state~\footnote{In most phase change material compounds only the amorphous phase is subject to drift. The crystalline phase has a time-invariant resistance.}. The resistance, encoding the stored information, and the threshold voltage, which must be overcome to rewrite the device, increase with log(time). This phenomenon is referred to as drift. At constant ambient temperature, the low-field (ohmic) resistance changes with 

\begin{equation}
	log(R) = log(R_0) + \nu_R \cdot log(t/t_0)
\end{equation} 

where R\textsubscript{0} denotes the resistance at time instance t\textsubscript{0} and $\nu$\textsubscript{R} is the resistance drift coefficient. The threshold voltage has been found to follow the empirical equation 

\begin{equation}
	V_{th} = V_{th,0} + \nu_{Vth} \cdot log(t/t_0)
\end{equation}  

where V\textsubscript{th,0} denotes the threshold voltage at t\textsubscript{0} and $\nu$\textsubscript{Vth} is the threshold voltage drift coefficient. 

The resistance drift coefficient of the amorphous \GST{} is on the order of 0.1 \cite{Wimmer2014b}. This means the resistance increases by 1.25x per order of magnitude in time. The threshold voltage increases at \unit[300]{K} ambient temperature by \unit[78]{mV} per order of magnitude in time \cite{Kersting2021}. Other materials like GeTe and doped Sb\textsubscript{2}Te exhibit slightly larger and smaller resistance drift coefficients, respectively \cite{Wimmer2014b}. 

Drift poses a significant challenge to device applications. Consider a scenario in which a PCM cell is operated as a 4-bit device. For this purpose, the dynamic resistance range is divided into 16 resistance bins. Eventually, at some time after programming, the device resistance will drift into another bin and the information is lost. The drift of a single device could be corrected relatively easily by updating the bounds of the resistance bins with time. However, such an approach is not feasible for large arrays in which the resistance of various devices changes at different rates. Drift coefficients are state-dependent, exhibit an inherent variability and cells may be programmed at different time instances. Additionally, the threshold voltage may drift beyond the highest voltages supported by the integrated circuit of a chip, making it impossible to reprogram the device. For this reason, drift remains one of the most pressing issues in PCM technology.

While the technological relevance of drift is one driving force for the extensive research, the scientific problem is of such richness and complexity, that the debate about the cause of drift is still ongoing and an answer to the question of how to mitigate drift is yet to be found. With few exceptions, the vast majority of theories attribute drift to structural relaxation of the amorphous phase. The amorphous phase is either in the supercooled liquid state or the glass state. The former denotes the disordered configuration with the lowest Gibbs free energy and the latter all other disordered configurations. Atomic vibrations and stochastic position changes of atoms result in a continuous change of the atomic structure towards arrangements with lower Gibbs free energy. Structural relaxation is inherent to the glass state and can be observed in any material. 

The main line of thought can be sketched as follows: With time the atomic arrangement evolves. These structural changes reflect on the electronic band structure. This in turn manifests in the drift of resistance and threshold voltage. In the following, the band structure of two archetype phase change materials (GeTe and \GST), its changes with time and the various theories proposed to explain these will be discussed. The section concludes with potential strategies to mitigate drift.
	
	\subsection{Evolution of the electronic band structure and its origins}
	\label{SubSect:EvolutionBandstructure}
	
\begin{figure}[bth!]
		\centering
		\includegraphics[width=1\linewidth]{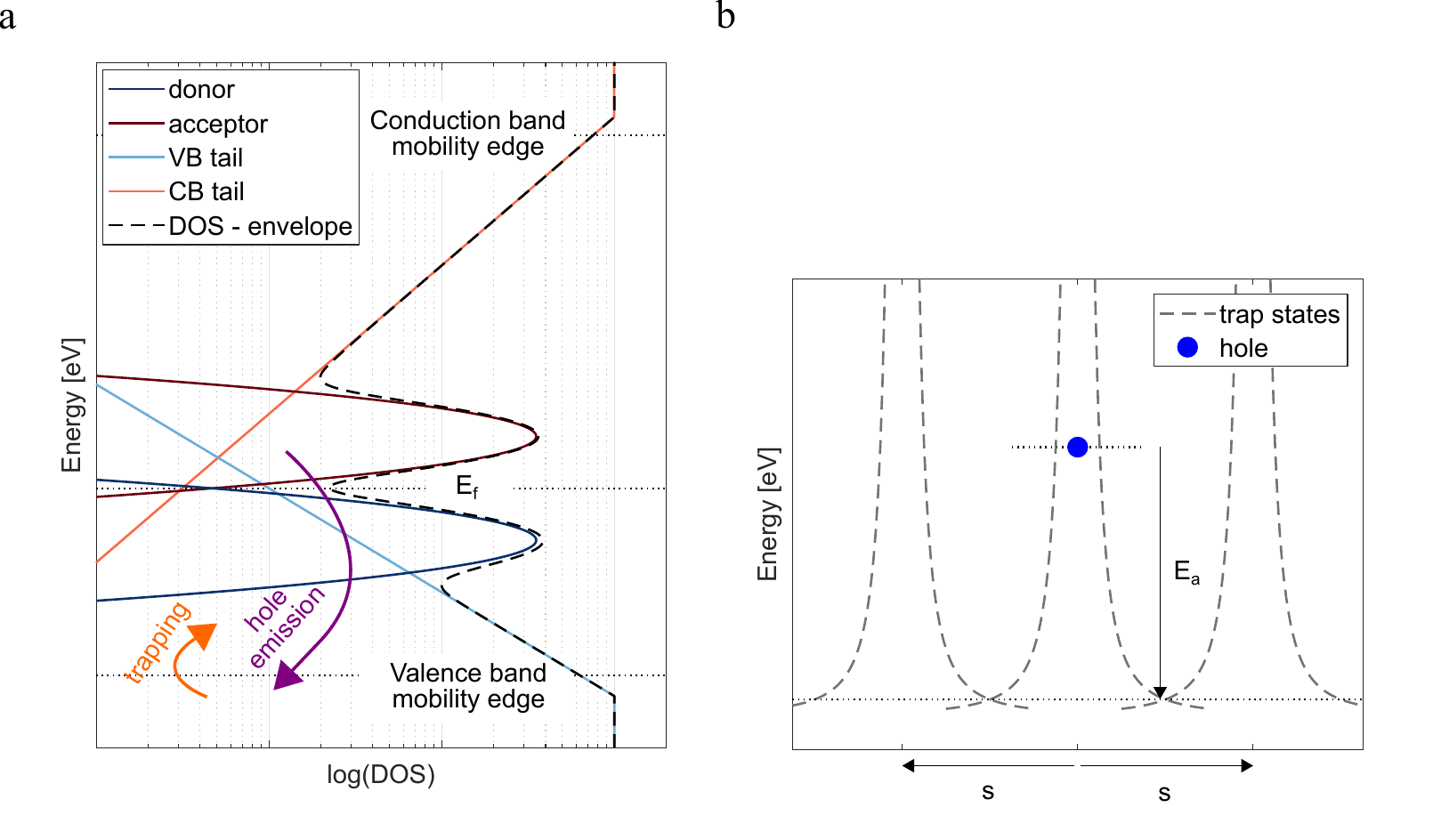}
		\caption[Electronic structure of the amorphous phase]{\textbf{Electronic structure of the amorphous phase. a}, Sketch of the density of states. The conduction and valence band exhibit extended tail states. The Fermi-level E\textsubscript{f}, pinned by a shallow donor and a deep acceptor state, is located closer to the valence band. \textbf{b}, Poole-Frenkel transport. Trapped holes must overcome the Coulomb potential of a negatively charged trap state to contribute to conduction. This activation energy E\textsubscript{a} is reduced by the electrostatic attraction of other negatively charged trap states in the local vicinity (s denotes the distance to a trap state that is not occupied by a hole).} 
		\label{fig:Intro_ElectronicStructure}
	\end{figure}

	Band structures with similar characteristic features have been proposed for amorphous GeTe and \GST{} (\Cref{fig:Intro_ElectronicStructure}a) \cite{Longeaud2012,Luckas2013}. Both conduction and valence band of the p-type semiconductors exhibit extended tail states. Photo-conductivity measurements \cite{Kaes2016, Krebs2014} and ab-initio simulations \cite{Konstantinou2019,Zipoli2016} suggest that the Fermi-level (E\textsubscript{f}) does not reside in the middle of the band gap but is shifted (notably) towards the valence band. Finally, deep acceptor and shallow donor states, potentially pinning the Fermi level, have been reported. While ab-initio simulations corroborate the presence of a deep acceptor state \cite{Konstantinou2019,Raty2015b,Zipoli2016}, they seem to show no clear indications of a shallow donor state. 
	
	Considering the discussed band structure, a variety of different changes and combinations thereof could result in a resistance increase. A wider separation of Fermi-Level and valence band mobility edge may be caused by a widening of the bandgap, a shift of the Fermi-level towards the middle of the bandgap or a reduction of the density of states in the valence band tail. Additionally, a decrease in the number of acceptor states could result in a reduced density of free charge carriers in the valence band. 
	
	The bandgap of amorphous phase change materials has been found to increase proportionally to the logarithm of time \cite{Rutten2015,Fantini2012}. It is generally assumed that this bandgap widening is caused by the structural relaxation of the amorphous phase. Different effects could be responsible for it. Namely, the release of compressive mechanical strain \cite{Karpov2007b} or the removal of certain structural defects \cite{Gabardi2015, Raty2015b}. 
	
	The hypothesis that a release of mechanical strain upon relaxation causes drift is appealing. It offers a universal mechanism applicable to any phase change material. In phase change nano wires, reduced drift coefficients with decreasing radius have been explained by reduced strain due to the large surface-to-volume ratio \cite{Mitra2010}. However, drift measurements of PCM cells with different degrees of built-in stress showed no strain-dependence of the drift coefficient \cite{Rizzi2011}. Also, experiments that measured the mechanical strain and resistance of blanket films over two orders of magnitude in time show that resistance drift and strain release don't follow the same time dependence \cite{Cho2012}. 
	
	Besides this widening or stretching of the band structure \cite{Krebs2012,Luckas2010} a shift of the relative position of the Fermi level has also been reported. It could result from a decay of the defect state density in the bandgap due to structural relaxation. A comparison of resistance drift measurements and the widening of the optical band gap suggests a shift of the relative position of the Fermi level towards the valence band in \AIST{} \cite{Rutten2015}. On the other hand, ab-initio simulations of GeTe indicate a shift of the Fermi level towards the bandgap center, while the bandgap remains constant \cite{Zipoli2016}. It should be noted, though, that other simulations \cite{Gabardi2015,Raty2015b} and experiments \cite{Rutten2015} indicate a bandgap widening and no shift of the Fermi level in GeTe. 
	
	In contrast to structural relaxation, an electronic mechanism that results in a shift of the Fermi level towards the middle of the band gap has also been discussed \cite{Pirovano2004a, Pirovano2004}. The authors postulated that Te\textsubscript{3}\textsuperscript{+} and Te\textsubscript{1}\textsuperscript{-} form valence-alternation-pair (VAP) defects~\footnote{The subscript denotes the coordination and the superscript the charge state}. During the RESET pulse, these charged defects get filled and form unstable Te\textsubscript{3}\textsuperscript{0} centers. The decay to the stable, charged state, causes a shift of the Fermi level back towards the mid gap. This theory, however, is challenged by numerous simulations showing that the defect states in GeTe and \GST{} are not simple VAP point defects \cite{Konstantinou2019,Raty2015b,Zipoli2016,Gabardi2015}.
	
	A second purely electronic mechanism, assuming a decay of the non-equilibrium electronic state created by the RESET process as well, has been proposed by Elliot \cite{Elliott2020}. The essential idea is that the defect states in the valence band are filled with electrons after the electrical excitation. Consequently, the density of holes in the valence band is increased. Over time electrons escape from the trap state and recombine with holes in the valence band. Consequently, the density of charge carriers (holes) reduces. However, such a mechanism can only explain the drift in a melt-quenched device. The author suggests that the drift observed in as-deposited amorphous samples at elevated temperatures must still be attributed to structural relaxation. 
	
	The last feature of the DOS remaining to be discussed is the change of defect states. To understand its particular importance we need to revise the transport mechanism in the amorphous phase. It has been identified as a trap-limited band transport with Poole-Frenkel emission \cite{Ielmini2007a,Gallo2015}~\footnote{Below 200 K a transition to variable-range hopping transport has been observed \cite{Krebs2014}}. Holes are emitted from negatively charged trap states into the valence band. The overlap of the Coulomb potential of an occupied trap state and its neighboring unoccupied trap state reduces the energy barrier (E\textsubscript{a}) that the hole needs to overcome to escape to free states (\Cref{fig:Intro_ElectronicStructure}b). It has been posed that the trap states could correspond to deep acceptor states in the band structure \cite{Gallo2015,Kaes2016}.  
	
	The ab-initio simulations of GeTe mentioned earlier consistently found a reduction of defect states upon relaxation. It has been attributed to structural changes in the material \cite{Raty2015b,Zipoli2016,Gabardi2015}. An intriguing model, linking resistance drift directly to the reduction of defect states, has been introduced by Ielmini et al. \cite{Ielmini2007c}. The spatial separation of Coulomb traps (s) in the Poole-Frenkel model is correlated with the density of (negatively charged) defect states (N\textsubscript{D}). As defect states vanish, the distance of neighboring Coulomb centers increases and their potentials overlap less. Thus, the activation energy a hole needs to overcome to escape the trap state increases. The observation that E\textsubscript{a} and s must increase upon relaxation to capture the change of the subthreshold current-voltage characteristic has been confirmed in succeeding studies \cite{Jeyasingh2011,LeGallo2018}. 
	
	Due to the uncertainty about the band structure of the amorphous material and its changes with time, resistance drift is described phenomenologically with the equation  
	
	\begin{equation}
		R(T,t) = R^*(T,t) \cdot exp(\frac{E_a(T,t)}{k_bT}).
	\end{equation}
	
	The aforementioned changes of the band structure can increase the activation energy for conduction E\textsubscript{a} and the pre-exponential factor R\textsuperscript{*}.  While in \GST{} only E\textsubscript{a} increases, both E\textsubscript{a} and R\textsuperscript{*} have been found to drift in GeTe and doped Sb\textsubscript{2}Te \cite{Wimmer2014b, Boniardi2009a, Oosthoek2012}. 
	
	In summary, a multitude of different and sometimes contradicting changes of the band structure with time have been reported. A fairly wide consensus has been reached, that the band gap widens upon relaxation and the density of defect states in the band gap reduces. Some experiments indicate that dependent on the material, the position of the Fermi-level may shift. These changes have been attributed to an alteration of the atomic configuration or a reduction of compressive strain, both resulting as a consequence of structural relaxation. Additionally, it has been proposed that the RESET process creates an electronic non-equilibrium occupation of defect states. Their slow decay to equilibrium could also result in resistance drift. The author's view is that the changing atomic configuration is the key process causing drift, but mechanical strain release and the decay of defect occupation could be additional, second-order, effects. 
		
	\subsection{Strategies to mitigate drift} 
	\label{Sect:MitigateDrift}
	A few ideas on how to mitigate drift or its impact on compute precision have been proposed. Here, they are subdivided into the categories of new materials, nano-confinement, new device concepts, and drift compensation.   	
		
		\begin{itemize}
			\item	\textbf{New materials:}  A significant effort has and still is being made to find better performing phase change materials. Studies range from experimental to simulated approaches and in recent years first demonstrations of simulation-guided material design have been presented \cite{Rao2017}. New materials can be either entirely new compounds, or established compounds with dopants, such as Ag, In, or N. 
			
			One approach to reduce drift could be to reduce the concentration of structural defects in the amorphous phase. It might allow minimizing the changes of the band structure upon relaxation. This mechanism has been proposed to explain the reduction of drift coefficients measured in GeTe when \unit[25]{\%} and \unit[50]{\%} are replaced by SnTe \cite{Luckas2013a,Chen2019}. In GeTe, drift has structurally been attributed to a reduction of homopolar Ge-Ge bonds and the concomitant tetrahedral structures \cite{Gabardi2015,Zipoli2016} and an increase of the Peierls distortion \cite{Raty2015b}. Due to a higher energy penalty, homopolar bonds and tetrahedral defects occur less frequently in amorphous SnTe. Additionally, contrary to the GeTe bonds, SnTe bonds show no Peierls distortions \cite{Chen2019}. Overall, fewer structural defects are observed in Ge\textsubscript{3}Sn\textsubscript{1}Te\textsubscript{4} and Ge\textsubscript{1}Sn\textsubscript{1}Te\textsubscript{2}. Instead of reducing the number of structural defects, it might also be possible to stabilize them. I.e., making them less prone to relax structurally. It has been speculated that the drift reduction reported for Nitrogen-doped \GST{}  \cite{Li2018} could be attributed to the formation of N-Ge bonds, hindering the relaxation \cite{Zhang2020}. However, it must be noted that both studies were performed on as-deposited amorphous samples \cite{Luckas2013a,Li2018}, which requires very careful analysis to not underestimate a materials drift coefficient \cite{Wimmer2014b}.  
			
			Another option could be to search for materials in which the changes of the band structure upon structural relaxation reflect little on R\textsuperscript{*} and E\textsubscript{a}. In \AIST, for example, the activation energy for conduction increases 4x less than the optical bandgap. It exhibits a smaller drift coefficient than GeTe and \GST, in which E\textsubscript{a} increases proportionally to E\textsubscript{g} \cite{Rutten2015, Wimmer2014b}. 
			
			Finally, it could be interesting to study materials in which the structural defects differ from traditional phase change materials. In the crystalline phase of \GST, GeTe, and \SbTe{} no Ge-Ge, Ge-Sb, and Sb-Sb bonds exist. These wrong bonds, present in the amorphous phase, are expected to vanish upon relaxation \cite{Gabardi2015}. In a pure material, wrong bonds must be of a different nature. This might also affect the resistance drift.
			
			\item	\textbf{Nano-confinement:} In the close vicinity of a rigid interface, atomic movements can be hindered and restricted. For example, the crystal growth rate in metallic glass nanorods decreases notably in constricted regions \cite{Sohn2015}. Additionally, the apparent viscosity increases dramatically, upon confinement below \unit[100]{nm} \cite{Shao2013}. This deceleration of crystallization kinetics and material flow upon deformation, i.e., certain atomic rearrangement occurring less frequently, should also reflect on the structural relaxation dynamics. Accordingly, confinement might be another approach to mitigate drift. 
			
			Crystallization studies of phase change material thin films give an idea of the length scales at which confinement effects begin to play a role. Numerous studies found deviations from the bulk material crystallization behavior for film thicknesses below a few tens of nanometers \cite{Raoux2008,Raoux2009, Cheng2010}. If confinement indeed affects drift, a drift reduction would be expected for a comparable degree of confinement. In fact, for a device with Sb-rich Ge\textsubscript{x}Sb\textsubscript{y}Te\textsubscript{z}, confined to a cross-section of \unit[7.5 $\cdot$ 17]{nm\textsuperscript{2}}, a drift coefficient of only 0.011 was measured \cite{Kim2010a}. While it is not mentioned how this value compares to the unconfined material it is a remarkably small drift coefficient. 
		
			\item	\textbf{New device concepts:} Since optimizing a material with respect to property A often implicates compromises on properties B and C, concepts to mitigate drift through novel device concepts gained traction, for example, the projected memory device \cite{Redaelli2010,Kim2013,Koelmans2015,Bruce2021} and the phase change heterostructure \cite{Ding2019}. 
			
			The idea of a projected memory device is to introduce a shunt resistor in parallel to the amorphous volume. The shunt resistor scales proportionally to the size of the amorphous volume. But the read current bypasses the amorphous phase. Thus, the device resistance is determined by the size of the amorphous volume, but the read current is decoupled from its resistance drift. 
			
			The phase change heterostructure resembles a mushroom cell. But instead of one thick phase change material film, alternating thin layers of \SbTe{} and TiTe\textsubscript{2} are deposited on the bottom electrode heater. These devices show at least a 10x smaller drift coefficient than the regular \SbTe{} reference device. This drift reduction was attributed to the \SbTe{} confinement by the TiTe\textsubscript{2} layers. It is assumed that structural rearrangements are restricted and an increase of the Peierls distortion, causing the band gap widening, is suppressed near the interfaces \cite{Ding2019}.    
			
			\item \textbf{Drift compensation:}	Besides improvements of the phase change material or device concept, the issue of drift can also be tackled by suitable device operation schemes. 
			
			At the beginning of this section, we saw that the low-field (ohmic) resistance has a power-law time dependence, whereas the threshold voltage shows a much weaker log(t) dependence. While the subthreshold high-field resistance still follows a power-law time dependence, its drift coefficient is notably smaller than in the ohmic regime. To reduce the impact of drift on the measured cell state it has been proposed to read the device at elevated fields, e.g., by defining a high, but subthreshold, read current (\unit[1]{$\mu$A}) \cite{Ielmini2009b,Sebastian2011}. It should be noted that this approach increases the power consumption of the readout. 
			
			Another approach, suitable for classical memory applications, is to combine multiple PCM devices into a codeword \cite{Papandreou2011}. The codeword information is defined by the relative order of the devices' resistance levels. It is resilient to drift, as long drift variability does not alter the resistance order. Importantly, not the array level drift variability is critical, but only the variability within the codeword.
			
			In the field of in-memory computing, a multiply-accumulate operation is executed on a crossbar column. The resistance of individual devices is not retrieved but drift can be corrected at the array level. It is compensated with a single scaling factor for the crossbar array. The correction is computed by periodically measuring the summed current I\textsubscript{array} for a defined input vector. The ratio of the initially measured value and the current at time instance t defines the scaling factor \mbox{A = I\textsubscript{array}(t\textsubscript{0})/I\textsubscript{array}(t)} \cite{LeGallo2018a,Joshi2020}. Like this, the mean drift of the array can be compensated, but still the compute precision degrades with time due to drift variability. 
			
		\end{itemize}

\section{Scope and structure of the thesis}

The question of how drift can be eliminated or at least substantially mitigated is still one of the most pressing ones in PCM technology. Drift limits the multi-level storage capability, leads to a loss of the stored information, and results in a degradation of the in-memory compute precision. The goal of this work is to broaden the understanding of how drift can be mitigated at the device level. In particular, the prospects and potential pitfalls of nano-confinement and the projected memory concept were studied. 

In a first step, drift is measured under unprecedented conditions, specifically at temperatures down to \unit[100]{K} and timescales as short as \unit[30]{ns}. These experiments distinctly show a time regime in which drift is absent and the transition to a continuous log(t) dependence. The new insights about the onset of drift are used to revise two phenomenological models, describing the structural relaxation dynamics of the amorphous phase. Namely, the Gibbs relaxation model and the collective relaxation model. \Cref{Chapt:Relax} concludes with a gedankenexperiment about the implications of extreme confinement for the Gibbs relaxation model. 

\Cref{Chapt:Confine} is dedicated to the consequences of device scaling and extreme material confinement. Anticipating its potential impact not only on drift but also on the crystallization dynamics, a new phase change material is selected for these studies. Despite its extreme proneness to crystallization, pure antimony can be melt-quenched into an amorphous state in a confined device structure. One-dimensional confinement, down to \unit[3]{nm}, makes it possible to study the antimony's phase change properties and the effect of confinement on device efficiency, stability against crystallization and drift.  

Next, the drift in a projected antimony device is characterized. Compared to previous studies of projected devices the measuring range was expanded by four orders of magnitude in time. Once again, an extension of the measurement range provides vital insights. Importantly, these measurements reveal that resistance drift no longer obeys a simple power-law in a projected device. To describe the state-dependent temporal evolution of the device resistance an equivalent circuit model is introduced. Here, the hitherto overlooked critical role of the interface between phase change material and projection layer becomes apparent.
  
The thesis concludes with a revision of the different strategies to mitigate drift. Based on the findings of this work it is assessed how drift could be reduced in future PCM devices. Furthermore, the drift reduction mechanism proposed for the phase change heterostructure device is discussed \cite{Ding2019}. The results of this thesis indicate that other effects than the originally proposed mechanism might play a key role in explaining the remarkably low drift coefficients measured in this device.

		\chapter{The origin of structural relaxation}
\label{Chapt:Relax}

\textbf{Preliminary remark:} The results presented in this chapter have been published as a journal article. \newline
\textit{Measurement of onset of structural relaxation in melt-quenched phase change materials - Kersting et al. - Advanced functional materials \cite{Kersting2021}} \\[6pt]

\begin{figure}[htb]
	\centering
	\includegraphics[width=0.6\linewidth]{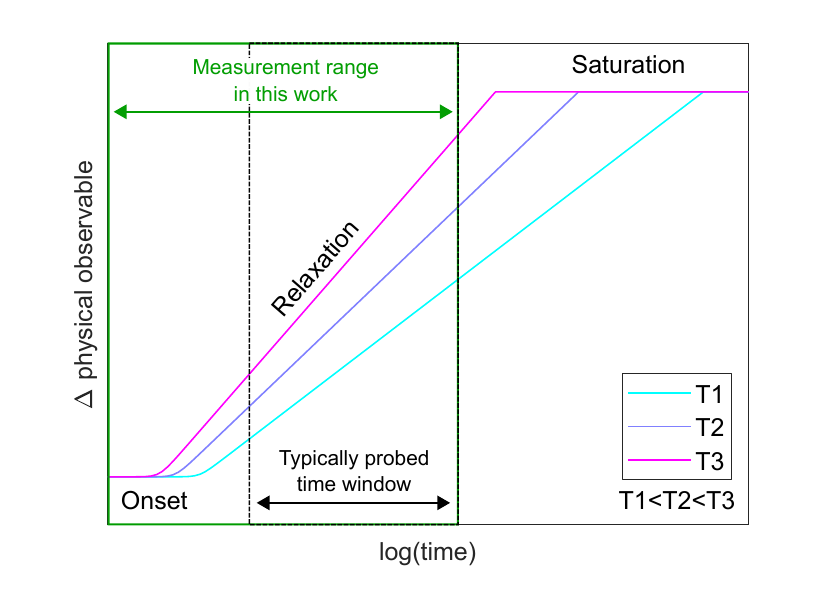}
	\caption[Sketch of the temporal evolution of material properties upon structural relaxation]{\textbf{Sketch of the temporal evolution of material properties upon structural relaxation.} Previous studies of phase change materials have been limited to timescales and temperatures in which drift obeys a log(t) dependence. In this work, the measurement range was expanded to the onset of relaxation and study its temperature dependence. Since relaxation is thermally activated, the onset and saturation of drift shift to shorter timescales with increasing ambient temperature.}
	\label{fig:RELA_Sketch3Regimes}
\end{figure}

When a liquid is quenched faster than its critical cooling rate, crystallization events can be overcome and atoms can structurally freeze into a disordered solid state \cite{Turnbull1969,Zhong2014a,Salinga2018}. Upon cooling the melt, the atomic mobility decreases. Eventually, the system can no longer assume the equilibrium structure of the supercooled liquid state within the timescale of the experiment, and a non-equilibrium glass state is created. The free energy difference between the super-cooled liquid and glass state results in structural relaxation, where the atomic configurations in the glass change over time. The intrinsic material properties, including viscosity, density, and the electronic bandgap, change due to relaxation \cite{Angell2000b,Priestley2005a}, and this is understood to occur in three phases (\Cref{fig:RELA_Sketch3Regimes})\cite{Chen2007,McKenna2003,Priestley2009}. For every rearrangement, a finite energy barrier must be overcome, and therefore, for some amount of time, during the onset phase, the properties do not change. In the second phase, where relaxation is most profound, the properties have been observed to change proportionally to log(t). Finally, approaching the supercooled liquid, the glass reaches a saturation phase, and the properties no longer continue to change. 

Tracking this structural relaxation process through all three phases is experimentally challenging. In a PCM device, this might be done by resistance and threshold voltage drift measurements. Although drift has been studied extensively on timescales ranging from hundreds of $\mu$s to \unit[3]{months}, all these experiments only revealed the log(t) dependence of resistance and threshold voltage \cite{Gorbenko2019,LeGallo2018,Salinga2018,Bruce2021}. A stand-alone sub-\unit[100]{ns} drift measurement by Ielmini et al. on \GST{}  shows a hint of the region where drift is absent \cite{Ielmini2007}. What remains to be investigated both qualitatively and quantitatively are the other two phases, namely, when does structural relaxation begin and when does it end? To this end, measurements that capture the relaxation from extremely short to long timescales at different temperatures are required. 

In the following, more light will be shed on the phase where relaxation is absent. Specifically, the goal is to quantify on what timescales the commonly observed log(t)-dependent drift begins. We employ \Vth{} as a means to observe the state of relaxation. The \Vth{} drift of \GST{} mushroom cells is probed at temperatures ranging from \unit[100]{K} to \unit[300]{K} over 9 orders of magnitude in time (tens of ns to \unit[10]{s}). The experimental data are fitted with two models, that have been used to describe structural relaxation and drift in phase change materials, namely the Gibbs model \cite{Gibbs1983,Karpov2007b,Ielmini2008a} and the collective relaxation model \cite{Knoll2009,Sebastian2015,Gallo2016,LeGallo2018}. Finally, it is discuss how the collective relaxation model should be modified, if structural relaxation is facilitated by many-body thermal excitations and the consequences of a limited number of structural defects for the relaxation dynamics.  

\section{State of the art}
		
	\subsection{Gibbs relaxation model}
	
	The relaxation model introduced by Gibbs in the 1980s describes the relaxation process by a spectrum of defect states with different activation energies (q(\Ed)). Each structural defect is described as a double-well potential with activation energy E\textsubscript{d}. Assuming the defect relaxation is thermally activated and obeys a first-order rate equation, the density of defect states changes with time as
	
	\begin{equation}
		\frac{dq(E_\text{d},t)}{dt} = -\nu_0 \cdot \exp\left(\frac{-E_\text{d}}{k_b T}\right) \cdot q(E_\text{d},t).
		\label{equ:RELA_DifferentialGibbs}
	\end{equation}
	
	The attempt to relax frequency $\nu$\textsubscript{0} is estimated on the order of the Debye frequency (\unit[10\textsuperscript{12}]{s\textsuperscript{-1}}) but could change if groups of atoms are involved in a relaxation process \cite{Gibbs1983}. Since the probability of relaxation depends exponentially on E\textsubscript{d}, defects with small E\textsubscript{d} relax first and only defects in a narrow range of activation energies relax at the same time. The activation energy that must be overcome for further relaxation increases monotonically. 
	
	Structural changes in the material, i.e., the relaxation of defects, reflect on its physical properties (P). The total change is given by 
	
	\begin{equation}
		\Delta P = \int_{0}^{E} c(E_d) \cdot q(E_d)~dE
		\label{equ:RELA_GibbsPropertyChange}
	\end{equation}
	
	with the coupling factor c(E\textsubscript{d}). A physical property does only change proportional to log(t) if $\mathrm{c(E_d) \cdot q(E_d)}$ is constant does . In general, the coupling factor could depend on the type of defect or relaxation process \cite{Gibbs1983}. Nevertheless, often times a constant coupling factor between structural relaxation and physical properties is assumed \cite{Friedrichs1989,Shin1993,Khonik2008,Knoll2009}, also in the field of phase change materials \cite{Karpov2007b,Ielmini2008a,LeGallo2018}. Making this assumption, it is possible to derive q(E\textsubscript{d}) from relaxation experiments. In this model, the relaxation onset is determined by the defect states with the lowest E\textsubscript{d} and the finite time it takes to overcome it. The saturation emerges when the super-cooled liquid state is approached. 
	
	Two models, based on these structural relaxation dynamics, have been proposed to describe drift in phase change materials. On the one hand, Karpov et al. assume that the relative volume change (dilatation), affecting \Vth{} and R, is proportional to the number of unrelaxed defect states  \cite{Karpov2007b}. On the other hand, Ielmini et al. argue that the concentration of localized electronic states decreases concomitant with the density of structural defects \cite{Ielmini2007c,Ielmini2008a} (see also \Cref{SubSect:EvolutionBandstructure}). 
	
	The major critique against this relaxation model questions the plausibility of a uniform $\mathrm{c(E_d) \cdot q(E_d)}$ (or $\mathrm{q(E_d)}$, assuming $\mathrm{c(E_d)}$ is constant), necessary to explain the log(t)-dependent change of physical properties \cite{Knoll2009,LeGallo2018}. Interestingly, a wide variety of phenomena in all kinds of glasses can be described by double-well potentials with uniformly distributed activation energies \cite{Karpov2007b}. This might be considered a convenient justification to assume a uniform distribution but by no means provides a physical explanation. Anderson noticed: "Why the statistics of these barriers is so uniform over a wide spectrum of types of glasses remains a mystery" \cite{Anderson1972}. In the context of structural relaxation, it might be explained by the idea that a relaxing defect does not simply vanish but could create new defective configurations with a higher E\textsubscript{d}. Already Gibbs concluded that the activation energy spectrum evolves with time and the spectrum deduced from relaxation measurements includes defects, that arose during the relaxation process \cite{Gibbs1983}. This concept is taken a step further and formulated more explicitly in the collective relaxation model. 
	
	\subsection{Collective relaxation model} 
	
	The collective relaxation model does not specify individual structural defects or relaxation states. Instead, the material is assumed to relax globally towards an ideal glass state \cite{Knoll2009}. With progressing relaxation, the whole system becomes structurally more stabilized. Larger groups or clusters of atoms relax collectively and the activation energy E\textsubscript{b} = E\textsubscript{s}$\cdot$(1-$\Sigma$) of these processes increases monotonically. It depends linearly on the state of relaxation $\Sigma$. $\Sigma$ = 1 is an infinitely unrelaxed state and $\Sigma$\textrightarrow 0 means that the system approaches equilibrium. E\textsubscript{s} is the activation energy as the system approaches equilibrium. Assuming structural transitions 'backward' to a less relaxed state can be neglected, the temporal evolution of the state variable $\Sigma$ is captured in the rate equation
	
	\begin{equation}
		\label{equ:RELA_DifferentialSigma}
		\frac{d\Sigma(t)}{dt}= - \nu_0 \Delta_\Sigma \cdot\exp\left(\frac{-E_\text{s}\cdot(1-\Sigma(t))}{k_\text{b}T}\right).
	\end{equation} 
	
	The attempt to relax frequency $\nu$\textsubscript{0} is expected to be on the order of phonon frequencies and $\Delta$\textsubscript{$\Sigma$} is the change of $\Sigma$ with each relaxation process. Consequently, $\Delta$\textsubscript{$\Sigma$}$\cdot$E\textsubscript{s} defines the increase of the activation energy for each subsequent relaxation step. For a constant ambient temperature, the differential equation can be solved analytically as 
	
	\begin{equation}
		\label{equ:RELA_SigmaOftime}
		\Sigma(t,T) = -\frac{k_\text{b}T}{E_\text{s}} \cdot \log\left(\frac{t+\tau_0}{\tau_1}\right).
	\end{equation}
	
	Here the relaxation onset of the initial glass state $\Sigma_0$  is defined by 
	
	\begin{equation}
		\label{equ:RELA_tau0}
		\tau_0 = \frac{k_\text{b}T}{\nu_0\Delta_\Sigma E_\text{s}} \cdot \exp\left(\frac{E_\text{s}\cdot(1-\Sigma_0)}{k_\text{b}T}\right)
	\end{equation}
	
	and $\mathrm{\tau_1 = \frac{k_\text{b}T}{\nu_0\Delta_\Sigma E_\text{s}} \cdot  \exp(\frac{E_\text{s}}{k_\text{b}T})}$ is the time at which the system reaches equilibrium. In the range $\mathrm{\tau_0<<t<\tau_1}$ the temporal evolution of $\Sigma$ follows log(t) and the change of $\Sigma$ depends linearly on the ambient temperature. 
	
	It is worth noting, that the two models resemble each other and are constructed similarly. In both models, the structural relaxation is described by a first-order rate equation with an Arrhenius temperature dependence on the activation energy. Even though the Gibbs model defines a defect distribution function q(E\textsubscript{d}), at any time the further progress of relaxation is limited by defects with activation energy in a narrow range, and effectively the activation energy of relaxing defects continuously increases, like in the collective relaxation model. 
	
	The defect distribution function of the Gibbs model, however, is a misleading picture. These defect states with different activation energies do not exist at the same time, they are merely an integration over time of the processes that limited the progression of relaxation. Already in the 1950s, it was conceived that relaxing defects can create new defect states with higher activation energy \cite{Primak1955,Gibbs1983}. The collective relaxation model appears to describe this transition of the system between states with gradually increasing activation energy more naturally.
		
	\subsection{Research questions}
	\begin{enumerate}
		\item	Is it feasible to probe the relaxation onset of a melt-quenched PCM device?
		\item	Can the Gibbs model and the collective relaxation model capture the drift dynamics with time (tens of ns to \unit[10]{s}) and Temperature (\unit[100]{K} to \unit[300]{K})?
		\item	How can the models be challenged further and when/why might they break down?
	\end{enumerate}
	\newpage
	
\section{Threshold voltage drift}
	\label{Sect:VthDrift_Experiment}
	
	Because structural relaxation processes are thermally activated, ambient temperature can be used as a parameter to shift the onset and saturation of drift to experimentally accessible timescales (see \Cref{fig:RELA_Sketch3Regimes}). However, the observation of the saturation phase by raising the ambient temperature (\textgreater \unit[400]{K}) is prohibited due to potential recrystallization of the amorphous phase. On the other hand, there is a potential for measuring the onset of drift by monitoring it at lower ambient temperatures. The challenge, however, is the inability to reliably measure electrical resistance at short timescales and low temperatures. Hence, \Vth{} is used as a means to observe the state of relaxation. 
	
	The study was conducted on \GST{} mushroom cell devices (\Cref{fig:Intro_DeviceConcepts}b), fabricated in the IBM AI hardware center. The multi-layer ring bottom electrode with a radius of $\sim$\unit[20]{nm} and a height of $\sim$\unit[40]{nm} was patterned with a sub-lithographic hardmask process. The sputter-deposited \GST{} film is $\sim$\unit[75]{nm} thick. An on-chip series resistor ($\sim$\unit[2]{k$\Omega$} at \unit[300]{K}), limits the device current at the moment of threshold switching. To assure a stable device operation throughout this study the cells were cycled at least 100,000 times in advance. All experiments presented in this thesis were performed in a JANIS ST-500-2-UHT cryogenic probe station (\Cref{Sect:ProbeStation}). 
	
	The \Vth{} drift is probed, by programming \GST{} mushroom cell devices (\Cref{fig:Intro_DeviceConcepts}b) repeatedly to a new RESET state and applying SET pulses with varying delay times (\Cref{fig:RELA_Experiment_VthDrift}a). The pause t\textsubscript{delay} before applying the SET pulse is increased in logarithmically spaced steps. This sequence with t\textsubscript{delay} ranging from \unit[10]{ns} to \unit[10]{s} is repeated 15 times to average out drift and threshold switching variability. Note that each \Vth{} measurement results in the erasure of the corresponding RESET state by recrystallization. The measured \Vth{} drift represents an averaged behavior of RESET states created in the device. Details on the algorithm to define \Vth{} are provided in \Cref{Sect:VthAlgorithm}. 
	
	\subsection{Evolution in three regimes}
	
	\begin{figure}[tbh]
		\centering
		\includegraphics[width=1\linewidth]{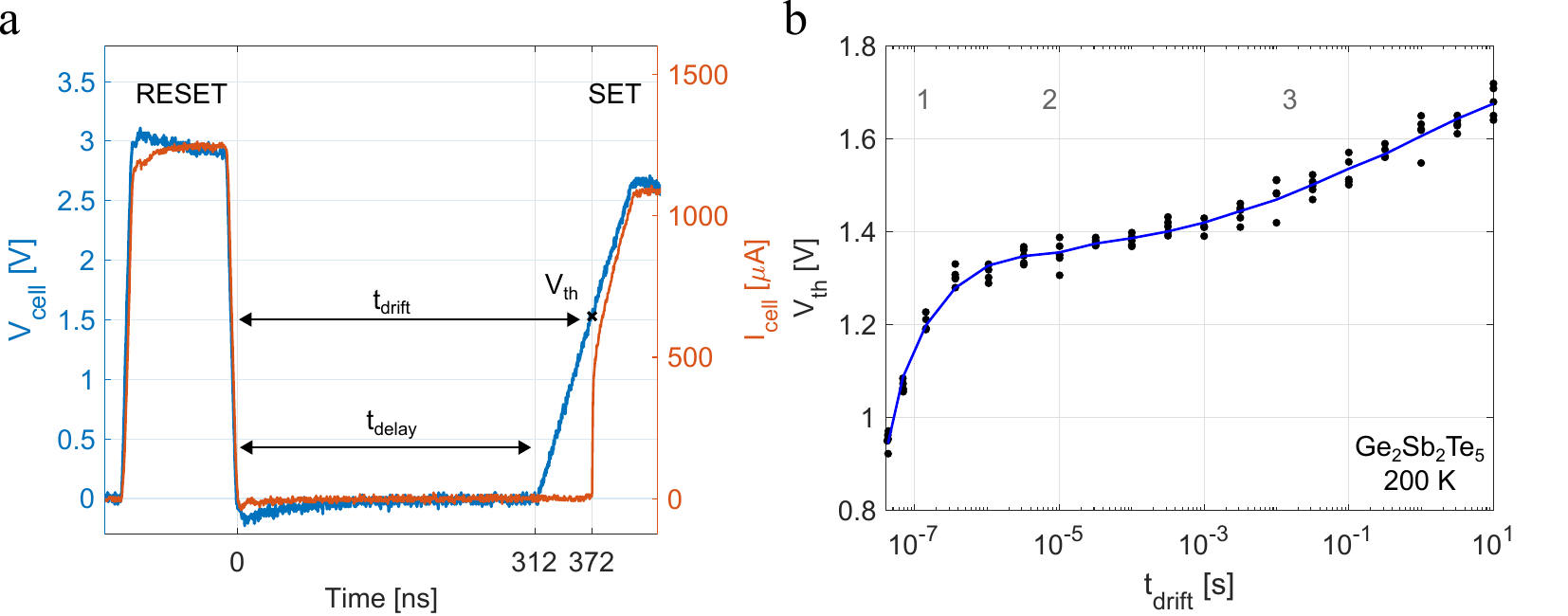}
		\caption[Threshold voltage drift experiment]{\textbf{Threshold voltage drift experiment. a}, Programming scheme. A RESET pulse with a \unit[3]{ns} falling edge programs the device to a melt-quenched amorphous state. A SET pulse with a \unit[100]{ns} leading edge is applied to probe the threshold voltage value. In the example shown here, the device drifts for \unit[372]{ns} before threshold switching occurs at \unit[1.5]{V}. The time interval (t\textsubscript{delay}) between the RESET and SET pulses is varied to probe the temporal evolution. \textbf{b}, V\textsubscript{th} drift in \GST. The \Vth{} evolution exhibits three distinct regimes (labeled in grey). First a steep increase up to \unit[$\sim$1]{$\mu$s}, second, a flattening to almost constant values, and third a transition to a monotonic increase approximately proportional to log(t).}
		\label{fig:RELA_Experiment_VthDrift}
	\end{figure}
	
	Three distinct regimes are apparent in the temporal evolution of \Vth{}  (\Cref{fig:RELA_Experiment_VthDrift}b). In regime~1, up to \unit[$\sim$1]{$\mu$s}, there is a steep increase of \Vth. This rapid change is most likely caused by the decay of the RESET excitation. On the one hand, a large number of excess charge carriers is generated when the device is molten at high fields, on the other hand, the temperature is locally raised to more than \unit[900]{K} before the applied power is switched off within \unit[3]{ns} (RESET pulse trailing edge). Both a decay of the excess charge carriers \cite{Elliott2020,Ielmini2007} and slow decay of the local temperature when the device is already close to ambient temperature would result in a continuous increase of the threshold voltage. Regime~2 shows a flattening of the curve and almost constant threshold voltage values. Finally, a continuous linear increase with log(t) is observable in regime~3. The temporal evolution in regimes~2 and 3 can be attributed to the structural relaxation of the amorphous phase with the transition between them marking the onset of relaxation. 
	
	Careful inspection of the data presented in \cite{Ielmini2007} shows the same three regimes. While a continuous drift with log(t) is observed from about \unit[$\mathrm{10^{-5}}$]{s} in both studies, regime~1 occurs at different timescales. In this study, it lasts up to \unit[1]{$\mu s$} and in the work by Ielmini et al. it ends after approximately \unit[30]{ns}. This could be attributed to the different device geometry and size (mushroom cell with a bottom electrode \unit[$\sim$706]{nm$^2$} vs. $\mu$-trench with a bottom electrode \unit[\textgreater 1500]{nm$^2$}) and the resulting difference in the thermal environment. Additionally, in \cite{Ielmini2007}, the bias of the RESET pulse is reduced to a lower level, sufficient to melt-quench and induce threshold switching, but it is not turned off between RESET and threshold switching. The experiment probes the decay time to a partially excited state (I \unit[$\sim$150]{$\mu$A}) defined by the bias applied at the end of the RESET pulse. Thus, the different bias scheme, where the bias is completely switched off at the end of the RESET pulse, is another potential reason why regime~1 lasts shorter.
	
\begin{figure}[htb]
		\centering
		\includegraphics[width=1\linewidth]{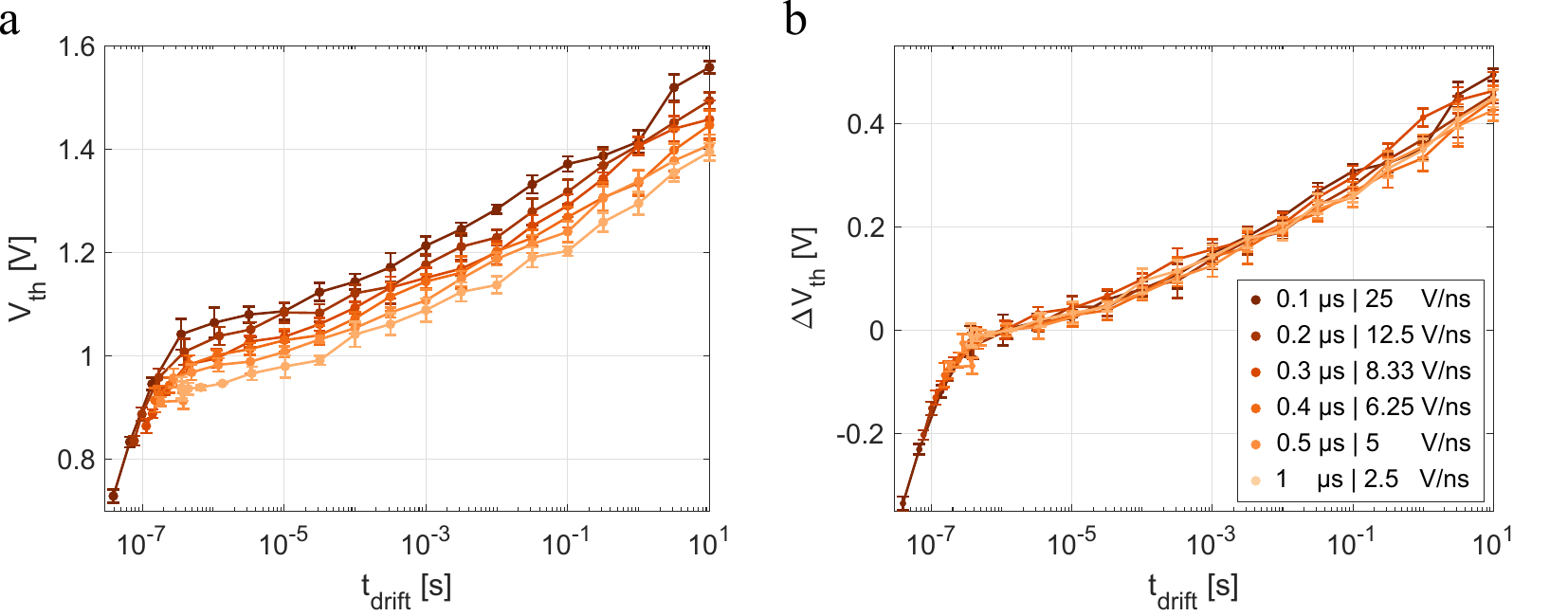}
		\caption[Impact of the SET pulse leading edge duration]{\textbf{Impact of the SET pulse leading edge duration. a}, With an increasing SET pulse leading edge the threshold voltage values decrease continuously. The value of t\textsubscript{drift} is the sum of t\textsubscript{delay} (the delay between the RESET pulse and the SET pulse) and the time elapsed until the applied voltage crosses V\textsubscript{th}. The second term changes with the SET pulse leading edge. The resulting change of t\textsubscript{drift} is most apparent on short timescales, where t\textsubscript{delay} is small. \textbf{b}, The threshold voltage changes with respect to a reference point, which is here the value measured \unit[1]{$\mu$s} after RESET, is independent of the SET pulse leading edge duration. In the legend, the leading edge duration and its slope are noted. All SET pulses have an amplitude of \unit[2.5]{V}.}
		\label{fig:RELA_Vth_SET_LeadingEdge}
	\end{figure}

	In the here presented experiments, threshold switching is induced by biasing the device with a trapezoidal voltage pulse. \Vth{} does change depending on the transient signal applied to the device \cite{Wimmer2014a,LeGallo2016}. With decreasing slope of the SET pulse leading edge \Vth{} decreases (\Cref{fig:RELA_Vth_SET_LeadingEdge}a). Both, the electrical stress, and joule heating in the device prior to switching may affect the relaxation dynamics. However, the absolute change of \Vth{} with time appears to be independent of the leading edge (\Cref{fig:RELA_Vth_SET_LeadingEdge}b). This suggests that the rise of the cell bias is fast enough to not notably alter the relaxation process of the glass state, prior to switching. First, the time for which the cell is biased before switching is short on absolute timescales. Second, it is at least in regimes~2 and 3, which are governed by the relaxation dynamics, much shorter than the time for which the material relaxes without any bias being applied.  
	
	\newpage
	\subsection{Temperature-dependent threshold voltage evolution}
	\label{Sect:VthDrift_Temperature}
	
\begin{figure}[htb]
		\centering
		\includegraphics[width=1\linewidth]{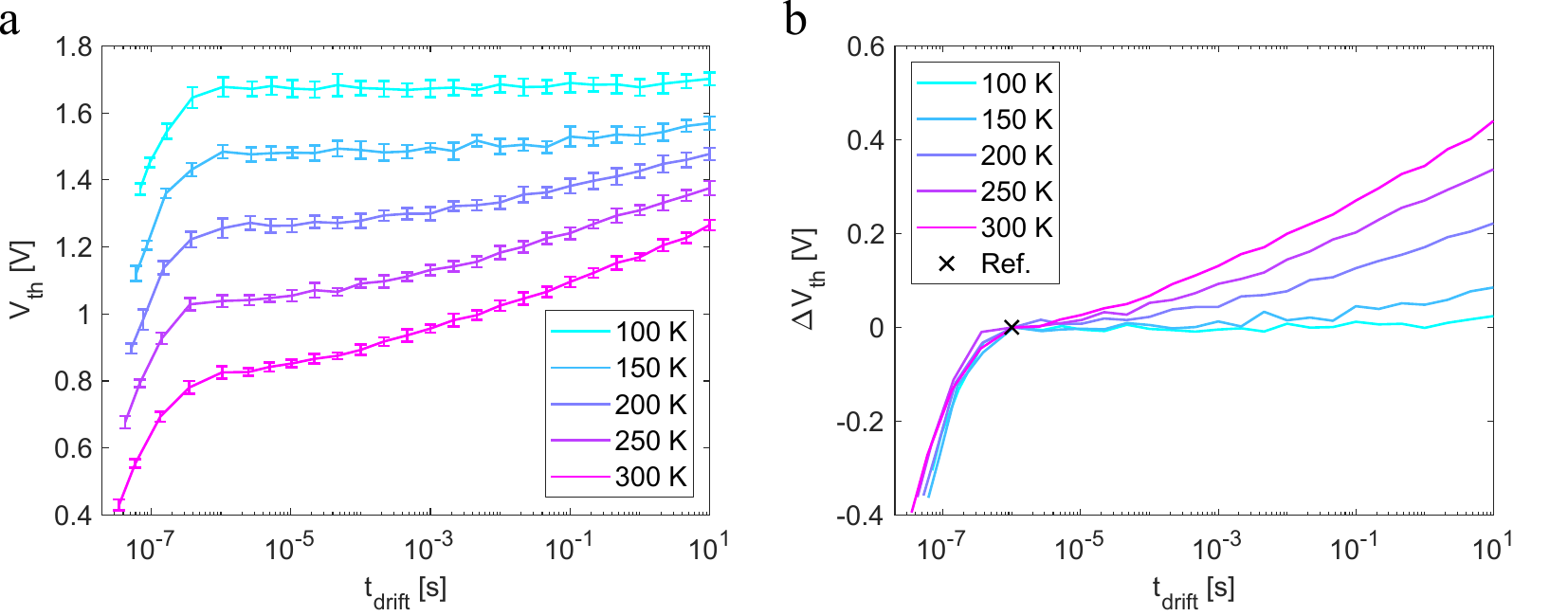}
		\caption[Temperature-dependent \Vth{} evolution]{\textbf{Temperature-dependent \Vth{} evolution. a}, With decreasing ambient temperature \Vth{} increases. The transient regime~1 is not impacted by temperature. Regime~2, in which \Vth{} reaches a plateau, extends to longer timescales and the drift in regime 3 reduces. \textbf{b}, Threshold voltage change with respect to \unit[1]{$\mu$s}. The evolution with respect to a common reference point highlights the impact of the ambient temperature.}
		\label{fig:RELA_Tdep_Vth_Drift}
	\end{figure}

	Next, the threshold voltage evolution with time is probed between \unit[100]{K} and \unit[300]{K} (\Cref{fig:RELA_Tdep_Vth_Drift}). To create comparable RESET states at different ambient temperatures, the programming power was scaled such that the initially molten volume remains approximately constant (\Cref{Sect:SimilarRESETstates}) and the RESET pulse trailing edge was kept constant. This approach is taken to minimize other potential sources of \Vth{} variations and different drift dynamics than the ambient temperature. 
	
	With increasing ambient temperature, the \Vth{} values decrease. This is due to the thermally activated transport of the amorphous phase. The \Vth{} evolution in regime 1 appears to be almost independent of the ambient temperature (\Cref{fig:RELA_Tdep_Vth_Drift}b). Between \unit[30]{ns} and \unit[1]{$\mu$s} after RESET the threshold voltage increases by \unit[0.4]{V}. Contrary to this, the slope (drift coefficient) in regime 3 decreases at lower ambient temperatures. The transition between regimes 2 and 3 marks the onset of relaxation. It changes as well. With decreasing slope in regime 3, the onset appears to be more gradual and is shifted to longer timescales. Both these effects, namely, the onset shift and the slope change are expected because the relaxation processes are decelerated with decreasing temperature.
	
	At \unit[100]{K}, \Vth{} drift appears to be absent. This could happen either because the drift coefficient is very small or because the onset has shifted outside of the measurement range. One possible reason for the former might be a change of the transport mode at very low temperatures. In phase change materials with trap states deep within the band gap, like \GST{} \cite{Luckas2010,Rutten2019,Konstantinou2019}, a transition from trap limited band transport to hopping type transport might happen. The activation energy for hopping would be defined by the distance between the Fermi level and the trap states, which may not necessarily change upon structural relaxation. 
	
	\subsection{Relaxation dynamics \& threshold voltage drift}
	
	In this subsection, based on experimental observations, the use of \Vth{} to monitor the state of relaxation will be motivated. Establishing an analytical relation between \Vth{} and the state of relaxation, \Glass, where \Thist{} captures the thermal history, poses a non-trivial problem. First, the exact mechanism of threshold switching is still debated, and second, it is not clear which material parameters change with relaxation. 
	
	It is assumed that \Vth{} can be defined by the sum of a temperature-dependent function f(T), a term proportional to \Glass{}, and an offset value C\textsubscript{2} that could change with the size of the amorphous dome for example. 
	
	\begin{equation}
		\label{equ:RELA_VariableSeparation}
		V_\text{th} = f(T) + C_1 \cdot \mathrm{Glass}(t,T_{hist}) + C_2
	\end{equation}
	
	A key basis for this assumption is the approximately linear change of V\textsubscript{th} with the activation energy for electrical conduction (E\textsubscript{a}) \cite{Pirovano2004,LeGallo2016}. E\textsubscript{a}, in turn, has been shown to increase with drift \cite{Boniardi2011,LeGallo2018,Wimmer2014b}, and a linear increase of E\textsubscript{a} with log(t) has been experimentally measured for different phase change materials \cite{Fantini2012,Rutten2015}. Thus, \Vth{} is expected to change proportional to \Glass. Finally, it is assumed that the change of \Vth{} upon relaxation is decoupled from the change with ambient temperature. Such a decoupling was previously deduced in \cite{Ciocchini2012} and \cite{Karpov2007b}. Dependent on the relaxation model, \Glass{} corresponds either to the state variable $\mathrm{\Sigma(t,T_{hist})}$ or the total number of defect states $\mathrm{\int q(E\textsubscript{d},t,T_{hist})~dE}$.  
	
	In the following, the temperature-dependent evolution of \Vth{} in regimes 2 and 3 is analyzed in more detail. From Equation \ref{equ:RELA_VariableSeparation}, it can be seen that the temporal change in \Vth{} with respect to a defined reference point, $t_\text{ref}$, is dependent only on \Glass, i.e., the state of relaxation. For the analysis, \Vth(1$\mu$s) serves as a common reference point, where drift is absent for all temperatures studied.
	
	\begin{equation}
		\label{equ:RELA_DeltVthDeltaGlass}
		\Delta V_\text{th} = V_\text{th}(t,T)-V_\text{th}(1 \mu s,T) = C_1 \cdot (\mathrm{Glass}(t,T_{hist})-\mathrm{Glass_0})
	\end{equation}
	
	It will be assumed that the initially created, unrelaxed glass state $\mathrm{Glass_0}$ and the coupling factor C\textsubscript{1} do not depend on the ambient temperature. Under this premise, the experiments at all temperatures can be fitted with a common set of parameters.  
		
	\subsection{Model analysis}
	
	The main challenge in probing the Gibbs model is to infer what spectrum of defect states the material has. In fact, experimental studies on other materials show bell-shaped or more complex distribution functions \cite{Shin1993,Chen1976,Tsyplakov2014,Friedrichs1989}. To capture a strict log(t) dependence, however, a rather flat $\mathrm{q(E_\text{d})}$ is required \cite{Gibbs1983,Karpov2007b,Knoll2009}. It has been estimated that the defect spectrum of phase change materials should be constant over more than \unit[0.8]{eV} \cite{LeGallo2018}.
	
\begin{figure}[htb]
		\centering
		\includegraphics[width=1\linewidth]{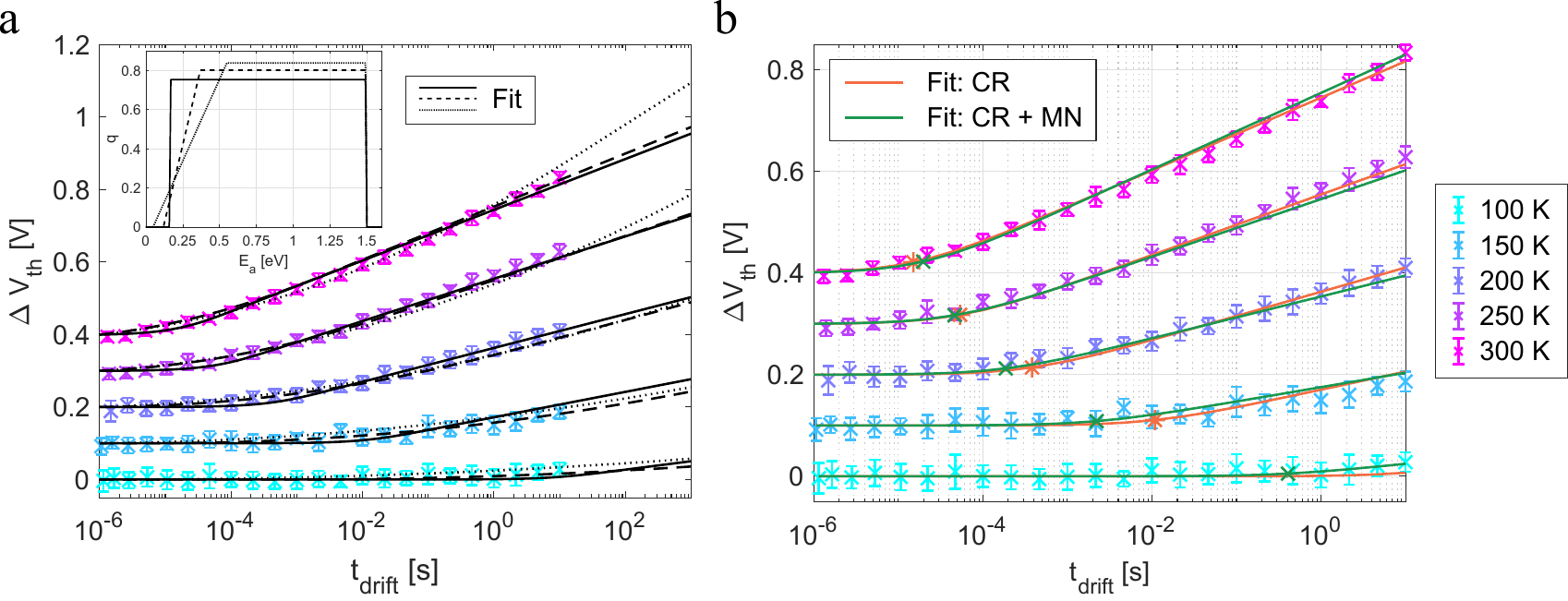}
		\caption[Model fits]{\textbf{Model fits.} The temporal evolution of V\textsubscript{th} is fitted to (\textbf{a}) Gibbs model and (\textbf{b}) collective relaxation model. \textbf{a}, Three different initial defect distributions (q\textsubscript{0}(E\textsubscript{d}); see inset) are simulated for the Gibbs model. With a widening of the range over which the defect distribution increases the onset gets stretched out more and the relaxation phase (i.e., a constant drift coefficient) is reached later. \textbf{b}, Fit to the collective relaxation model (CR) and the extended model, including the Meyer-Neldel rule (CR+MN). For better visibility, the experiments at different temperatures are shifted along the y-axis by \mbox{(\unit[T-100]{K})~$\cdot$~\unit[0.2]{V/K}}.}
		\label{fig:RELA_Gibbs_Collective_Fit}
	\end{figure}

	The onset of relaxation is defined by the defects with the smallest activation energy \mbox{(min-E\textsubscript{d})}. How sharp it is, depends on the energy range over which the defect density increases until it becomes rather flat and the attempt to relax frequency. With a wider energy range and smaller frequency, the onset gets stretched out more. Khonic et al. characterized the relaxation onset of a metallic glass (Zr\textsubscript{52.5}Ti\textsubscript{5}Cu\textsubscript{17.9}Ni\textsubscript{14.6}Al\textsubscript{10}) and found an increase in the defect concentration over a range of about \unit[0.5]{eV} (with \nuo{}~=~\unit[10\textsuperscript{13}]{s\textsuperscript{-1}}) \cite{Khonik2008}. The experiment was analyzed assuming three different initial defect distributions $\mathrm{q_0(E_\text{d})}$. One with a step-like transition and two with a linear transition over an energy range of \unit[0.25]{eV} and \unit[0.5]{eV} (\Cref{fig:RELA_Gibbs_Collective_Fit}a~-~Inset). The defect with the highest activation energy is set to \unit[1.5]{eV}, which is well beyond the activation energies that can be overcome on the probed timescales and temperatures. Free fitting parameters are \mbox{min-E\textsubscript{d}}, the attempt to relax frequency \nuo{}, and the coupling factor C\textsubscript{1} (\Cref{tab:RELA_FittingParameters}). 
	
	All three distributions give a fairly good fit. An extrapolation to longer timescales, however, shows that the onset is stretched out too much for the \unit[0.5]{eV} wide transition. The relaxation onset seems to be captured best by a $\mathrm{q_0(E_\text{d})}$ with a sharp transition. During the melt-quenching within nanoseconds, defects with small E\textsubscript{d} can relax. Also, during quenching, the transition between \unit[100]{\%} and \unit[0]{\%} probability of relaxation is within a narrow energy range at a given time. Thus, it is sensible, that the defect distribution created in the melt-quench process $\mathrm{q_0(E_\text{d})}$ has a sharp transition.  
	
	Compared to the Gibbs model, the collective relaxation model has fewer degrees of freedom. The initial defect spectrum $\mathrm{q_0(E_\text{d})}$, which gives enough degrees of freedom to notably control or alter the evolution of the relaxation process (e.g., the stretching out of the relaxation onset), is replaced by $\Sigma_0$, a single state variable. By definition, relaxation begins at $\tau_0$ (\Cref{equ:RELA_SigmaOftime}) and subsequently obeys a log(t) dependence. Despite these stricter constraints, the collective relaxation model captures the shift of the relaxation onset to longer timescales and the decrease of the drift coefficient with decreasing ambient temperature neatly (\Cref{fig:RELA_Gibbs_Collective_Fit}b).
	
	In this analysis, four model variables are unknown $\Sigma_0$, C\textsubscript{1}, \Es, and $\nu_0 \Delta_\Sigma$. But since they are interdependent the experiment can be fitted with three parameters (\Cref{tab:RELA_FittingParameters}). The drift coefficient depends on  $\mathrm{C_1/E_s}$ (\Cref{equ:RELA_SigmaOftime,equ:RELA_VariableSeparation}), the exponential prefactor of $\tau_0$ on $\nu_0 \Delta_\Sigma \mathrm{E_s}$, and the exponential term on $\mathrm{(1-\Sigma_0 )\cdot E_s}$ (\Cref{equ:RELA_tau0}). It would be necessary to determine also, when drift saturates ($\tau_1$), in order to obtain the four model variables separately. Based on the fitting parameter  $\mathrm{\nu_0 \Delta_\Sigma E_s}$ and the longest times for which drift has been reported, a lower limit for E\textsubscript{s} can be estimated. In \GST{} drift was measured for more than \unit[8e6]{s} at \unit[300]{K} \cite{Gorbenko2019}. This requires E\textsubscript{s} to be larger than \unit[0.95]{eV}. The upper limit has been estimated on the order of the activation energy for crystallization \cite{LeGallo2018}, which is \unit[3.2]{eV} for \GST{} \cite{Jeyasingh2014a}. 
	
	\begin{table}[htb]
		\centering
		\begin{tabular}{ m{2.2cm} || m{3.5cm} m{3.5cm} m{3.5cm} }
			\toprule[.2em] 
			\multicolumn{4}{c}{} \\
			\multicolumn{4}{c}{Gibbs:   $ \frac{dq(E_d,t)}{dt} = -\nu_0 \cdot exp\left(\frac{-E_d}{k_b T}\right)\cdot q(E_d,t)$} \\ 
			\multicolumn{4}{c}{} \\
			\hline \hline
			Transition q\textsubscript{0} & min E\textsubscript{d} [eV] & C\textsubscript{1} [V] & $\nu_0$ [s\textsuperscript{-1}] \\ \hline
			step & 0.16 & -1.57 & 3.66e7 \\
			\unit[0.25]{eV} & 0.12 & -1.54 & 7.33e8 \\
			\unit[0.5]{eV} & 0.05 & -2.33 & 8.64e7 \\
			\midrule[.2em] 
			\multicolumn{4}{c}{} \\
			\multicolumn{4}{c}{Collective relaxation:   $\frac{d\Sigma(t)}{dt}= - \nu_0 \Delta_\Sigma \cdot\exp\left(\frac{-E_\text{s}\cdot(1-\Sigma(t))}{k_\text{b}T}\right)$}\\
			\multicolumn{4}{c}{} \\
			\hline \hline
			& (1-$\Sigma_0$)$\cdot$E\textsubscript{s} [eV] & C\textsubscript{1}/E\textsubscript{s} [V/eV] & $\nu_0\Delta_\Sigma E_s [eV/s]$ \\  \hline
			& 0.19 & -1.2 & 2.48e6 \\
			\midrule[0.2em]
			\multicolumn{4}{c}{} \\
			\multicolumn{4}{c}{Collective relaxation + MN:   $\frac{d\Sigma(t)}{dt}= - \nu_{00}\cdot \exp\left(\frac{E_\text{s}\cdot(1-\Sigma(t))}{k_\text{b}T_\text{MN}}\right) \Delta_\Sigma \cdot\exp\left(\frac{-E_\text{s}\cdot(1-\Sigma(t))}{k_\text{b}T}\right)$} \\
			\multicolumn{4}{c}{} \\
			\hline \hline
			& (1-$\Sigma_0$)$\cdot$E\textsubscript{s} [eV] & C\textsubscript{1}/E\textsubscript{s} [V/eV] & $\nu_{00}\Delta_\Sigma E_s [eV/s]$ \\ \hline
			& 0.15 & -0.77 & 6.82e4 \\  
			\bottomrule[0.2em]
		\end{tabular}
		\caption[Fitting parameters - relaxation models]{\textbf{Fitting parameters - relaxation models.} The threshold voltage drift is fitted to the Gibbs model, collective relaxation model, and the collective relaxation model including the Meyer-Neldel rule. Comparable model parameters are summarized in the same columns of the table ("smallest activation energy", "coupling factor", "attempt to relax frequency").}
		\label{tab:RELA_FittingParameters}
		
	\end{table}
	
	Interestingly, the relaxation of the glass created at ambient temperatures ranging from \unit[100]{K} to \unit[300]{K} can be fitted with an identical initial glass state $\mathrm{q_0(E_\text{d})}$ and $\Sigma_0$ for Gibbs and Collective relaxation model respectively. The degree of relaxation of the initially created glass state does not notably change. Note, that the relaxation onset $\tau_0$ in the collective relaxation model, depends exponentially on $\Sigma_0$. Variations of the degree of relaxation would have a pronounced impact on $\tau_0$. Faster quenching has been observed to create less relaxed glass states \cite{Greer1982}. In the PCM device, the cooling profile during the melt-quenching process changes with ambient temperature. But seemingly, the part of the cooling profile that defines the initial state of relaxation (it should be defined by the cooling rates at elevated temperatures just below the glass transition temperature), does not change too much. Developing an understanding of the quench rates as well as glass transition temperature in the device and how these determine the initial degree of relaxation could be the subject of future work.    

\section{Discussion}

	Compared to drift measurements in the log(t)-dependent time regime, the value of the here presented experiments lies in the revelation of the deviation from this regime. This new feature of the drift dynamics and in particular its temperature dependence constraints the (effective) attempt to relax frequency (Gibbs: $\nu_0$; collective relaxation: $\nu_0\Delta_\Sigma$) and the lowest defect activation energy (Gibbs: min-\Ed; collective relaxation: \mbox{(1-$\Sigma_0$)\Es)}. It becomes evident in the representation of the onset in the collective relaxation model (\Cref{equ:RELA_tau0}). The linear temperature dependence of $\tau_0$ is almost negligible compared to the exponential term. In a first approximation, the shift of the relaxation onset with temperature is defined by the smallest activation energy for relaxation and the effective attempt to relax frequency is a scaling factor defining the timescale of the onset. 
	
	In both models, the smallest activation energy of defects is on the order of \unit[100]{meV}. In the Gibbs model, it changes slightly depending on the assumed shape of the defect spectrum. In a PCM device, a melt-quenched state is created with extremely high cooling rates on the order of $\sim$\unit[10\textsuperscript{10}]{K\,s\textsuperscript{-1}}. It thus allows studying relaxation from an extremely unrelaxed glass state, which manifests in the low activation energy for relaxation and consequently an onset of relaxation at short timescales.
	
	Previous studies on phase change materials estimated the attempt to relax frequency in the typical phonon frequency range of 10\textsuperscript{13} to \unit[10\textsuperscript{14}]{s\textsuperscript{-1}} \cite{LeGallo2018,Ielmini2008a}. Oftentimes used as a free fitting parameter in the Gibbs model, values ranging from \unit[10\textsuperscript{6}]{s\textsuperscript{-1}} \cite{Stutzmann1986} to \unit[10\textsuperscript{24}]{s\textsuperscript{-1}} \cite{Roura2009} have been reported. These extreme deviations from the typical vibration frequency of atoms in solids are surprising and require further explanation. It has been argued that the attempt frequency may change dependent on the defect activation energy \Ed{} and the number of atoms involved in a relaxation process.  
	
	With increasing \Ed{} the number of possibilities for a transition over \Ed{} increases \cite{Kruger1992}. Yelon et al. described this increase in the 'number of possibilities for a transition' more specifically as the number of ways of assembling excitations (phonons) with total energy \Ed{} \cite{Yelon1992}. This results in an increase in the apparent attempt to relax frequency \nuo{} in accordance with the Meyer-Neldel rule \cite{Meyer1937}. It states that the exponential prefactor (here $\nu_0$) of processes that obey an Arrhenius temperature dependence, increases with \Ed. More precisely ln(\nuo) scales linearly with \Ed. This dependence can be described as $\mathrm{\nu_0 = \nu_{00}\cdot exp(\frac{\Ed}{\kb\TMN})}$, where \TMN{} denotes the Meyer-Neldel temperature and \kb\TMN{} is the energy of a single excitation event (annihilation of a phonon). Taking the Meyer-Neldel rule into account, the fundamental attempt to relax frequency is \nuoo{}, and the apparent attempt to relax frequency \nuo{} might be larger than typical phonon frequencies. 
	
	A decrease in the fundamental attempt to relax frequency \nuoo{} on the other hand has been attributed to configurational changes in large groups of atoms, i.e., cooperative, and hierarchically constrained atomic rearrangements \cite{Kruger1992, Hygate1990, Gibbs1983}. Assume for example a relaxation process that requires 5 randomly vibrating atoms to move in a coordinated way. This coordinated motion may happen at a notably lower frequency than the atomic vibrations. Let us consider the consequence for \nuo{} during relaxation. On the one hand, \Ed{} increases with time, which would cause an increase of \nuo{}. On the other hand, with time larger clusters and groups of atoms may relax collectively, which would result in a decrease of \nuoo{} and counter the increase of \nuo. While the assumption of a constant \nuo{} is a crude approximation, introducing an \Ed{}-dependent spectrum for \nuo{} seems hardly to be possible \cite{Friedrichs1989}, especially since this parameter cannot be measured directly.  
	
	In this analysis, \nuo{} was assumed to be constant. For the Gibbs model \nuo{} is on the order of 10\textsuperscript{7} to \unit[10\textsuperscript{8}]{s\textsuperscript{-1}}, notably smaller than phonon frequencies. This indicates that the relaxation processes are defined by a cooperative rearrangement of atoms. The corresponding fitting parameter in the collective relaxation models is $\mathrm{\nu_0\Delta_\Sigma E_s}$ = \mbox{\unit[2.46$\cdot$10\textsuperscript{6}]{eV\,s\textsuperscript{-1}}}. The variable $\Delta_\Sigma$ is expected to be $\ll$1; 1/$\Delta_\Sigma$ is the hypothetical number of different configurational states that the system could assume. Accordingly, the collective relaxation model requires orders of magnitude higher attempt to relax frequencies than the Gibbs model. While this is the most striking difference between both models, the general vagueness of the variable makes it difficult to decide if it is an unreasonable value for either of them. 
	
	In the following, it is shown how the Meyer-Neldel rule can be implemented in the collective relaxation model and the consequences. Many processes in various materials have been observed to obey the Meyer-Neldel rule. For example, electronic transport \cite{Jackson1988,Overhof1989} as well as defect relaxation in amorphous semiconductors \cite{Crandall1991} and aging in polymers \cite{Crine1991}. Interestingly, Ielmini et al. found that in \GST{} both, relaxation, and crystallization of the glass state, follow the Meyer-Neldel rule with a common \TMN{} = \unit[760]{K} \cite{Ielmini2009}. The same value was later used to model the band gap widening with time \cite{Fantini2012}. 
	
	To incorporate the Meyer-Neldel rule into the collective relaxation model  \nuo{} is replaced with $\mathrm{\nu_{00}\cdot exp\left(\frac{(1-\Sigma(t))\Es}{\kb\TMN} \right)}$. With this revised attempt to relax frequency, \Cref{equ:RELA_DifferentialSigma} can be rewritten as 
	
	\begin{equation}
		\label{equ:RELA_DifferentialSigma_MN}
		\frac{d\Sigma(t)}{dt}= - \nu_{00} \Delta_\Sigma \cdot\exp\left[\frac{-E_s\cdot(1-\Sigma(t))}{k_b}\cdot\left(\frac{1}{T}-\frac{1}{\TMN}\right)\right].
	\end{equation}
	
	Compared to the version of the model without the Meyer-Neldel rule, \nuo{} is replaced by \nuoo{} and T by $\left(\frac{1}{T}-\frac{1}{\TMN}\right)^{-1}$ in all equations. Note that the Meyer-Neldel rule is only applicable if the activation energy for relaxation is larger than k\textsubscript{b}\TMN = \unit[65]{meV}, i.e., more than one fundamental excitation is required to exceed the energy barrier \cite{Yelon1992}. With (1-$\Sigma_0$)$\cdot$\Es = \unit[190]{meV}, this criterion is met even for the onset of relaxation of the \GST{} device. 
	
	\begin{figure}[htb]
		\centering
		\includegraphics[width=1\linewidth]{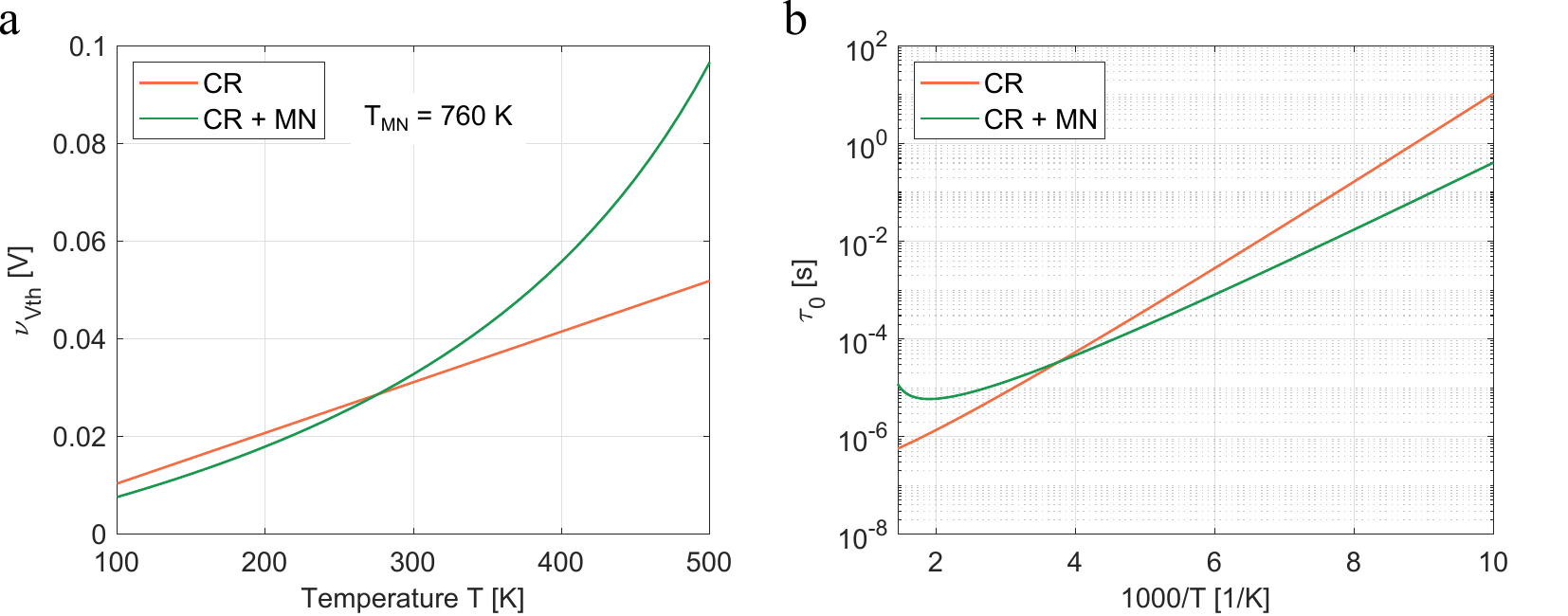}
		\caption[Collective relaxation model - impact of the Meyer-Neldel rule]{\textbf{Collective relaxation model - impact of the Meyer-Neldel rule. a}, on the threshold voltage drift coefficient and \textbf{b}, on the relaxation onset. The temperature dependence of both parameters is calculated based on the fits to the experiment (\Cref{fig:RELA_Gibbs_Collective_Fit}b \& \Cref{tab:RELA_FittingParameters}). The drift coefficients predicted above \unit[300]{K} deviate notably and $\tau_0$ begins to depart from an Arrhenius temperature dependence when the Meyer-Neldel rule is incorporated in the model.}
		\label{fig:RELA_Impact_MN}
	\end{figure}
	
	The \Vth{} drift data is fitted to the extended model with \TMN{}~=~\unit[760]{K} (\Cref{fig:RELA_Gibbs_Collective_Fit}b). Within the here studied temperature range both versions of the model capture the experiment. Extrapolated to higher temperatures, however, approaching \TMN, the model predictions begin to deviate notably \Cref{fig:RELA_Impact_MN}. When the Meyer-Neldel rule is taken into account, the drift coefficient increases notably above \unit[300]{K}. With increasing temperature $\tau_0$ decreases less and at around \unit[500]{K} begins to decrease again. Evidently, \Vth{} drift experiments at even higher temperatures could help to probe the validity of the Meyer-Neldel rule further. However, particular care must be taken to exclude a corruption of the drift measurements by gradual recrystallization of the material.

\section{Gedankenexperiment - a limited number of defects}	
	
	Last but not least a gedankenexperiment is employed to show, that drift cannot be explained if one assumes a defined number of structural defects is created during melt-quenching and relaxes/vanishes over time, without creating new defects. 
	
	Studying a map of Switzerland, one notices that only a few main roads through the valleys and over the lowest mountain passes cross the country. All traffic flows along these main connections and falters at the passes. Rizzi et al. nicely discussed that the same picture is valid for the current flow through the amorphous volume in a PCM device. In the disordered state, the activation energy for conduction varies locally and the vast majority of read current will flow through the amorphous volume along a valley with the lowest \Ea{} and be limited by one 'pass' \cite{Rizzi2013}. The idea that effectively only the largest \Ea{} within a conduction path (E\textsubscript{a,limiting}) defines the electrical transport of the amorphous device was also adopted by others \cite{Gallo2015}. A direct consequence of this idea is that drift results mostly from structural rearrangements that alter E\textsubscript{a,limiting}, i.e., relaxation processes in its vicinity. 
	
	Under the premise, that the relaxation of a defect must not create new defects or influence others (i.e., the defect spectrum is created during melt-quenching and does not change) only a limited number of relaxation processes is possible near E\textsubscript{a,limiting}. The rearrangement of one atom in a defective position must reflect on its direct neighbors. Thus, those may not be structural defects. Under these constraints, the maximum number of structural defects can be estimated to correspond to \unit[10]{\%} of the atoms in a volume. Let us assume further that the relaxation of defects up to \unit[3]{nm} away from E\textsubscript{a,limiting} can change it. For \GST{} with 9 atoms per unit cell of volume \unit[272]{$\AA^3$} this would correspond to approximately $\frac{4}{3} \pi\cdot 30^3 \cdot \frac{9}{272} = $ \unit[3742]{atoms}, surrounding E\textsubscript{a,limiting}, and only \unit[374]{defects}. 
	
\begin{figure}[hbt]
		\centering
		\includegraphics[width=0.6\linewidth]{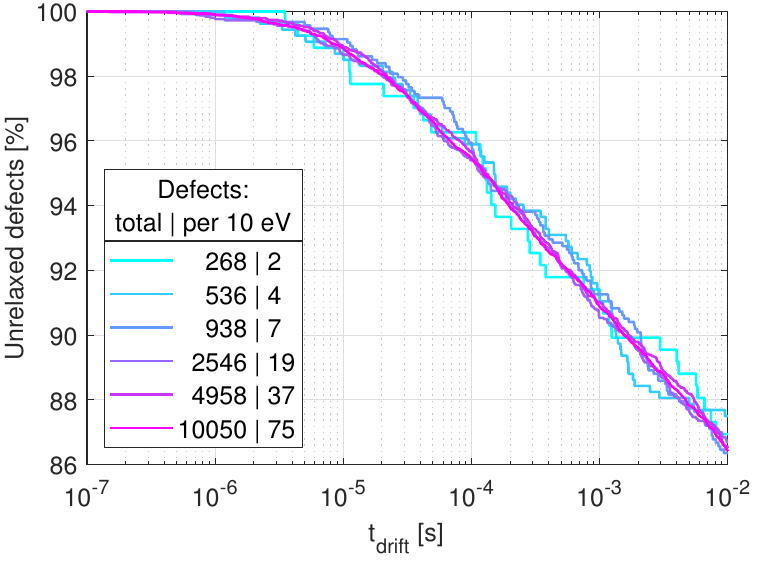}
		\caption[Monte Carlo Simulation - finite number of defects]{\textbf{Monte Carlo Simulation - finite number of defects.} The defect relaxation is simulated for different total numbers of defects. With decreasing number of defects, the relaxation dynamics become increasingly stochastic and begin to scatter around the behavior of a system with many defects. For these simulations, the parameters obtained with the Gibbs model for a uniform defect distribution were used (\Cref{tab:RELA_FittingParameters}).}
		\label{fig:RELA_DefectsFiniteNumber}
	\end{figure}

	The consequences of such a limited number of defects for the Gibbs model were investigated in Monte Carlo simulations (\Cref{fig:RELA_DefectsFiniteNumber}). In the range from \unit[0.16]{eV} to \unit[1.5]{eV} the defects are distributed uniformly in \unit[10]{meV} bins. Every 1/\nuo{} $\sim$\unit[27]{ns} each defect can relax with the probability $\mathrm{exp\left(\frac{-E\textsubscript{d}}{k_bT}\right)}$ (\Cref{tab:RELA_FittingParameters,equ:RELA_DifferentialGibbs}). With decreasing number of defects, the temporal evolution becomes less smooth. If the initial spectrum is made up of less than $\sim$1000 defects, the defect evolution begins to exhibit discontinuities. In a resistance drift measurement, this would manifest in jump-like increases in the device resistance. To the author's best knowledge, no indications for such discontinuities have been observed. Admittedly, they might be hidden by the device read noise. 
	
	Altogether this picture of a defined number of defects that vanish with time seems odd. And even though it was assumed that the relaxation of defects in fairly large volume results in an increase of E\textsubscript{a,limiting}, the Monte Carlo simulations indicate jump-like increases of the resistance during the drift. It provides further evidence, that the defect spectrum used in the Gibbs model does not exist at any time but is an integration, of the defects limiting the progression of structural relaxation, over time. The rearrangement of a defective configuration results in another, energetically more favorable defective configuration, which has larger activation energy for relaxation. Like this multiple relaxation processes can occur in one and the same spot within the material, specifically where E\textsubscript{a,limiting} is located. The conceptual idea behind the collective relaxation model describes the relaxation process more naturally.  
	
\section{Summary \& Outlook}

The evolution of \Vth{}, between \unit[$\sim$30]{ns} and \unit[10]{s} after melt-quenching the device, exhibits three distinct regimes. Initially, within the first $\mu$s \Vth{} increases steeply, next the value stagnates and finally begins to increase again proportional to log(t). Regime 1 can be attributed to the decay of the RESET excitation (electronic and/or thermally). Regimes 2 and 3 arise from the dynamics of structural relaxation. The shift of the transition between these regimes with temperature is captured by the Gibbs model and the collective relaxation model. The excellent match between experimental data and the models can be considered another indication that drift in phase change materials must be attributed to structural relaxation.  

Concerning the first two research questions, we can state: Yes, it is feasible to probe the onset of relaxation in a PCM device by \Vth{} measurements. No, the relaxation models alone cannot capture the full experiment, but only the time-evolution beginning from one $\mu$s. On shorter timescales other effects, most likely the transient decay of the RESET excitation must be taken into account. This phenomenon requires further investigation. It might be of particular importance when read and write operations are executed within nanoseconds, i.e., on an integrated chip with a clock rate of \unit[1]{GHz}.

Regarding the limitations and potential shortcomings of the Gibbs model, it should be noted that a uniform defect distribution appears to capture the experiments best. The sensibility of such a distribution, in particular across all kinds of materials exhibiting a log(t)-dependent change of their physical properties, has been questioned frequently. Additionally, Monte Carlo simulations of a system with a fixed number of defects show that in this scenario resistance drift would probably exhibit step-like jumps. In conclusion, the relaxation of one defect must be expected to create new defects that relax later. The defect spectrum used in the Gibbs model to describe the relaxation does not exist at any time but is actually the result of an integration of defects that limit the further progress of relaxation over time.

Last but not least, it should be noted that both relaxation models must be modified if structural relaxation is facilitated by many-body thermal excitation. The implications were discussed for the collective relaxation model. Within the studied temperature range, both versions of the model capture the experiments with subtle differences. At even higher temperatures $\sim$\unit[350]{K} and above, however, the model predictions begin to deviate. As k\textsubscript{b}T approaches the energy of a single excitation process, the modified model predicts a significantly larger $\nu$\textsubscript{Vth} and a weaker temperature dependence of $\tau_0$. As long as the phase change material does not begin to crystallize, drift measurements at higher temperatures could allow differentiating which model prediction is correct; an essential step to identifying a model that allows simulating reliably the impact of time, but also temperature variations on \Vth{} and device resistance.

\textbf{Key findings:}
\begin{itemize}
	\item	In a \GST{} mushroom cell, the onset of structural relaxation is around \unit[$\sim$15]{$\mu$s} after melt-quenching the device, at \unit[300]{K} ambient Temperature. At this time, defects with activation energy on the order of \unit[150]{meV} begin to relax. 
	
	\item	Assuming a fixed, finite number of defect states is created during melt-quenching and vanishes upon relaxation cannot plausibly explain drift in phase change materials. Instead, the relaxation of a structural defect must be expected to result in another defective configuration that can relax further. The collective relaxation model is motivated by this very idea. In this regard, the defect distribution defined in the Gibbs model is misleading. 
	
	\item	The Meyer-Neldel rule can be incorporated into the collective relaxation model. If structural relaxation is enabled by thermal many-body excitations, as other studies indicate, this refined model should be used to describe drift. Both versions of the model capture the experimental data, but for higher ambient temperatures the model predictions begin to deviate notably. 
	
\end{itemize}

		\chapter{Exploiting nanoscale effects in phase change memories}
\label{Chapt:Confine}

\textbf{Preliminary remark:} The results presented in this chapter have been published as two journal articles. \newline
\textit{Monatomic phase change memory - Salinga, Kersting, et al. - Nature materials \cite{Salinga2018} \newline
Exploiting nanoscale effects in phase change memory - Kersting and Salinga - Faraday Discussions \cite{Kersting2019a}} \\[6pt]

Switching in phase change memories is an electrically initiated but ultimately thermally driven process in both directions: amorphization via melt-quenching and crystallization through annealing. The progress towards nanoscale cell dimensions is not only following requirements regarding device density, in the case of PCM it is also incentivized by the promise to achieve chips with lower power consumption \cite{Fong2017,Xiong2017,Kang2011,Raoux2008a}. Smaller volumes of phase change material have a smaller heat capacity, i.e., need less energy to be molten or annealed to a given temperature. Also, a smaller volume can be fully crystallized in less time without the need to increase the crystal growth velocity. As a consequence, electrical pulses may be less intense and shorter, both reducing the required energy per switching event. Fully confined cell structures \cite{Kang2011} do not only allow the shrinking of, in particular, the cross-sectional area of the electrodes so that the absolute currents required for switching the cell can be reduced, their thermal interfaces with dielectric materials also keep the Joule heat concentrated within the phase change material. Hence, the increase in the interface-to-volume ratio of a PCM cell has an immediate impact on its thermal properties and its electrical demands.

In the endeavor to improve device performance, one main field of research is phase change material optimization by adding dopants and fine-tuning the stoichiometry of the compound. This approach however disregards two critical consequences of the miniaturization required for future technology nodes. First, it becomes increasingly challenging to achieve and preserve a material composition in each of the billions of devices across a silicon wafer. Small variations in the number of atoms of a certain element within the device change the stoichiometry. Additionally, phase segregation, thermal diffusion, and field-assisted motion during device operation can result in a changing composition within the active volume, causing device stochasticity and eventually failure \cite{Debunne2011,Xie2018}. Second, interface effects become increasingly important. Upon confinement to only a few nanometers, the crystallization kinetics of phase change materials have been observed to change notably dependent on the material interface \cite{Raoux2008,Raoux2009,Simpson2010,Chen2016}. Thus, when interfacial and confinement effects begin to alter the material properties, compounds with optimized bulk properties may lose their advantageous behavior. 

In this chapter, the influence of confinement on phase change material properties will be discussed. Considering its consequences for the crystallization dynamics and the challenges of precisely achieving a compound's stoichiometry in a tiny volume motivates us to study a single element, pure antimony. Experimentally bridge cell devices with Sb thicknesses ranging from \unit[3]{nm} to \unit[10]{nm} are studied. First, it is verified that despite its extreme proneness to crystallization, making it seemingly unfit for a device, Sb can be melt-quenched to an amorphous state. Next, we asses under which boundary conditions (ambient temperature, pulse power, pulse trailing edge) the device can be amorphized. Finally, the impact of confinement on energy efficiency, stability against crystallization, and, particularly, resistance drift is addressed.  

\section{State of the art}

	\subsection{Confinement effects}
	
	When the actual volume of phase change material is confined more and more, even properties supposedly characteristic for a material begin to change. Most noticeably, both structural phase transitions, melting and crystallizing, have been reported to take place at significantly different temperatures than in bulk (for this purpose bulk means in \unit[100]{nm} thin films) \cite{Raoux2008a,Raoux2008,Sun2007,Gabardi2017}.
	
	Fundamentally, the divergence from bulk material characteristics originates from an increasingly dominating influence of interfacial effects. While subject to the exact choice of cladding material, the general trend in phase change materials is a reduction of T\textsubscript{m} and increased stability of the disordered states, i.e., slowed-down crystallization kinetics, upon confinement. 
	
	The reduced T\textsubscript{m} upon confinement can be attributed to the effect of premelting (or surface melting) \cite{Frenken1985,Dash1989}. Melting near interfaces is facilitated  if a thin layer of quasi-liquid material reduces the system's energy compared to the direct contact of two solid materials. A reduction of the T\textsubscript{m} compared to the bulk material has been observed for phase change material thin films, nanowires, and nanoparticles \cite{Raoux2008b,Sun2007b,Yarema2018}.  
	
	A variety of effects has been proposed to explain the slowed-down crystallization kinetics. Three contributing factors were identified: the influence of interfacial energies on crystal formation and growth, a reduction of the atomic mobility near interfaces, and mechanical strain. It should be noted that all factors must be expected to alter crystallization upon confinement, but their relative contribution may change. 
	
	The recrystallization of an amorphous device can be either driven by crystal growth from the crystal-glass interface or the formation of new crystalline clusters, either inside the amorphous matrix (homogeneous nucleation) or at the interface to the cladding material (heterogeneous nucleation). To form a stable nucleus the bulk free energy gained by the transition from an amorphous to the crystalline structure must be larger than the interfacial energy of the formed crystal. This criterion is met if the randomly formed nucleus has a critical size. Beyond this point, the crystal will not dissolve again but grow further.  
	
	Dependent on the cladding material, the size of a critical nucleus changes and nucleation at the interface can be enhanced or suppressed. If the crystal-cladding interfacial energy is smaller than the amorphous-cladding interfacial energy, nucleation becomes more favorable. Even crystal growth along such an interface can be sped up or slowed down compared to its bulk values depending on the interfacial energies \cite{Pandian2006}. 
	
	In general, crystal growth becomes more dominant in the crystallization process of nano metric PCM cells, not only compared to homogeneous but even to heterogeneous nucleation. As soon as a crystal-to-glass interface exists only a few nanometers away from any group of disordered atoms, they are much more likely to be incorporated into that growing crystal than to reorganize themselves into a new crystal nucleus. Since only a fraction of the phase change material is amorphized inside a device this interface exists inherently.   
	
	Besides a reduced nucleation probability in a smaller volume and the altered heterogeneous nucleation at the interface, the confinement of a material can have an impact on the ease with which the atoms can reconfigure within a nano metric volume. Researchers at Yale nicely discussed these competing effects when studying the crystallization of metallic glass nanorods of different sizes \cite{Shao2013,Sohn2015,Sohn2017}. The apparent viscosity is reduced as the confinement is narrowed. As a consequence, crystallization kinetics are slowed down in smaller structures. In a series of molecular dynamics simulations Scheidler et al. \cite{Scheidler2004,Scheidler2002} investigated cooperative motion in supercooled liquids in close proximity to confining walls. The roughness or smoothness of these surfaces turned out to play a decisive role, pointing to a physical quantity that might turn out to be a relevant specifier for comparing different claddings in the future.
	
	A third factor that can alter the crystallization behavior upon confinement is mechanical strain. Typically, the amorphous phase has a \unit[5-10]{\%} lower density than the crystalline phase \cite{Weidenhof1999,Njoroge2002}; i.e., requires a larger volume. Thus, compressive stress can induce crystallization of phase change materials \cite{Eising2013}. Though not yet experimentally studied, it seems plausible that tensile strain can likewise hinder crystallization \cite{Orava2012a}. Whether a memory device has built-in compressive or tensile strain depends on the state of the phase change material when it is encapsulated by a cladding. In a device encapsulated in the crystalline state compressive strain will build up upon amorphization. In a device encapsulated in the amorphous state tensile strain will build up upon crystallization. 
	
	Strain measurements on as-deposited amorphous thin-films, crystallized by annealing, show a) that about \unit[90]{\%} of the \unit[1-2]{GPa} tensile strain is accommodated plastically, by viscous flow of the amorphous phase, and b) that with decreasing film thickness the strain increases, which is attributed to the viscosity increase upon confinement \cite{Guo2008, LeervadPedersen2001}. Some authors argue that the increased strain upon confinement is not the origin of the changing crystallization kinetics, but merely a by-product of the viscosity increase \cite{Kooi2020}. Contrary to this, it has also been argued that crystal growth and viscous flow are coupled stronger due to the crystallization-induced strain \cite{Orava2012a}. 
	
	An important difference of nano-scale devices, compared to thin-film samples, is that the phase change material bridge or pillar is surrounded by a relatively rigid cladding. Thus, differently from a thin-film sample, a volume change of the phase change material cannot be accommodated by a thickness reduction and in-plane viscous flow. To accommodate strain plastically, the cladding material also needs to be deformed. For this reason, larger mechanical strain and potentially a more pronounced impact on crystallization must be expected in devices. 

	At the end of this section, critical questions from an application point of view will be answered. What are undesirable consequences of reduced T\textsubscript{m} and slowed-down crystallization dynamics and how can they be countered? 
	
	In principle, a lower melting point is desirable, since it would reduce the power required to melt-quench the device. Another consequence, however, would be a smaller difference between T\textsubscript{m} and the temperature range of fast crystallization or potentially even a reduction of the fast crystallization range. If this happens, the array level write operation would become less reliable. This effect can be avoided by choice of cladding material, such that T\textsubscript{m} does not change too much, or a phase change material, with a T\textsubscript{m} well separated from the temperature range of fast crystallization.     
	
	Slowed-down crystallization dynamics around room temperature are beneficial for device retention. However, if the crystallization at elevated temperatures is also hindered the write speed (SET operation) would be reduced and it might even become impossible to crystallize the device reliably. A simple solution to counter this is to choose materials that are extremely prone to crystallization. They might even be difficult to be amorphized in bulk. In the following, it will become evident that pure antimony fulfills these criteria. 
	
	\subsection{Amorphous antimony}	
	
	First studies of as-deposited amorphous antimony thin-films date back to the late 1930s. Antimony was evaporated in a vacuum tube and condensated on a substrate cooled to \unit[80]{K}. Upon annealing the semiconducting \unit[250]{nm} thick film crystallized at \unit[270]{K}. The phase transition manifested in an irreversible resistance decrease by three orders of magnitude and increased optical reflectivity \cite{Suhrmann1939}. 
	 
	Scaling studies later suggested that the crystallization temperature changes proportionally to 1/thickness \cite{Kinbara1976b} and found a reduction of the crystal growth rate \cite{Hashimoto1984}. Both observations were attributed to reduced growth rates at the substrate and vacuum interfaces. Extreme confinement to a thickness of \unit[4]{nm} results in an increase of the crystallization temperature to \unit[142]{$^\circ$C} \cite{Krebs2009a}, showing that, in principle, it is possible to retain the amorphous phase at room temperature and above. 
	
	Besides the phase transformation at relatively low temperatures, the antimony's proneness to crystallization can also manifest in an "explosive" phase transformation. Once the crystallization is locally triggered, its heat release facilitates the rapid propagation of the crystal growth front and the complete sample crystallizes \cite{Aymerich1975, Eising2014}. Explosive crystallization has been reported for Sb samples deposited by electrolysis \cite{Aymerich1975} and sputter-deposited (In, Ga)Sb compounds \cite{Wickersham1978a}.
	
	The insights gained from as-deposited amorphous samples indicate that pure antimony could be a suitable phase change material for extremely small, confined device nodes. Yet, its properties, beneficial for avoiding slow write speed (SET operation) or even RESET stuck failure upon device scaling, are also the main challenge. Due to its extreme proneness to crystallization the creation of a glass by quenching from the melt, the essential process in switching a PCM, has never been accomplished for pure Sb. 
	
	\subsection{Amorphization in a device structure}
	
	Recently, however, it has been shown, that even pure metals can be quenched to and stabilized in a glassy state by ultrafast cooling from the melt (\unit[10\textsuperscript{14}]{K\,s\textsuperscript{-1}}). Protruded nano-tips of Ta, V, W, and Mo were brought into contact in a TEM and melt-quenched by applying square electrical pulses (\unit[$\sim$3.7]{ns}). The rapid quenching allowed to suppress the recrystallization from the crystal-glass interface. At room temperature the growth rates of the studied metals are infinitesimally small; the glasses were stable \cite{Zhong2014a,Schroers2014,Greer2015a}. 
	
	The seminal work by Zhong et al. has shown that pure metals can be melt-quenched. In these experiments, the metal substrate on which the nano-tips were patterned serves as a heat-sink, enabling to freeze the melt within tens of ps. Still, some metals (Au, Ag, Cu, Pa, Al, Rh, Ir), crystallizing in a face-centered cubic structure, could not be quenched to a glass state. Either they recrystallized during the quenching process, or the observation method was too slow to detect the glass before it recrystallized at room temperature. It is yet to be seen whether sufficiently high cooling rates to melt-quench antimony (rhombohedral crystalline structure) can be achieved in a bridge cell patterned on a silicon wafer. The primary heat loss path of a bridge is through the dielectric insulation. Assuming an infinite oxide media surrounding the bridge, the thermal time constant was estimated to be on the order of nanoseconds \cite{Sadeghipour2006}. Compared to the nano-tip experiment, significantly lower cooling rates must be expected in a device structure. This could be compensated for by confinement effects, hindering recrystallization and thus enabling amorphization of the Sb device. 
	
	Assuming it is possible to quench the molten antimony fast enough to suppress nucleation and avoid complete recrystallization from the crystal-melt interface, the created state must still not be mistaken for an as-deposited amorphous sample. In the device, growth from the crystalline antimony nuclei at the rim of the amorphous mark must be suppressed, while crystallization of the as-deposited sample is additionally hindered by the requirement to form a crystalline nucleus. Furthermore, due to different amorphization processes (gas-phase deposition vs. melt-quenching) and thermal histories the glasses are differently relaxed and can have different concentrations of subcritical nuclei \cite{Lee2009a}. Compared to an as-deposited amorphous state, the melt-quenched state is highly unrelaxed. Thus, its atomic mobility is higher (it has a smaller viscosity) and exhibits orders of magnitude higher crystal growth velocities \cite{Salinga2013}. Overall, the melt-quenched state created in a device is more prone to crystallization than the as-deposited state.       
	
	\subsection{Research questions}
	\begin{enumerate}
		\item	Is it possible to create melt-quenched amorphous Sb and suppress recrystallization sufficiently to study the amorphous state?
		\item	How does extreme confinement alter the PCM characteristics? 
	\end{enumerate}
	\newpage
	
\section{Melt-quenched amorphous antimony}
	Addressing the second question requires a positive answer to the first. After a brief presentation of the bridge cell device, it will be discussed if and under which conditions devices with \unit[5]{nm} thick Sb can be amorphized.
	  
	\subsection{Antimony bridge cells}
	\label{Sect:FabricationBridge}
\begin{figure}[bth!]
		\centering
		\includegraphics[width=1\linewidth]{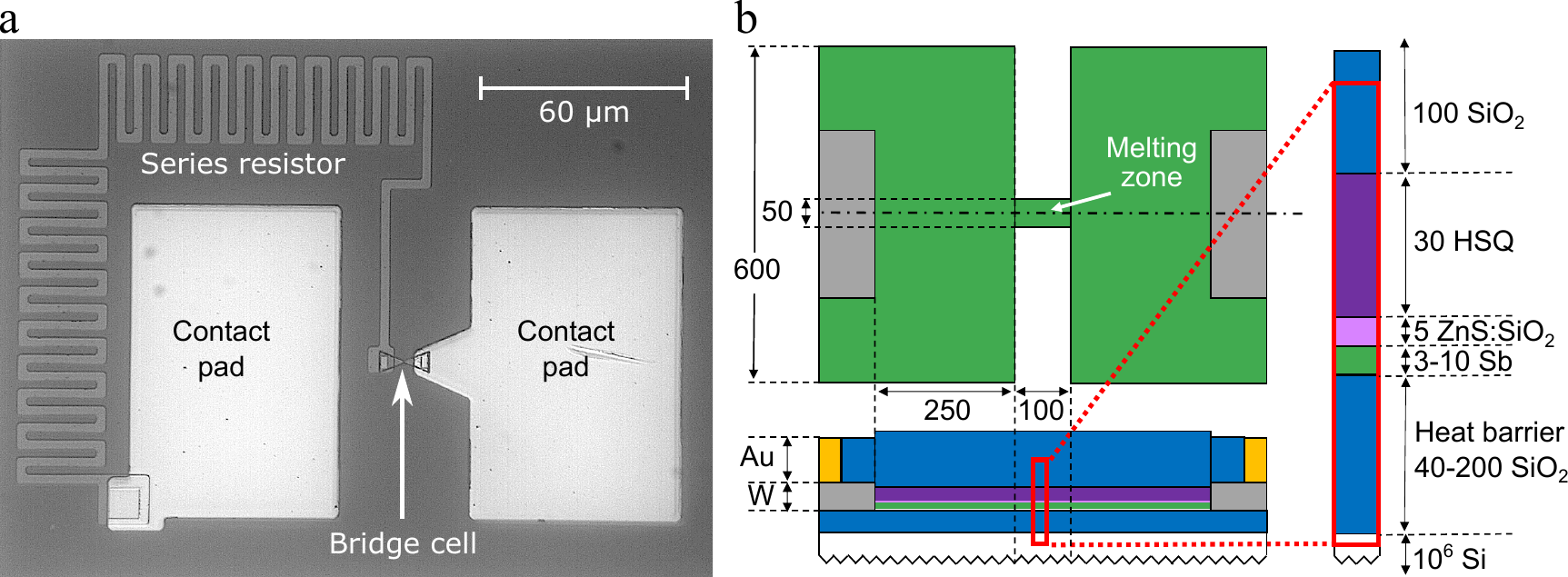}
		\caption[Antimony bridge cell]{\textbf{Antimony bridge cell. a}, Optical microscopy image of the device. Gold contact pads ensure stable electrical contact even at ambient temperatures down to \unit[100]{K}. The Tungsten series resistor limits the device current at the moment of threshold switching. \textbf{b}, Top view and cross-sectional sketch of the bridge cell with dimensions in nm. HSQ, hydrogen silsesquioxane e-beam resist. Devices with different material thicknesses were fabricated to study the influence of one-dimensional antimony confinement (\unit[3]{nm}, \unit[5]{nm}, and \unit[10]{nm} thick) and the thermal insulation (\unit[40]{nm}, \unit[100]{nm}, and \unit[200]{nm} heat barrier).}
		\label{fig:Nano_DeviceSketch}
	\end{figure}

	In the bridge cells studied here, Sb is patterned to a structure with a \unit[50]{nm} wide and \unit[100]{nm} long bridge in its center (\Cref{fig:Nano_DeviceSketch}). Three elements of the device structure are essential. First, the degree of one-dimensional confinement is controlled by the Sb thickness. Second, the thermal insulation determining how efficiently the material can be molten but also limiting the quench rates is defined by the thickness of the SiO\textsubscript{2} heat barrier. Third, the on-chip series resistor, limiting the device current at the moment of threshold switching, improves the switching stability.  
	
	The bridge cells were fabricated on a silicon substrate with a thermally grown SiO\textsubscript{2} top layer. The SiO\textsubscript{2} functions as electrical and thermal insulation (heat barrier). The Sb (\unit[3, 5, 10]{nm}) and a (ZnS)$_{80}$:(SiO$_2$)$_{20}$ capping layer (\unit[5]{nm}) were sputter deposited with a rate of \unit[\textless 0.1]{nm\,s\textsuperscript{-1}}. The small deposition rates enable better control of the deposited film thickness. Atom-probe tomography measurements show that the deposited Sb has a purity \textgreater \unit[99.9]{\%}. The thin films were patterned to the device structure depicted in \Cref{fig:Nano_DeviceSketch}b by electron-beam lithography with a hydrogen silsesquioxane resist and ion milling. The sidewalls, exposed after the etching process, were immediately passivated with $\sim$\unit[18]{nm} SiO\textsubscript{2} deposited by sputtering. To add the W vias, the capping was removed locally by ion-milling, next \unit[50]{nm} of W were deposited and subsequently patterned to lateral leads by ion-milling. The structures for both ion-milling steps were patterned by electron-beam lithography. The titanium series resistor (2 to \unit[4]{k$\Omega$}) was fabricated in a lift-off process (optical lithography). Next, the chip was encapsulated with \unit[80]{nm} SiO\textsubscript{2}. The encapsulation was opened again locally to add the gold contact pads (\unit[200]{nm}). Those were fabricated in a lift off-process (optical lithography)~\footnote{The Sb-based devices were fabricated by Vara Prasad Jonnalagadda and Xuan Thang Vu. Oana Cojocaru-Mir\'edin analyzed the integrity of the deposited Sb via atom-probe tomography}. 
	
	\subsection{Fingerprints of melt-quenched amorphous antimony}
	
	The initial step of this study is to increase the device resistance by electrical excitation and verify an amorphous mark has been created. To suppress crystal growth after melt-quenching and increase the temperature gradient while quenching, the experiment is performed at \unit[100]{K} ambient temperature. Additionally, the Sb is quenched with the highest cooling-rates achievable in our setup (\unit[40]{nm} heat barrier and \unit[3]{ns} pulse trailing edge). 
	
	\begin{figure}[bth!]
		\centering
		\includegraphics[width=1\linewidth]{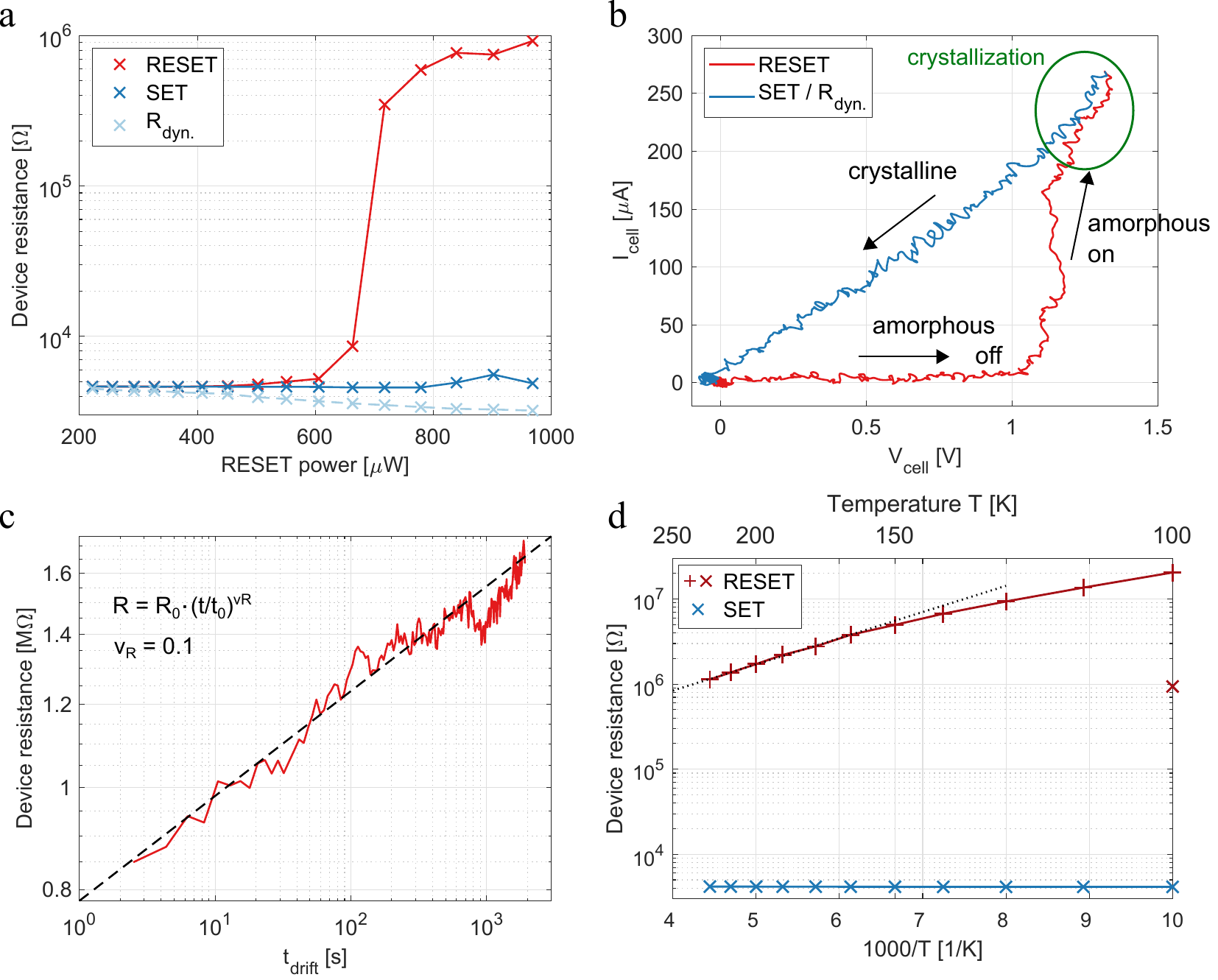}
		\caption[Creation of melt-quenched amorphous Sb]{\textbf{Creation of melt-quenched amorphous Sb.} \textbf{a}, Programming curve. The device is switched repeatedly between the SET and RESET state. The RESET power is continuously increased. SET and RESET resistances are read in the ohmic regime with a cell bias of \unit[100]{mV}. R\textsubscript{dyn} denotes the transient device resistance measured during the RESET pulse, i.e., in the crystalline or partially molten state at elevated fields and high temperatures. \textbf{b}, Threshold switching I-V characteristic. A triangular SET pulse with \unit[200]{ns} leading and trailing edge is applied to a \unit[531]{k$\Omega$} RESET state. \textbf{c}, RESET resistance drift. The dashed line is a fit of the power law conventionally used to describe resistance drift. \textbf{d}, Temperature-dependent device resistance. In the SET state, the device is fully crystalline. The RESET state was created at \unit[100]{K} with a programming power of \unit[957] {$\mu$W} (red x). The temperature-dependent resistance of the amorphous state is measured after annealing at \unit[225]{K} (red +). 
		The experiments \textbf{a-c} are performed at 100 K ambient temperature. Amorphization was induced by a trapezoidal voltage pulse with a \unit[50]{ns} plateau duration and \unit[3]{ns} trailing edge. The heat barrier thickness is \unit[40]{nm}. The thickness of Sb is \unit[5]{nm}.}
		\label{fig:Nano_PCMcharacteristics}
	\end{figure}
	
	The device programming curve gives the first indications of melting and amorphization of the antimony. The device is stressed with trapezoidal voltage pulses (\unit[3-50-3]{ns}) of increasing amplitude (\Cref{fig:Nano_PCMcharacteristics}a). Melting is indicated by the continuous decrease of the dynamic resistance (R\textsubscript{dyn.}), measured during the pulse plateau. While the crystalline device resistance is temperature-independent (\Cref{fig:Nano_PCMcharacteristics}d), the molten phase exhibits a lower resistivity than crystalline thin films (\unit[$\sim$114]{$\mu\Omega\,cm$} \cite{Wang2000,Newport1980} vs. \unit[\textgreater 150]{$\mu\Omega\,cm$} \cite{Elfalaky1995}). Thus, the gradually larger deviation from the SET resistance can be ascribed to molten volumes of increasing size~\footnote{The initial difference of less than \unit[150]{$\Omega$} could result from a weak field dependence of the transport or an offset in the R\textsubscript{dyn.} measurement.}. Consequently, the RESET resistance increases also as larger amorphous volumes are created in the device. Below \unit[650]{$\mu$W} the resistance increase is only subtle, suggesting a crystalline percolation path shunts the amorphous volume. Once the device cross-section is blocked by an amorphous volume the resistance increases sharply (around \unit[700]{$\mu$W}). As the amorphous volume grows larger and larger the resistance increases further. 
	
	The RESET state is characterized in terms of its electronic switching dynamics, variability with time, and temperature dependence. Triangular voltage pulses with \unit[200]{ns} trailing and leading edges induce threshold switching (\Cref{fig:Nano_PCMcharacteristics}b) and a reliable transition to the SET state (\Cref{fig:Nano_PCMcharacteristics}a). While abrupt switching at elevated fields alone could also result from other phenomena, e.g., the breakdown of a Schottky barrier, the reproducible transition to the SET state is better compatible with PCM switching. 
	
	The temporal evolution of the device resistance with time shows drift, another hallmark of a melt-quenched amorphous state. The drift coefficient of 0.01$\pm$0.02 is remarkably similar to that reported for conventional, multi-elemental phase change materials such as \GST{} \cite{Boniardi2010}. The structural defects in amorphous Sb must be of a different nature than those reported for amorphous \GST{} (tetrahedrally coordinated Ge; Ge-Ge, Ge-Sb, and Sb-Sb bonds). Nevertheless, structural relaxation and the resulting changes of the electronic band structure reflect quite comparably on the resistance evolution with time. 
	
	Last but not least the impact of ambient temperature on the RESET state is characterized (\Cref{fig:Nano_PCMcharacteristics} d). The device is annealed for \unit[15]{minutes} at \unit[225]{K} to accelerate and settle structural relaxation. The resistance series measured subsequently during cooling to \unit[100]{K} and heating back to \unit[225]{K} are reproducible. Neither did the sample crystallize, nor did it relax further. The measurement series reveals thermally activated electrical transport with an apparent activation energy of \unit[0.065]{eV} between \unit[225]{K} and \unit[175]{K} (dotted black line). Due to the structural relaxation during annealing, the device resistance increased by 20x compared to the initial state measured at \unit[100]{K} (red cross). 
	
	Each of these properties alone could potentially be explained by another resistive switching effect. But, the combination of huge resistance contrast, threshold switching, reproducible SET with \unit[200]{ns} long pulse trailing edges, resistance drift, and semiconducting transport provides pressing evidence that Sb was melt-quenched to an amorphous state in the device. 
		
	\subsection{Parameter space that allows melt-quenching}
	
\begin{figure}[bth!]
		\centering
		\includegraphics[width=1\linewidth]{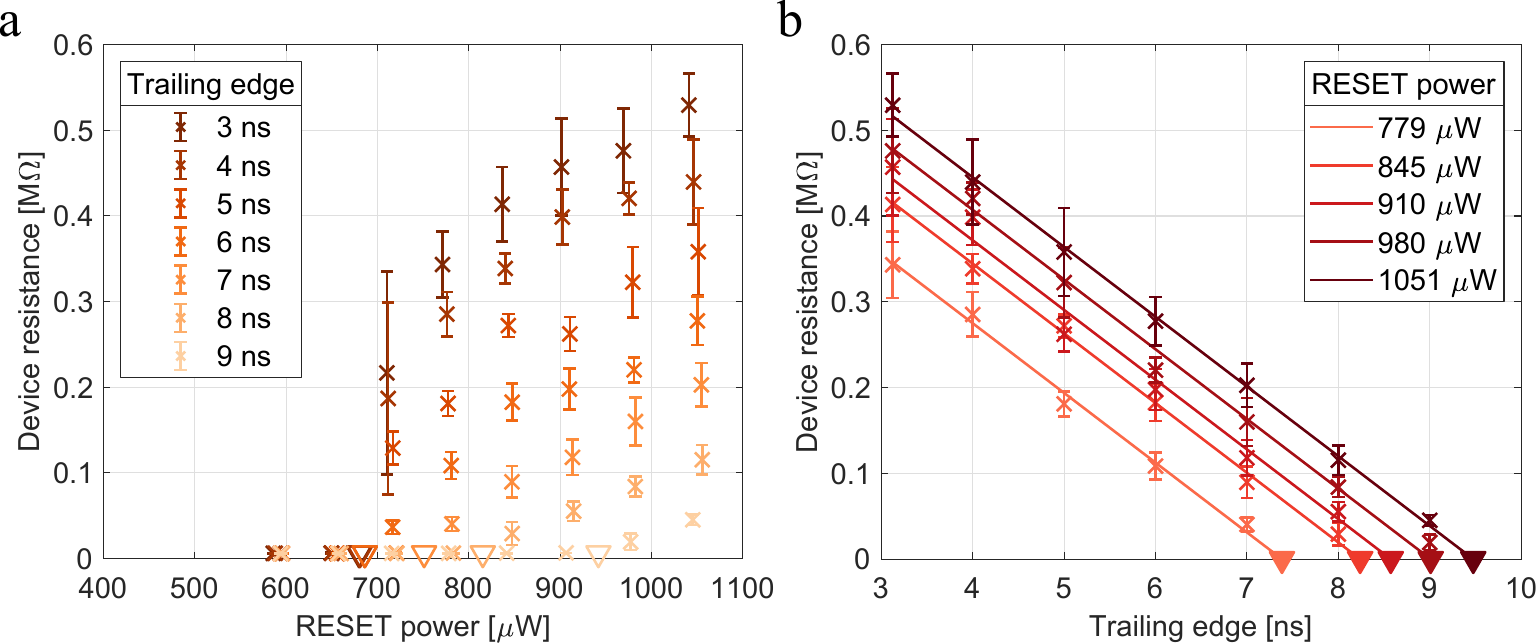}
		\caption[Limits of RESET programming]{\textbf{Limits of RESET programming.} The quench rate is controlled by changing the RESET pulse trailing edge. \textbf{a}, Programming curves. With increasing trailing edge duration the programming onset increases (more power is required to achieve an increase in the device resistance). Triangles mark the estimated programming onset, which is defined as the mean of the highest power that did not yet increase the device resistance and the smallest power that raises the resistance above the crystalline resistance. \textbf{b}, Impact of the trailing edge on the RESET resistance. For a defined RESET power the device resistance decreases linearly with the trailing edge. Triangles mark the trailing edge for which the molten volumes recrystallize completely during the melt-quench process.
		Error bars denote the standard deviation of five identical electrical excitations. The experiments are performed at \unit[100]{K} ambient temperature. The heat barrier thickness is \unit[40]{nm}. The thickness of Sb is \unit[5]{nm}.}
		\label{fig:Nano_ProgExpAmorphizationWindow}
	\end{figure}

	A successful amorphization of the Sb increases the device resistance. Thus it can be used to study under which conditions amorphous material can be created in the device. The experimentally altered parameters are the dissipated heat and the quench rate. While the temperature and cooling profile vary spatially in the device, the pulse power (\unit[590]{$\mu$W} to \unit[1050]{$\mu$W}), and trailing edge (\unit[3]{ns} to \unit[9]{ns}), i.e., how fast the Joule heating ends, are well-defined control parameters in the experiment (\Cref{fig:Nano_ProgExpAmorphizationWindow}). 
	
	For the highest programming power studied here it was not possible to RESET the device with trailing edges longer than \unit[9]{ns}. Even confined to \unit[5]{nm} Sb can still crystallize extremely fast. The Sb device must be quenched much faster than the mushroom cell presented in the Introduction (\Cref{fig:Intro_Programming}). The doped \GST device can be amorphized with hundreds of nanosecond long trailing edges. 

	Since variations of the pulse trailing edge duration in \unit[1]{ns} steps notably change the device resistance, the cooling time must be on a comparable timescale. The temperature in the device decreases within a few nanoseconds from above T\textsubscript{m} to base temperature. Compared to the studies on pure metals by Zhong et al. \cite{Zhong2014a}, the Sb in the bridge cell is quenched with approximately 3 orders of magnitude slower rates on the order of \unit[10\textsuperscript{11}]{K\,s\textsuperscript{-1}}.   
	
\begin{figure}[bth!]
		\centering
		\includegraphics[width=1\linewidth]{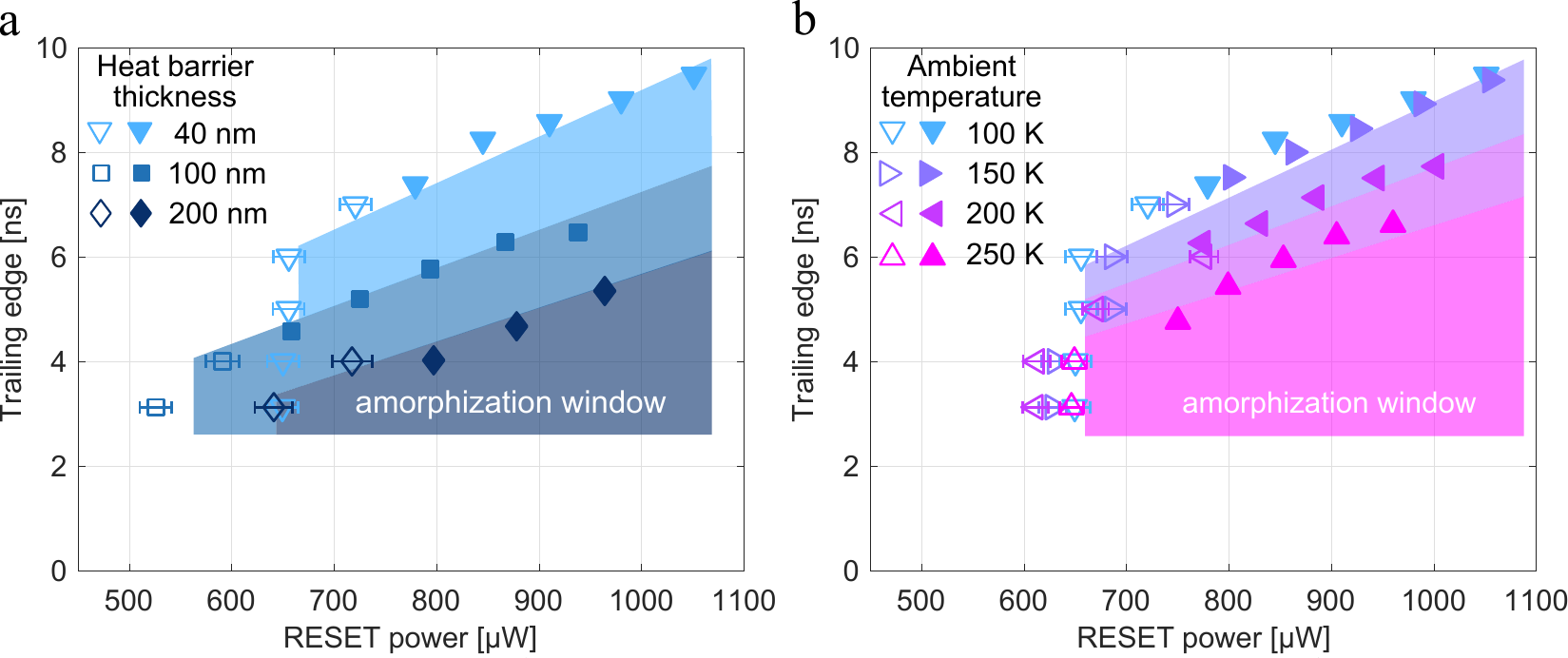}
		\caption[Controlling the amorphization window]{\textbf{Controlling the amorphization window.} Dependent on how efficient heat can be dissipated into the thermal environment (\textbf{a}) and the ambient temperature (\textbf{b}) the amorphization window changes. Filled symbols mark the limit defined in \Cref{fig:Nano_ProgExpAmorphizationWindow}b (forming an upper bound of the amorphization window) and empty symbols the limit defined in \Cref{fig:Nano_ProgExpAmorphizationWindow}a (forming the left border of the amorphization window). Error bars range from the highest power that did not yet increase the device resistance and the smallest power that did raise the resistance above the crystalline resistance. The base temperature in \textbf{a} is \unit[100]{K}. The heat barrier thickness in \textbf{b} is \unit[40]{nm}. The thickness of Sb is \unit[5]{nm} in \textbf{a} and \textbf{b}. As the amorphization window does not change considerably going from \unit[100]{K} to \unit[150]{K} (\textbf{b}), it is colored only once (purple). The successfully created amorphous states are stable for much longer than the few seconds it takes to measure their resistance after the melting pulse.}
		\label{fig:Nano_AmorphizationWindow}
	\end{figure}

	To create an amorphous volume the peak power of the RESET pulse must be sufficiently large to melt the Sb and the trailing edge must be fast enough to suppress the recrystallization of the molten mark. In the programming curves, the device resistance begins to increase once this condition is met (triangles in a). It shifts to larger powers with an increasing pulse trailing edge. For a given size of the initially molten volume (RESET power), the programmed device resistance scales quite linearly with the trailing edge (b). With increasing trailing edge more time is spent in the temperature regime of fast crystallization and thus a larger molten volume can recrystallize. The extrapolation to the crystalline resistance defines for a given RESET power the trailing edge required to avoid complete recrystallization of the molten mark. These combinations of RESET powers and trailing edges define the border of the amorphization window (\Cref{fig:Nano_AmorphizationWindow}). Only parameter combinations within the amorphization window allow the creation of amorphous Sb in the device.  
	
	Not only the RESET power and pulse trailing edge but also the thermal environment of the bridge cell determine the temperature and cooling rates. In the device, the SiO\textsubscript{2} layer underneath the Sb functions as a heat barrier. With increasing heat barrier thickness the Sb gets better thermally insulated. For a given programming power higher temperatures and larger molten volumes are reached. Concurrently, the quench rates reduce and more of the molten material can recrystallize. Increasing the heat barrier thickness from \unit[40]{nm} to \unit[100]{nm} and further to \unit[200]{nm} shifts the programming window to shorter trailing edges. The volume size that recrystallizes during quenching increases stronger than the size of the molten volume. To operate the Sb device fast quenching by an effective heat dissipation into the environment and short pulse trailing edges is of utmost importance. 
	
	Next, the impact of the ambient temperature on the programming window is examined (\Cref{fig:Nano_AmorphizationWindow}b). Increasing the ambient temperature has an influence similar to an increase in the heat barrier thickness on the amorphization process. At higher base temperatures the same RESET power gives rise to higher temperatures in the device and thus larger molten volumes. Due to decreasing temperature gradients, the melt is quenched less rapidly and more material can recrystallize. Again, the prolonged recrystallization dominates over the larger molten volume. Even though the amorphization window narrows with increasing temperature Sb can still be amorphized at \unit[250]{K}. Between \unit[150]{K} and \unit[250]{K} the programming window shrinks only gradually, suggesting that amorphous Sb can be created at even higher temperatures.   
	
	\section{Effect of confinement on the cell characteristics}
	
	After this extensive study of the amorphization in devices with \unit[5]{nm} Sb, we aim to quantify how application critical device metrics are impacted by confinement. More precisely, the stability against crystallization, drift, and programming efficiency are characterized. 
	
\begin{figure}[bth!]
		\centering
		\includegraphics[width=1\linewidth]{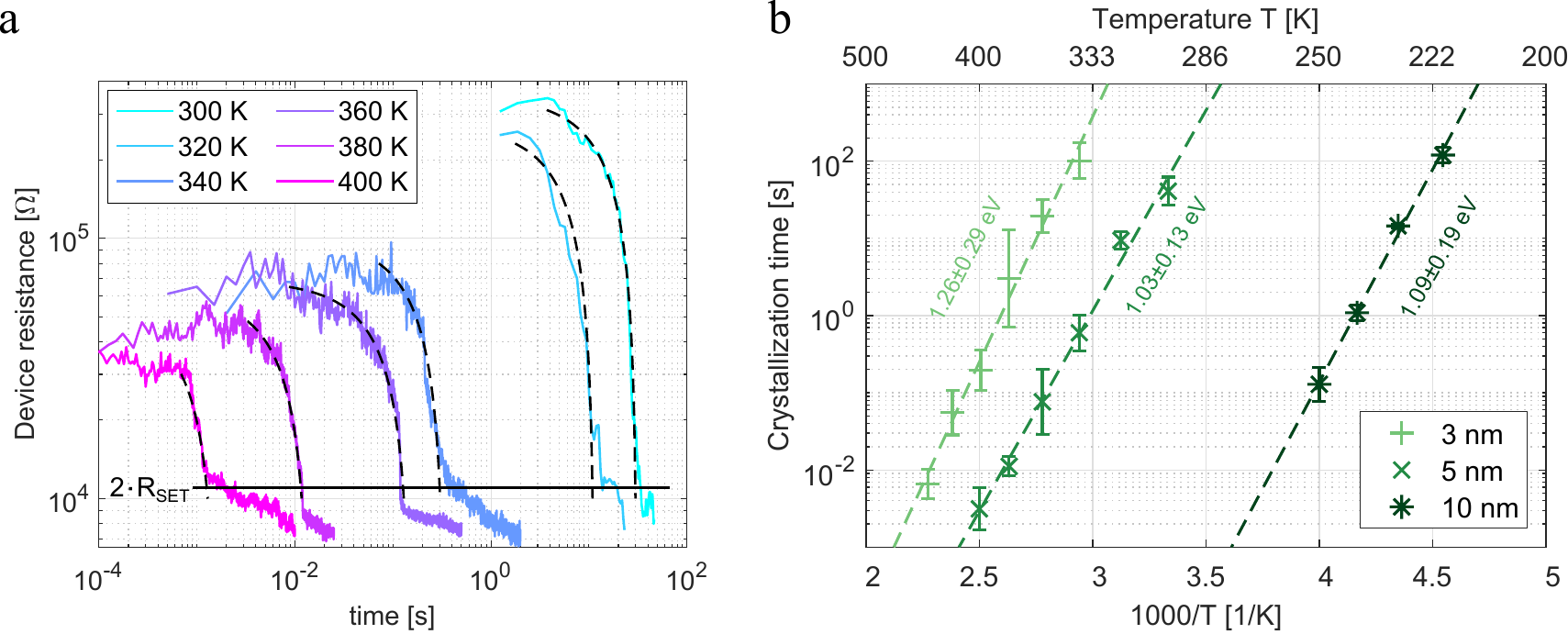}
		\caption[Improving robustness against crystallization by confinement]{\textbf{Improving robustness against crystallization by confinement. a}, Crystallization time measurements (\unit[5]{nm} Sb). The resistance evolution with time is probed at different ambient temperatures. Upon recrystallization, the device resistance gradually decreases to the SET resistance. The resistance decrease between 0.9$\cdot$max(device resistance) and 2 $\cdot$ R\textsubscript{SET} is fitted linearly (black dashed lines). The intercept of the fit with 2$\cdot$R\textsubscript{SET} marks the crystallization time.  
		\textbf{b}, Thickness-dependent crystallization time. The time until a device recrystallizes from a RESET state to 2$\cdot$R\textsubscript{SET} appears to follow an Arrhenius temperature dependence. The 3 nm sample exhibits a slightly steeper temperature dependence. This change, however, is still within the margin of error. To obtain an average value for the crystallization time at a given temperature, devices were melt-quenched (and recrystallized) at least five times. Error bars correspond to the standard deviation of log\textsubscript{10}(crystallization time).}
		\label{fig:Nano_CrystallizationTime}
	\end{figure}

	In the bridge cells, Sb can be melt-quenched to an amorphous state. However, due to the extreme proneness to crystallization, a limited retention time must be expected at elevated temperatures. To quantify this, the evolution of the device resistance after melt-quenching is measured (\Cref{fig:Nano_CrystallizationTime} a \& \Cref{Sect:FastResMeas}). Initially, the resistance evolution with time is dominated by structural relaxation, but eventually recrystallization results in a continuous resistance decrease. In this experiment, the time elapsed after programming until the device resistance decreases to 2$\cdot$R\textsubscript{SET} is defined as the crystallization time. This crystallization time appears to depend on the temperature in an Arrhenius way (\Cref{fig:Nano_CrystallizationTime} b). 
	
	Confining the antimony has a striking impact on the stability against crystallization. The thickness reduction from \unit[10]{nm} to \unit[5]{nm} shifts comparable crystallization times to almost \unit[100]{K} higher temperatures. Even stronger confinement to \unit[3]{nm} boosts it by another two orders of magnitude. While a certain systematic error due to differently sized amorphous marks cannot be excluded, these differences alone can certainly not explain the increased crystallization times. Instead, they demonstrate that the crystal growth velocity is significantly reduced in the more narrowly confined material~\footnote{Sb-rich phase change materials are fast-growth materials with already low nucleation rates in much larger volumes \cite{Wang2000}. A small device volume like ours reduces the probability of nucleation even further and crystalline nuclei exist at the rim of the amorphous volume. Thus, the recrystallization can be assumed to be purely growth-driven.}. The amorphous phase gets stabilized against crystallization. 
	
	Besides the thickness reduction, different material configurations in the as-deposited state might contribute to the significantly lower crystallization times of the \unit[10]{nm} sample. Preliminary experiments showed that \unit[3]{nm} and \unit[5]{nm} Sb thin films were as-deposited amorphous but the \unit[10]{nm} thin film was as-deposited crystalline. Accordingly, compressive strain might facilitate the crystallization of the device with \unit[10]{nm} Sb, whereas tensile strain might hinder crystallization in the device with \unit[3]{nm} and \unit[5]{nm} Sb. Note that all devices were crystalline after fabrication. 
	
\begin{figure}[bth!]
		\centering
		\includegraphics[width=0.6\linewidth]{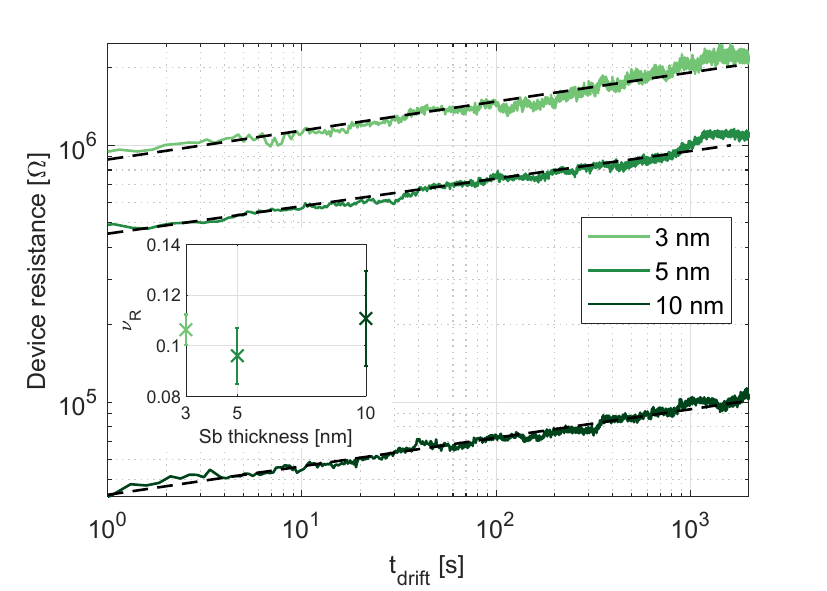}
		\caption[Unaltered resistance drift upon confinement]{\textbf{Unaltered resistance drift upon confinement.} The resistance drift shows no notable reduction with reduced Sb thickness. Even the \unit[3]{nm} thick films have a drift coefficient comparable to conventional phase change materials. The resistance drift was measured at \unit[100]{K} ambient temperature. The inset shows the mean drift coefficient obtained in three measurements.}
		\label{fig:Nano_Drift_Confinement}
	\end{figure}

	Upon confinement, the atomic rearrangements to a crystalline structure are slowed down by several orders of magnitude. Clearly, structural changes happen at a significantly different speed. Yet, the resistance drift coefficients hardly change (\Cref{fig:Nano_Drift_Confinement}). Intuitively these two findings may appear incompatible, but they are not. To crystallize the atoms must rearrange into one defined structure. The activation energy for these specific structural changes may very well increase. When the material relaxes, on the other hand, it changes only from one defective configuration to another. The activation energy for the next rearrangement increase with every relaxation step, but it does not define how rapidly the state of relaxation changes. Besides the activation energy for crystallization, confinement might also have an impact on the attempt to relax (or rearrange) frequency. It defines ultimately at what time processes with certain activation energy happen. Changes would reflect on the relaxation onset, the time when the glass approaches the super cooled liquid, and the crystallization time. But it would not influence how much the degree of relaxation changes per order of magnitude in time (i.e., the drift coefficient). 
	
	Nevertheless, one cannot infer from the thickness-independent drift coefficients that structural relaxation is not altered near the interfaces. Drift is also determined by how strong structural relaxation and electronic transport are coupled. In polymer glasses both, slower and faster relaxation upon confinement has been reported. Several studies report slowed down relaxation for polymers exhibiting an attractive interaction with the substrate and accelerated relaxation for suspended films. Additionally, also the chemical structure appears to play a significant role in these relatively large molecules \cite{Priestley2009}. At this point, it is not possible to differentiate whether both, structural relaxation and its coupling with the electrical transport, remain constant or counterbalance each other upon confinement in Sb.  
	
\begin{figure}[bth!]
		\centering
		\includegraphics[width=0.6\linewidth]{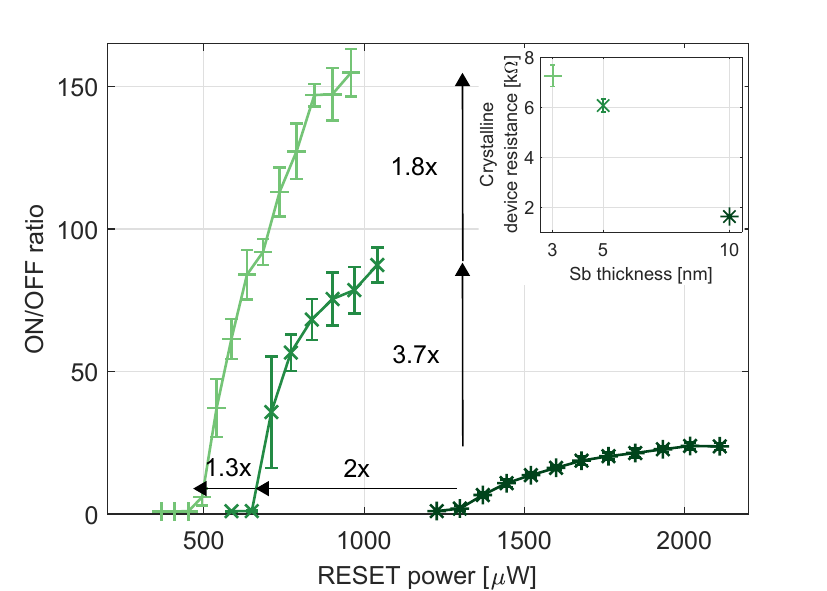}
		\caption[Improving cell efficiency and dynamic range by confinement]{\textbf{Improving cell efficiency and dynamic range by confinement.} The programming onset decreases fairly proportionally to the thickness reduction. Additionally, the ON/OFF ratio (R\textsubscript{RESET}/R\textsubscript{SET})increases, indicating larger amorphous volumes can be created in the device. Error bars denote the standard deviation of five identical electrical excitations. The devices are programmed with a \unit[3]{ns} trailing edge at an ambient temperature of \unit[100]{K}. The heat barrier thickness is \unit[40]{nm}.}
		\label{fig:Nano_ProgCurves_Confinement}
	\end{figure}

	Last but not least the devices' programming characteristics are compared. With reduced Sb thickness a smaller material volume must be molten to program the device (\Cref{fig:Nano_ProgCurves_Confinement}). Consistent with previous studies a reduction of the programming power with shrinking device dimensions can be observed \cite{Fong2017a}. In fact, the programming onset shifts almost proportionally to the thickness reduction. 
	
	At the same time, the maximum ON/OFF ratio increases. The crystalline device resistance scales linearly with the Sb thickness. Thus, the increased dynamic range must be attributed to differences in the RESET state. The analysis of the crystallization times indicates that structural  rearrangements happen orders of magnitude slower in a more confined device. Thus, a smaller fraction of the molten volume can recrystallize during melt-quenching. Additionally, the more confined glass might be less relaxed and therefore less resistive. Considering both effects, one can suppose that the length of the amorphous mark in the device increases by at least 3.7x and 1.8x as the Sb thickness is reduced to \unit[5]{nm} and \unit[3]{nm}, respectively.

\section{Summary \& Outlook}

	The first research question can be answered positively. It is possible to melt-quench Sb to an amorphous state. Efficient heat dissipation and abrupt ending of the electrical excitation proved to be instrumental parameters to change the device temperature by hundreds of kelvins within nanoseconds, sufficiently fast to avoid complete recrystallization of the molten volume. 
	
	Concerning the second question, it was discovered that one-dimensional confinement of the Sb through a thickness reduction does not alter the resistance drift coefficient, but it improves the device efficiency and increases the stability against crystallization by orders of magnitude. The metric for the stability against crystallization defined here only gives an estimate for how long two states (SET and RESET) can be distinguished. Thus, it should at best be considered a very optimistic estimate of the achievable retention times. Nevertheless, it is worth noting that the crystallization time of the \unit[3]{nm} Sb sample (\unit[32]{s} at \unit[80]{$^\circ$C}) already exceeds the DRAM refresh interval by 500x \cite{JEDEC2003}. An additional improvement of the stability against crystallization could be achieved by a further well-controlled reduction of the thickness or the use of other cladding materials. Recently published experiments by Cheng et al. indicate that optically melt-quenched Sb (\unit[3]{nm}), sandwiched between indium-tin-oxide might be more stable against crystallization \cite{Cheng2021}.  
	
	Three consequences of this study for future research and device optimization should be highlighted. 
	
	First, with ongoing device scaling interfacial and confinement effects will eventually outweigh bulk material properties. Instead of tuning phase change material compositions, the optimization of device properties can be achieved through the careful engineering of interfaces. To this end, a better understanding of how phase change material properties vary near different interfaces, e.g., due to their atomic-scale roughness, rigidity, nature of the chemical bond, or mechanical strain, is required. Quantitative analyses in theory, simulations, and experiments studying these dependencies under careful control of all mentioned aspects are lacking. This study demonstrates the impact such research could have for realizing reliable phase change-based devices integrated with the highest spatial density. 
	
	Second, the simplest material imaginable, a pure element, can become a viable alternative to established phase change materials when confined in extremely small volumes. In principle, any material must be expected to exhibit some resistivity contrast between its crystalline and amorphous phase. Other elements, besides Sb, could also be suitable for PCM devices. In particular, As has been identified as a potential candidate \cite{Suhrmann1939,Kooi2020}. Resistive switching with a relatively small contrast between crystalline and amorphous phase (\textless 2x) has been reported for devices based on two metallic glass formers, Ta and Mo \cite{Nakanishi2019,Nakanishi2019a}. 
	
	Third, with Sb, yet another material is added to the list of phase change materials exhibiting resistance drift. It must be stressed, once again, that structural relaxation is inherent to the glass state and there seems to be no good reason to assume it will not reflect on the electronic band structure. To suppress drift one must face the challenging endeavor of designing a material in which the density and mobility of free charge carriers do not change, despite a changing band structure. Additionally, we have seen that material confinement neither eliminates nor notably suppresses drift, at least in pure Sb. In the light of these findings, novel device concepts appear to be one of the most promising approaches to mitigate resistance drift. One of these ideas will be discussed in the next chapter, the projected memory cell. 
	
	\textbf{Key findings:}
	\begin{itemize}
		\item	Melt-quenched amorphous Sb exhibits the same characteristic properties as established phase change material compounds, namely semiconducting transport, a resistivity orders of magnitude higher than the crystalline phase, threshold switching and drift.
		
		\item	Material confinement improves the robustness against crystallization by orders of magnitude. Still, the Sb must be melt-quenched within a few nano-seconds to create an amorphous state, indicating that confinement might have a much weaker impact on crystal growth in the temperature regime of fast crystallization.
		
		\item	While confinement stabilizes the amorphous phase significantly against crystallization it has no notable impact on the resistance drift coefficient. Interestingly, a rearrangement to the crystalline structure is hindered, but drift, which also results from atomic rearrangement, appears not to differ.
		
	\end{itemize}

		\chapter{State-dependent drift in a projected memory cell}
\label{Chapt:Proj}	

\textbf{Preliminary remark:} The results presented in this chapter have been published as a journal article. \newline
\textit{State dependence and temporal evolution of resistance in a projected phase change memory - Kersting et al. - Scientific reports \cite{Kersting2020}} \\[6pt]

The idea of the projected memory is to decouple the device readout from the unstable electrical properties of the amorphous phase. For this purpose, an electrically conducting material, called the projection layer, is placed in parallel to the phase change material. The sheet resistance of the projection layer is chosen to lie between the sheet resistance of the crystalline and the amorphous states of the phase change material. In a projected memory device, the majority of the read current bypasses the amorphized volume in the device and flows in the projection layer instead. Hence, the resulting device resistance can be viewed as a projection of the length of the amorphous region onto the projection layer. In contrast to the read operation, most current flows through the phase change material during the write operation. When the amorphous phase undergoes threshold switching, its resistance decreases to a fraction of the projection layer's resistance. Like in an unprojected device, the Joule heating required to write the cell is localized inside the phase change material. Ideally, the stable resistance of the projection layer improves the device's memory performance, while the write operation does not change \cite{Redaelli2010,Kim2013,Koelmans2015}. 

In this chapter, a comprehensive device model is developed to capture the behavior of the projected memory bridge cell for any arbitrary device state. Here, device state refers to the phase configuration of the phase change material, i.e., the amount of crystalline and amorphous volume in the device. Particularly the interface resistance between the phase change material and projection layer is identified as a decisive parameter. It hinders the current flow into the projection layer and consequently determines the fraction of read current that bypasses the amorphous volume. Thus, it defines how effectively the projection works. Dependent on the interface resistance, drift characteristics, and scaling of the device resistance with the size of the amorphous volume change significantly. The model is experimentally validated on Sb bridge cells (\unit[3]{nm} thick), both with and without a projection layer. Finally, the modeling framework is extended to identify guidelines for materials selection to optimize projected PCM device characteristics.

\section{State of the art}
	
	\subsection{Projected memory - proof of concept}

	 A material that exhibits no drift, lower read noise and weaker temperature dependence than the amorphous phase change material is chosen for the projection layer. Consequently, drift and read noise reduce by more than one order of magnitude \cite{Kim2013,Koelmans2015}. Additionally, weak, linear temperature dependence of the resistance has been observed in projected devices, as opposed to the exponential dependence of the amorphous phase change material \cite{Giannopoulos2019}. Last but not least, it has been shown that projected devices can be recovered after a destructive write (void formation in the phase change material) \cite{Xie2018}. First studies of the ramifications on the in-memory compute performance report significantly improved multiplication accuracy (quantified by the reliability of a single device readout) \cite{Giannopoulos2019} and slower decay of neural network inference accuracy with time \cite{Bruce2021}. 
	
	Clearly, the proof of concept has been demonstrated. A comprehensive understanding of the device, however, is still lacking. To guide the design of future device generations, a deeper understanding beyond the idealized picture is required. In particular, the interface resistance between phase change material and projection layer must be expected to be finite. Additionally, the resistance ratio between the projection layer and the amorphous phase changes with time, and the projection efficiency increases. The consequences of these two aspects have not yet been addressed. To reduce drift and read noise by projection, one must compromise on the dynamic range. In the projected device it is defined by the resistance ratio of the projection layer and crystalline phase change material. Last but not least, it is essential to assess if the resistance drift is state-independent or depends on the size (length) of the amorphous volume.
	
	
	\subsection{The challenge of state-dependent drift coefficients}

	Among the strategies to mitigate drift, the idea of rescaling the results of the matrix-vector-multiply operation executed with the PCM crossbar array was discussed. In the case of resistance independent drift coefficients, this scheme would work perfectly. Since all values (resistances) stored in the crossbar increase by a constant factor, the elements of the result vector (column currents) decrease by the same factor, i.e., a global scaling factor could correct drift. In reality, however, drift coefficients vary. Thus, the relative contribution of individual devices to the column current changes with time. For a single matrix column, this variability could still be compensated by row-wise adjusting the read signal. However, since the complete matrix with 'randomly' distributed resistances is multiplied in parallel, the state-dependent drift cannot be corrected perfectly. The global drift correction only compensates for the mean drift behavior of the array. The changing relative contribution of individual devices to the column current, due to different drift coefficients, is not corrected and causes a continuous decrease in the precision with time \cite{LeGallo2018a,Joshi2020}. 
	
	The amorphous phase might exhibit varying drift coefficients. Device-to-device variability, for example, can result from differences in the local phase change material stoichiometry. Within a single device, the coupling between relaxation and electrical transport might differ in subsequent amorphizations due to the finite size effects. The amorphous volume in the device is so small, that variations in the disordered structure are not averaged out but reflect on the drift. Additionally, the drift coefficient changes depending on the device state, i.e., the phase configuration of amorphous and crystalline material in the device. As long as an amorphous volume fills the complete device cross-section, the average drift coefficient of an unprojected cell is constant \cite{Kersting2020,GhaziSarwat2021}. This value corresponds to the drift coefficient of the amorphous phase. Intermediate device states exhibit lower drift coefficients. The current path through the device is no longer blocked by a proper amorphous volume. Instead, (crystalline) percolation paths, exhibiting no or much less drift contribute to the device current. With decreasing device resistance the contribution of the percolation path to the device current increases and the drift coefficient decreases to the value of the crystalline state. Typically the crystalline state does not drift. In some doped phase change materials, however, the device SET state has been reported to exhibit drift \cite{Boybat2019,Bruce2021}. It might be caused by the structural relaxation of disordered material at crystal grain boundaries.  
	
	A parallel conduction path to the amorphous phase change material exists in the projected memory cell by design, even when no percolation paths exist in the amorphous volume. However, the resistance ratio between amorphous volume and parallel projection layer remains constant. Thus the projected device is expected to have a constant, yet reduced drift coefficient (when the amorphous volume has no percolation paths). If percolation paths exist in intermediate states, the drift will decrease further. In the scenario of an ideal projected memory cell, both, absolute drift coefficients and their state dependence should be smaller than in an unprojected cell. 
		
	\subsection{Research questions}
	\begin{enumerate}
		\item	Is it possible to design a tractable model of a projected memory bridge cell?
		\item	How do the drift coefficient and device resistance depend on the length of the amorphous volume and the interface resistance between projection and phase change material? 
		\item	How should projection and phase change materials be selected to achieve ideal device characteristics? 
	\end{enumerate}
\newpage
\section{A compact model for projected bridge cells}
	
	In this section, an equivalent circuit model for projected bridge cells will be introduced. With this model, the implications of an interface resistance (\Rint) between the phase change material and projection layer are studied. Additionally, the equivalent circuit model is compared to FEM simulations and the model input parameters for the projected Sb bridge cell are defined experimentally.    
	
	\subsection{Equivalent circuit model}
	
\begin{figure}[bth!]
		\centering
		\includegraphics[width=1\linewidth]{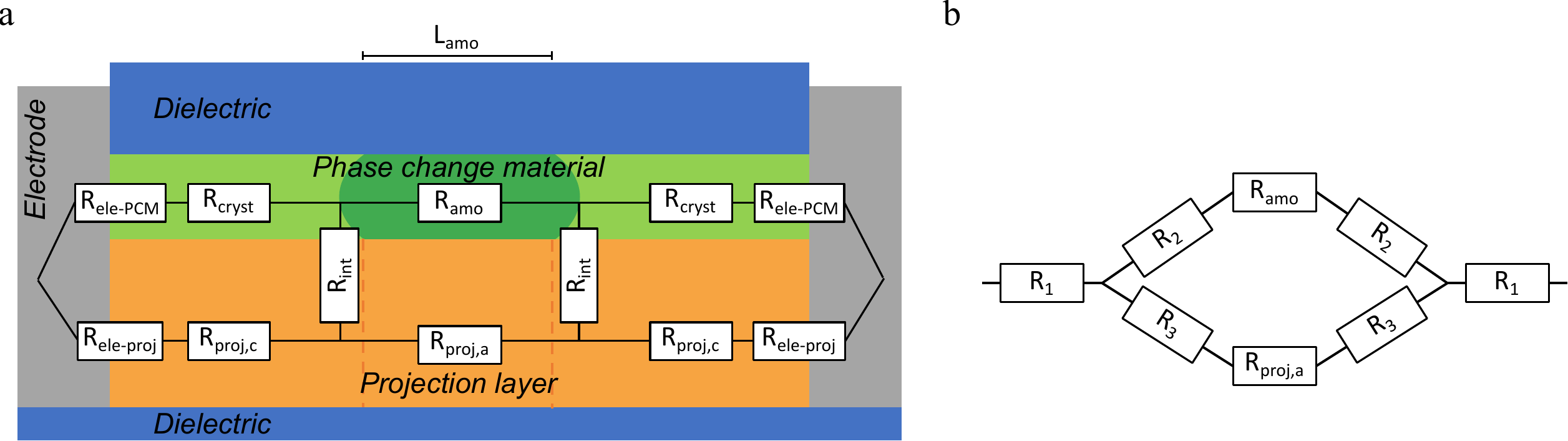}
		\caption[Equivalent circuit model of a projected bridge cell]{\textbf{Equivalent circuit model of a projected bridge cell.} The projected bridge cell is abstracted to a resistor network. Central building blocks are the crystalline and amorphous resistors with their parallel projection elements. Additionally, contact resistances to the electrodes and, crucially, an interface resistance between the phase change material and projection layer at the crystalline-amorphous interface are taken into account in this model.}
		\label{fig:PROJ_DeviceModel}
	\end{figure}

	In the model of a projected bridge cell, it is assumeed that during RESET the amorphous regions are created close to the center of the device (illustrated in \Cref{fig:PROJ_DeviceModel} a). This has been observed for lateral device structures in both, FEM simulations, as well as TEM studies \cite{Koelmans2015,Castro2007,Goux2009}. Minor shifts from the center of the confined device region can occur due to thermoelectric effects, but such shifts typically remain within the confined region of the device \cite{Castro2007,Oosthoek2015a}. An equivalent electrical circuit of this device during the low-field read operation can be described as a network of resistors, corresponding to the resistances of the crystalline and the amorphous region (\Rc{} \& \Ra) in the device, and the projection layer resistors in parallel to them (\Rprojc{} \& \Rproja). These resistors will be determined by the material’s sheet resistances, the device length (L\textsubscript{line}) and width (w), and the length of the amorphous region (L\textsubscript{amo}) in the device. They are defined by the following equations
	
	\begin{equation}
		R_{cryst} = \frac{1}{2} \cdot R_{s,cryst} \cdot \frac{L_{line}-L_{amo}}{w}
		\label{equ:ProjPCM_Rcryst}
	\end{equation}
	
	\begin{equation}
		R_{proj,c} = \frac{1}{2} \cdot R_{s,proj} \cdot \frac{L_{line}-L_{amo}}{w}
		\label{equ:ProjPCM_Rprojc}
	\end{equation}
	
	\begin{equation}
		R_{amo} = R_{s,amo} \cdot \frac{L_{amo}}{w}
		\label{equ:ProjPCM_Ramo}
	\end{equation}
	
	\begin{equation}
		R_{proj,a} = R_{s,proj} \cdot \frac{L_{amo}}{w}
		\label{equ:ProjPCM_Rproja}.
	\end{equation}
	\newpage
	
	Here, \Rsc{} denotes the sheet resistance of the crystalline phase change material, \Rsa{} the sheet resistance of the melt-quenched amorphous phase change material, and \Rsp{} the sheet resistance of the projection material. Additional circuit elements are the contact resistances of the metal electrodes to the projection and phase change material (\RePCM{} and \RePROJ) and the interface resistance (\Rint) between the phase change material and projection layer. These resistors may be non-negligible if a Schottky barrier forms like it has been reported at the interface between amorphous \GST{} and TiN \cite{Huang2014}, or could simply be due to the nanoscopic size of the contact. Last but not least, the atomic-scale structure (e.g. smooth interfaces, material mixing in an interface layer, or poor adhesion and local delamination) will have an impact on the contact resistance.
	
	Another representation of this resistor network is shown in \Cref{fig:PROJ_DeviceModel}b. With the $\Delta$-Y transformation, the $\Delta$ formed by the resistors \RePCM{}~+~\Rc, \Rint, and \RePROJ{}~+~\Rprojc{} (\Cref{fig:PROJ_DeviceModel}a) is translated to a Y formed by R\textsubscript{1}, R\textsubscript{2}, and R\textsubscript{3} (\Cref{equ:ProjPCM_R1,equ:ProjPCM_R2,equ:ProjPCM_R3}). 
	
	\begin{equation}
		R_{1} = \frac{(R_{ele-PCM} + R_{cryst})\cdot(R_{ele-proj}+R_{proj,c})}{R_{ele-PCM}+R_{cryst}+R_{ele-proj}+R_{proj,c}+R_{int}}
		\label{equ:ProjPCM_R1}
	\end{equation}
	
	\begin{equation}
		R_{2} = \frac{(R_{ele-PCM} + R_{cryst})\cdot R_{int}}{R_{ele-PCM}+R_{cryst}+R_{ele-proj}+R_{proj,c}+R_{int}}
		\label{equ:ProjPCM_R2}
	\end{equation}
	
	\begin{equation}
		R_{3} = \frac{(R_{ele-PCM} + R_{proj,c})\cdot R_{int}}{R_{ele-PCM}+R_{cryst}+R_{ele-proj}+R_{proj,c}+R_{int}}
		\label{equ:ProjPCM_R3}
	\end{equation}
	 \newpage 
	The total device resistance is given by 
	
	\begin{equation}
		R_{total} = 2 R_1 + \frac{(2 R_2 + R_{amo})\cdot(2 R_3 + R_{proj,a})}{2 R_2 + R_{amo}+2 R_3 + R_{proj,a}}.
		\label{equ:ProjPCM_Rtotal}
	\end{equation}
	
\begin{figure}[bth!]
		\centering
		\includegraphics[width=1\linewidth]{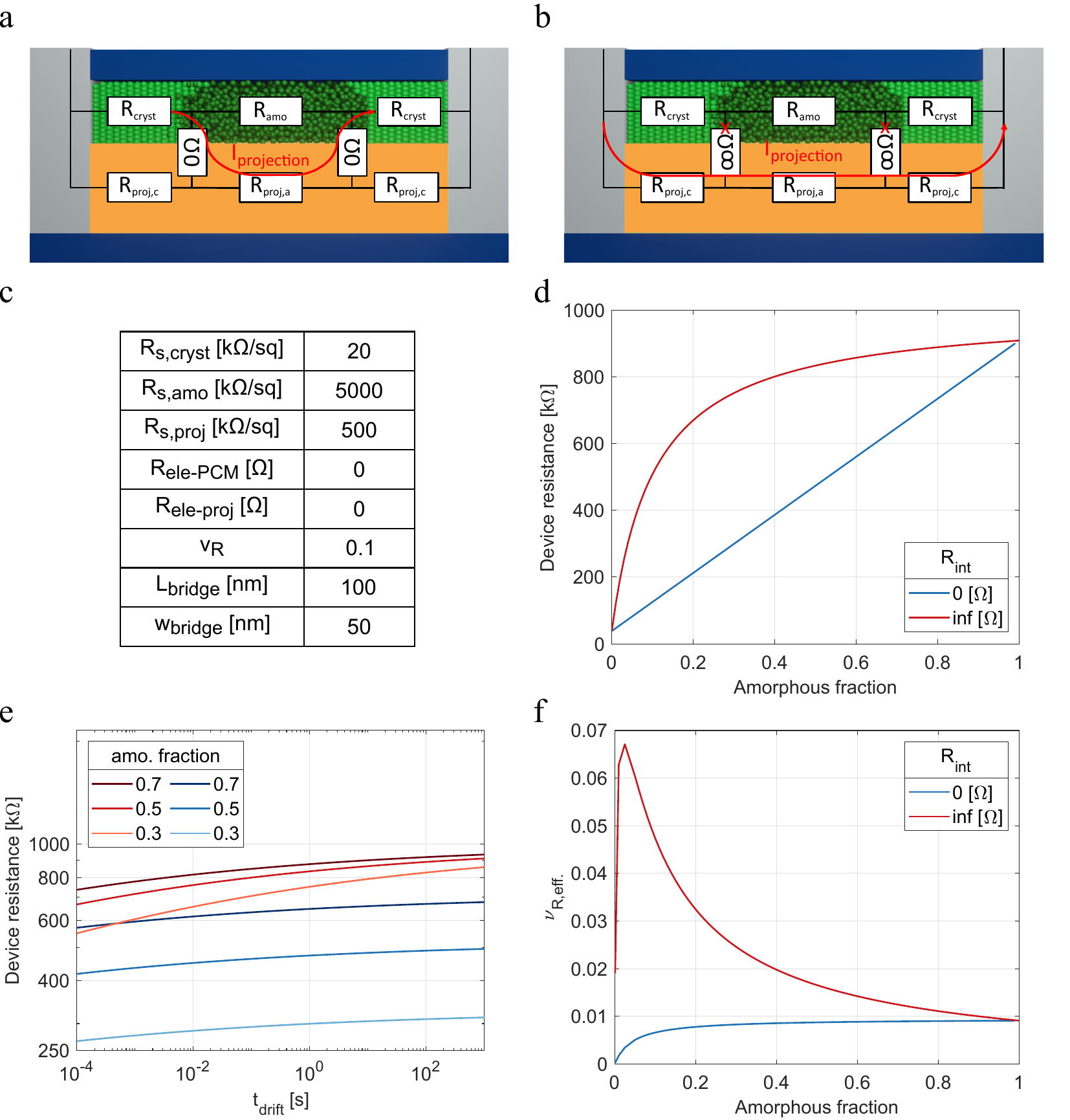}
		\caption[Model assessment for two borderline cases]{\textbf{Model assessment for two borderline cases.} \textbf{a}, For the scenario \Rint{}~=~\unit[0]{$\Omega$}, the projection current bypasses only the amorphous region. \textbf{b}, If \Rint{} is infinite, the projection current must bypass the entire phase change layer. \textbf{c}, Material parameters assumed for the model assessment. R\textsubscript{s} denotes the material sheet resistances. The contact resistance to the electrodes is assumed to be \unit[0]{$\Omega$} in order to study only the effect of the interface resistance between phase change material and projection layer. \textbf{d}, Device resistance as a function of the amorphous fraction. \textbf{e}, Temporal evolution of the device resistance for different RESET states. \textbf{f}, Drift coefficient one second after programming. Both, device resistance and drift coefficient are sensitive to the interface resistance. In a projected device drift decreases with time (\textbf{e}).}
		\label{fig:PROJ_Model_ExtremeRinterface}
	\end{figure}

	The model is employed to gain an insight into how the interface resistance influences the state dependence and temporal evolution of the device resistance. Initially, the two most extreme scenarios of zero and infinite \Rint~are compared. The material parameters assumed for this calculation are summarized in \Cref{fig:PROJ_Model_ExtremeRinterface}c. The resistance values are chosen such that the typical requirements for CMOS integration in large-scale crossbars are satisfied. For these purposes, the device SET resistance should be on the order of several tens of k$\Omega$ and the maximum RESET resistances close to \unit[1]{M$\Omega$} or larger in order to minimize the ‘IR’ voltage drop on the interconnects, which reduces the computational precision when performing in-memory analog computing \cite{Ielmini2018a}. Here, the contact resistances to the metal electrodes are disregarded in order to focus on the effect of the interface resistance between the phase change material and projection layer. The contact resistances for the experimental validation of the model will be included in the next section. The drift coefficient associated with the amorphous phase change material, $\nu_R$, is assumed to be 0.1.
	
	First, the state dependence of the resistance is studied. In the case of no interface resistance (R\textsubscript{int}~=~\unit[0]{$\Omega$}), the device comprises three serial elements of parallel resistors (see \Cref{fig:PROJ_Model_ExtremeRinterface}a). Two are crystalline regions parallel to the projection layer and one is the amorphous region parallel to the projection layer. All resistors in this equivalent circuit scale linearly with the length of the amorphous region in the bridge cell. Thus, the device RESET resistance scales linearly with the amorphous fraction of the device L\textsubscript{amo}/L\textsubscript{line} (blue trace in \Cref{fig:PROJ_Model_ExtremeRinterface}d). In the case of infinite interface resistance (\Rint~=~\unit[Inf]{$\Omega$}), the device comprises two parallel resistors (\Cref{fig:PROJ_Model_ExtremeRinterface}b). The entire projection layer is parallel to the phase change layer due to the infinite interface resistance. In other words, a fixed resistor (the projection layer) is connected in parallel to a resistor that changes with the amorphous fraction (the phase change layer). Consequently, the device resistance no longer scales linearly with the amorphous fraction as previously observed (red trace \Cref{fig:PROJ_Model_ExtremeRinterface}d).
	
	Next, the temporal evolution of device resistance is studied. In a projected PCM device, drift deviates from the standard power law for resistance drift $R=R_0 \cdot (t/t_0)^{\nu_R}$ (\Cref{fig:PROJ_Model_ExtremeRinterface}d) irrespective of the amorphous fraction or the interface resistance. The reason is that as the resistance of the amorphous phase increases with time, the current flowing through the amorphous plug decreases, whereas the current flowing through the element \Rproja{} remains constant (assuming a negligible change of the voltage drop over the amorphous segment). Hence over time, the drift suppression in a projected device becomes stronger and the effective drift coefficient is time-dependent. The temporal dependence of the drift coefficient is, however, influenced by the interface resistance.
	
	In the zero-interface resistance scenario, the resistance ratio between amorphous volume (\Ra) and parallel projection segment (\Rproja) is constant, since both elements scale with L\textsubscript{amo} (\Cref{equ:ProjPCM_Ramo,equ:ProjPCM_Rproja}). The current fraction flowing in the amorphous material, affected by drift, is independent of L\textsubscript{amo}. For small amorphous volumes, when the resistance of the crystalline fraction is non-negligible compared to \Ra$\parallel$\Rproja, the drift coefficient is reduced further. The more the overall device current is determined by the crystalline segment in series with the drifting circuit element (\Ra$\parallel$\Rproja), the less apparent is the effect of drift on the overall device current. Thus, device states with smaller amorphous fractions exhibit a slightly lower effective drift coefficient (blue trace in \Cref{fig:PROJ_Model_ExtremeRinterface}f).
	
	On the other hand, in a device with infinite interface resistance, the projection layer and phase change layer are electrically separated. The projecting segment here is constant \linebreak (R\textsubscript{proj}~=~\Rproja~+~\Rprojc~=~\unit[1]{M$\Omega$}). Instead of bypassing only the amorphous volume, the read current must bypass the entire phase change layer. Compared to the zero interface-resistance scenario, the fraction of read current flowing in the amorphous volume will always be larger. Moreover, the ratio of current flowing in the projection layer and the phase change layer will be strongly state-dependent. The smaller the amorphous region, the larger the fraction of read current passing through it. Accordingly, smaller amorphous regions lead to a larger effective drift coefficient and drift suppression is worse than in the zero interface-resistance scenario (red trace in \Cref{fig:PROJ_Model_ExtremeRinterface}f).
	
	\subsection{Assessment of model assumptions with FEM simulations}
	
	The proposed device model is an approximate picture of the real device. The goal is to provide an easily comprehensible and tractable model that is capable of capturing key device metrics. The model has two potential shortcomings. 
	
	First, the field-dependent transport of the amorphous phase is neglected. Hence, the results of this study are valid only in the case when the device is read in the ohmic regime. The antimony device studied in this work show an ohmic regime for read voltages smaller than \unit[$\sim$0.15]{V}. For doped \GST{} mushroom cells, ohmic transport has been reported for read voltages smaller than \unit[$\sim$0.2]{V} \cite{Gallo2015}. It can be assumed that it is usually feasible to perform the device read in the ohmic regime. 
	
	Second, in this model the interface resistor between phase change material and projection layer is localized at the boundary of the crystalline and amorphous phase change material. In reality, the device current will flow from the crystalline material to the projection layer in an extended range around the boundary of crystalline and amorphous phase change material. For metal to semiconductor contacts, this range is defined by the transfer length L\textsubscript{t} = ($\rho$\textsubscript{c}/R\textsubscript{sh})\textsuperscript{0.5}, where $\rho$\textsubscript{c} is the contact resistivity and R\textsubscript{sh} is the sheet resistance of the material from which the current is flowing to another material\cite{Schroder1991}. Accordingly, the material sheet resistances and the interface resistivity determine how localized the current flow from one material to another is. 
	
\begin{figure}[bth!]
		\centering
		\includegraphics[width=1\linewidth]{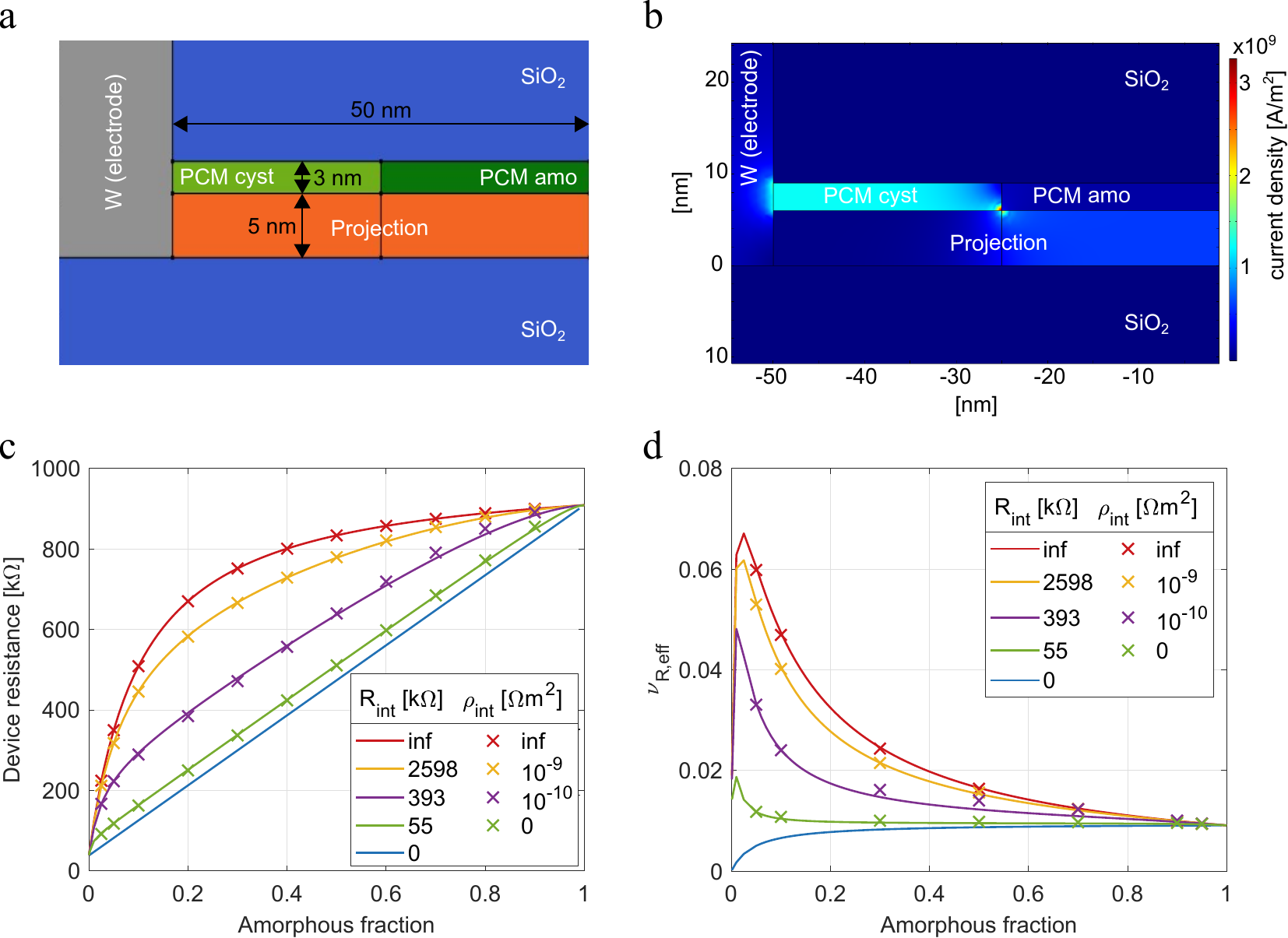}
		\caption[FEM vs. Equivalent circuit model]{\textbf{FEM vs. Equivalent circuit model. a,} FEM device geometry. The device read is simulated in a 2D model. In the FEM simulation, the spacing of mesh grid points is set to \unit[1]{nm}. \textbf{b}, Current density during a read with 100 mV. The interface resistivity between the projection layer and phase change material is $\rho$\textsubscript{int} = \unit[0]{$\Omega$\,m\textsuperscript{2}}. The current flow into the projection layer is localized at the crystalline-amorphous interface. \textbf{c},\textbf{d}, Resistance and drift coefficient as a function of the amorphous fraction. FEM simulations are marked by crosses, the equivalent circuit model by full lines. The interface resistance \Rint{} is fitted to match the FEM simulation. All FEM simulations can be reproduced with the simple equivalent circuit model.}
		\label{fig:PROJ_Model_FEM_study}
	\end{figure}

	The potential error introduced by this model simplification is assessed with 2D FEM simulations (\Cref{fig:PROJ_Model_FEM_study}). The device geometry and material sheet resistances are identical to those defined in \Cref{fig:PROJ_Model_ExtremeRinterface}c. In the FEM model, the current flow between phase change material and projection layer is hindered by the interface resistivity ($\rho$\textsubscript{int}), instead of the localized interface resistance. 
	
	First indications for the soundness of the equivalent circuit model become clear from the simulated current densities (\Cref{fig:PROJ_Model_FEM_study}b). For $\rho$\textsubscript{int} = \unit[0]{$\Omega$\,m\textsuperscript{2}}, i.e., unhindered current flow from the phase change material into the projection layer, the current flow is strongly localized at the crystalline-amorphous interface. 
	
	The equivalent circuit model is further corroborated by comparing the device resistance and drift coefficient as a function of the amorphous fraction with the FEM model. Besides the two scenarios $\rho$\textsubscript{int} \unit[0]{$\Omega$\,m\textsuperscript{2}} and \unit[$\infty$]{$\Omega$\,m\textsuperscript{2}} also two finite $\rho$\textsubscript{int} are simulated (\unit[10\textsuperscript{-9}]{$\Omega$\,m\textsuperscript{2}} and  \unit[10\textsuperscript{-10}]{$\Omega$\,m\textsuperscript{2}}). To identify the corresponding interface resistance, the equivalent circuit model is fitted to the device resistance as a function of the amorphous fraction obtained in the FEM simulation. The only fitting parameter is \Rint. For each $\rho$\textsubscript{int}, a corresponding \Rint{} allows reproducing the scaling of device resistance and drift coefficient with the amorphous fraction (\Cref{fig:PROJ_Model_FEM_study}c\&d). Interestingly, the two models don't match for $\rho$\textsubscript{int} = \unit[0]{$\Omega$\,m\textsuperscript{2}} and \Rint{}~=~\unit[0]{$\Omega$}. Instead, the FEM simulation with zero-interface resistivity is reproduced with an interface resistance of \unit[55]{k$\Omega$}. The extreme confinement of current flow from the phase change material to the projection layer results in high current densities and current crowding at the interface, leading to a finite effective \Rint.  
	\newpage
	It becomes clear that the distributed interface resistivity in a real device, as well as the current flow from one layer to another over an extended range, can be appropriately described by a localized effective interface resistance located at the boundary between amorphous and crystalline phase change material. However, the link between material interface resistivity and \Rint{} may be described by a non-trivial functionality that has not been defined as a part of this work.  
	\newpage
	\subsection{Projected antimony bridge cell}
	
	Experimentally unprojected and projected bridge cells based on pure Sb are studied. The quantitative assessment of the model should rely on as few unknown fitting parameters as possible. For this reason, the device geometry is inspected by Scanning Transmission Electron Microscopy (STEM) and unprojected bridge cells and thin-film reference structures of the metal nitride are electrically characterized. By this means all model input parameters except for \Rint{} and L\textsubscript{amo} can be obtained in preliminary experiments \Cref{tab:PROJ_ModelParams}. 
	
	\begin{table}[h!]
		\centering
		\begin{tabular}{ c c}
			\hline
			\Rsp{} [k$\Omega$/sq] & 21.8 \\
			\Rsc{} - Sb [k$\Omega$/sq] & 1.26 \\
			\Rsa{} - Sb [k$\Omega$/sq] & 410 $\pm$ 60 \\
			R\textsubscript{W-Sb} [k$\Omega$] & 1.6 \\
			R\textsubscript{W-proj} [k$\Omega$] & 78 to 202 \\ 
			\hline
		\end{tabular}
		\caption[Experimentally obtained model input parameters]{\textbf{Experimentally obtained model input parameters}: The table summarizes the material sheet resistances and contact resistances to the W electrode at an ambient temperature of \unit[200]{K}. The sheet resistance of the amorphous state corresponds to the melt-quenched state one second after device RESET. The contact resistance W to projection material was measured on macroscopic reference structures and extrapolated to the nanoscopic contact area in the device. Accordingly, R\textsubscript{W-proj} is estimated with lower and upper bounds of \unit[78]{k$\Omega$} and \unit[202]{k$\Omega$}, respectively. The errors of the other parameters are negligible, and thus excluded in the analysis.}
		\label{tab:PROJ_ModelParams}
	\end{table}  
	
\begin{figure}[bth!]
		\centering
		\includegraphics[width=1\linewidth]{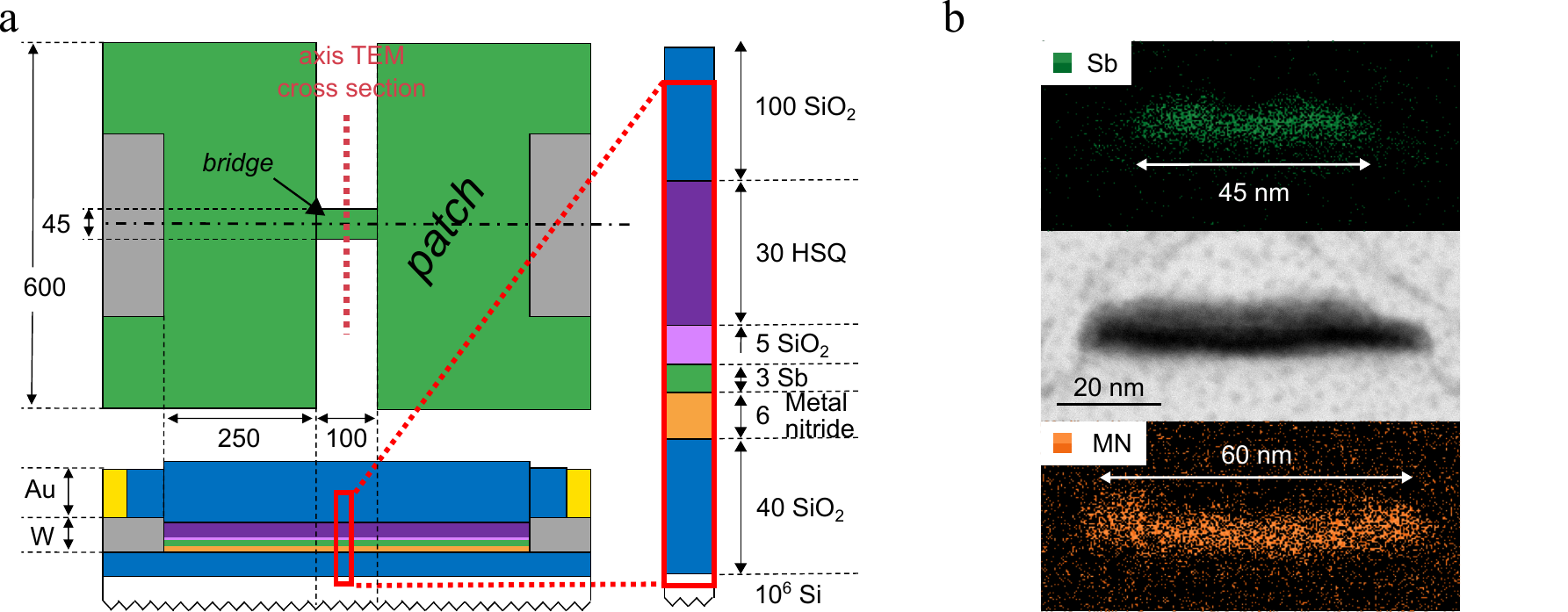}
		\caption[Projected antimony bridge cell]{\textbf{Projected antimony bridge cell. a}, Top view and cross-sectional sketch of the bridge cell with dimensions in nm. HSQ, hydrogen silsesquioxane e-beam resist.  \textbf{b}, STEM and EDX analysis of the active region cross-section along the axis marked in \textbf{a}. The top and bottom panels show an EDX-map for antimony (green) and the metal nitride (orange). These images give an estimate of the bridge width. The central panel is a bright-field STEM image.}
		\label{fig:PROJ_AntimonyDevice}
	\end{figure}

	In the projected device (\Cref{fig:PROJ_AntimonyDevice}), \unit[6]{nm} metal nitride, functioning as projection material, is added to the initially deposited layer stack (metal nitride, Sb, cladding). It is sputter-deposited without a vacuum break. The subsequent fabrication process is identical to the unprojected bridge cell (see \Cref{Sect:FabricationBridge})~\footnote{Both projected and unprojected bridge cells studied in this chapter have a SiO\textsubscript{2} cladding. The devices studied in \Cref{Chapt:Confine} have a ZnS:SiO\textsubscript{2} cladding instead.}. The cross-section of the \unit[100]{nm} long bridge was investigated by STEM and energy dispersive X-ray spectroscopy (EDX) (\Cref{fig:PROJ_AntimonyDevice}b). In this active region that switches between the crystalline and amorphous state, the Sb is \unit[45]{nm} wide. The metal nitride stripe underneath is \unit[60]{nm} wide. Differences in the widths of projection and phase change lines result from the mismatch of the etching rates of the phase change and projection layer. During fabrication, the width of the phase change material is reduced by \unit[15]{nm} due to over-etching. The width of the unprojected device is reduced by approximately \unit[8]{nm}, due to over-etching. Note, that a reasonable estimate of the device width is essential to correctly determine the resistors in the equivalent circuit model~\footnote{Vara Prasad Jonnalagadda fabricated the devices and prepared TEM lamellas. Marilyne Sousa performed the TEM measurements}. 
	
\begin{figure}[tbh!]
		\centering
		\includegraphics[width=1\linewidth]{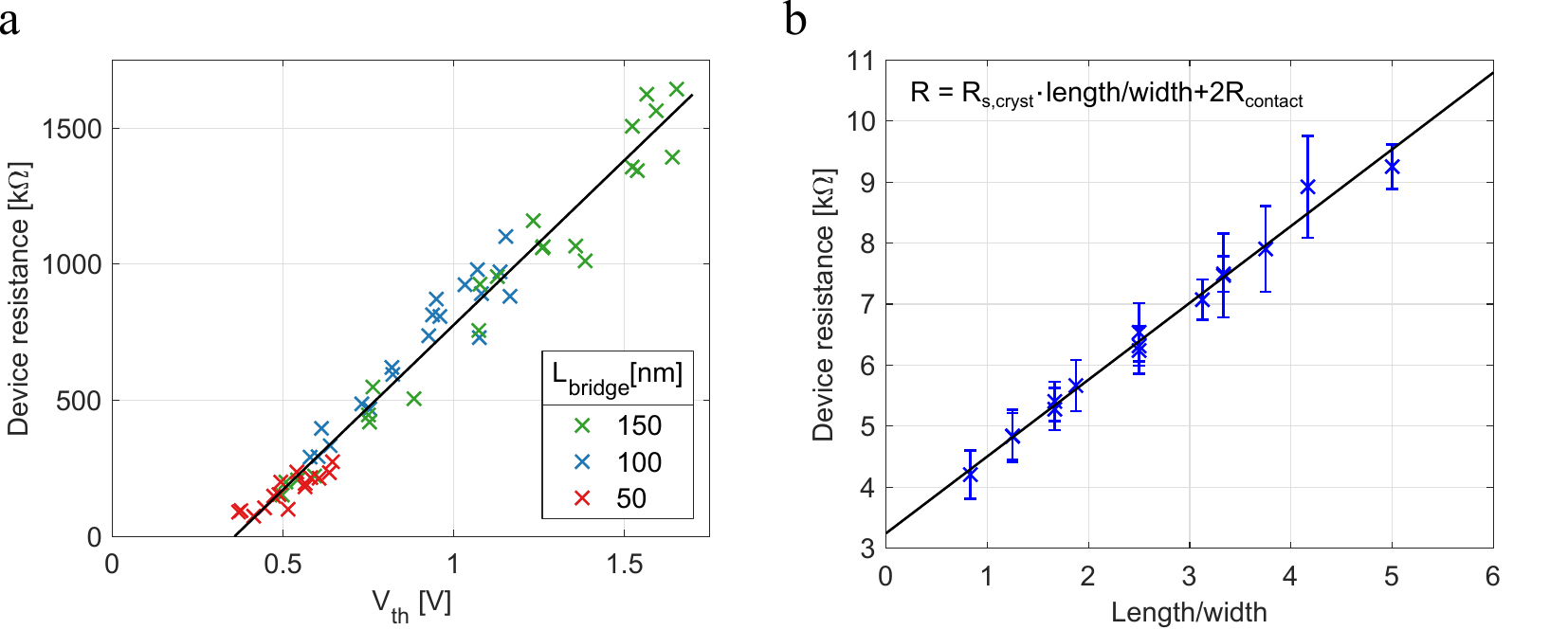}
		\caption[Resistance of amorphous and crystalline Sb]{\textbf{Resistance of amorphous and crystalline Sb. a}, RESET resistance and threshold voltage of devices with different bridge lengths obtained from a series of programming experiments. Both metrics scale linearly with L\textsubscript{amo} \cite{Krebs2009a}. The increase of the achieved RESET resistance with the device length affirms that amorphous volumes of increasing length are created. The bridge of these devices is \unit[32]{nm} wide. \textbf{b}, Resistance in the crystalline, pristine device state. Error bars denote the standard deviation measured across multiple devices of the same bridge geometry.}
			
		\label{fig:PROJ_AntimonySheetResistance}
	\end{figure}

	The electrical characterization of the devices is performed at an ambient temperature of \unit[200]{K}, to exclude potential recrystallization of the Sb in these measurements. An estimate for the sheet resistance of the melt-quenched amorphous state (\Rsa -Sb) is obtained from RESET resistances of devices with different bridge lengths (\Cref{fig:PROJ_AntimonySheetResistance}a). It is based on two assumptions about the geometrical constraints of L\textsubscript{amo}. First, the maximum of L\textsubscript{amo} is defined by the bridge length (i.e., the amorphous volume does not extend beyond the confined region in the device). If the highest RESET resistance measured for the \unit[100]{nm} and \unit[150]{nm} long bridges corresponds to a fully amorphous bridge, the sheet resistance would be \unit[350]{k$\Omega$/sq}. This defines the lower limit of \hbox{\Rsa-Sb}.
	Second, it is assumed that L\textsubscript{amo} of the least resistive RESET state (\unit[74]{k$\Omega$}) is \unit[$\sim$5]{nm}. This corresponds to a sheet resistance of \unit[470]{k$\Omega$/sq}. The amorphous length must be smaller than \unit[6.75]{nm}, otherwise the highest resistive RESET states would be larger than the device length. The \unit[74]{k$\Omega$} RESET state is still more than 10x larger than the SET resistance. If L\textsubscript{amo} gets too small, the probability of a crystalline percolation path increases, and consequently, only a very subtle resistance increase would be expected. Based on these considerations \Rsa{} is estimated in the range \unit[410$\pm$60]{k$\Omega$/sq}.
	
	The sheet resistance of crystalline Sb is calculated from the pristine resistance of devices with different geometries (\Cref{fig:PROJ_AntimonySheetResistance}b). The device resistance scales linearly with the aspect ratio (length/width) of the bridge. The crystalline sheet resistance is \hbox{\Rsc-Sb~=~\unit[1.26]{k$\Omega$/sq}} and the contact resistance is R\textsubscript{contact} = \unit[1.6]{k$\Omega$}. R\textsubscript{contact} is equivalent to the model parameter R\textsubscript{W-Sb}. It is the sum of the actual contact resistance between the W electrode and Sb and the resistance of the Sb which is not part of the confined bridge, labeled patch in \Cref{fig:PROJ_AntimonyDevice}. Details on the experimental procedure to determine \Rsp, R\textsubscript{W-Sb}, and R\textsubscript{W-proj} are summarized in \Cref{Sect:ProjBridge_preliminary}.  
	
\section{Experimental verification of the model}
	
	In this section, drift and RESET resistance of unprojected and projected Sb bridge cells and their dependence on L\textsubscript{amo} are compared. Measurements of the unprojected device provide the last reference points required to fit the characteristics of the projected device to the model. All experiments are performed at \unit[200]{K} ambient temperature.  
	
	\subsection{Drift characterization}
	
\begin{figure}[bth!]
		\centering
		\includegraphics[width=1\linewidth]{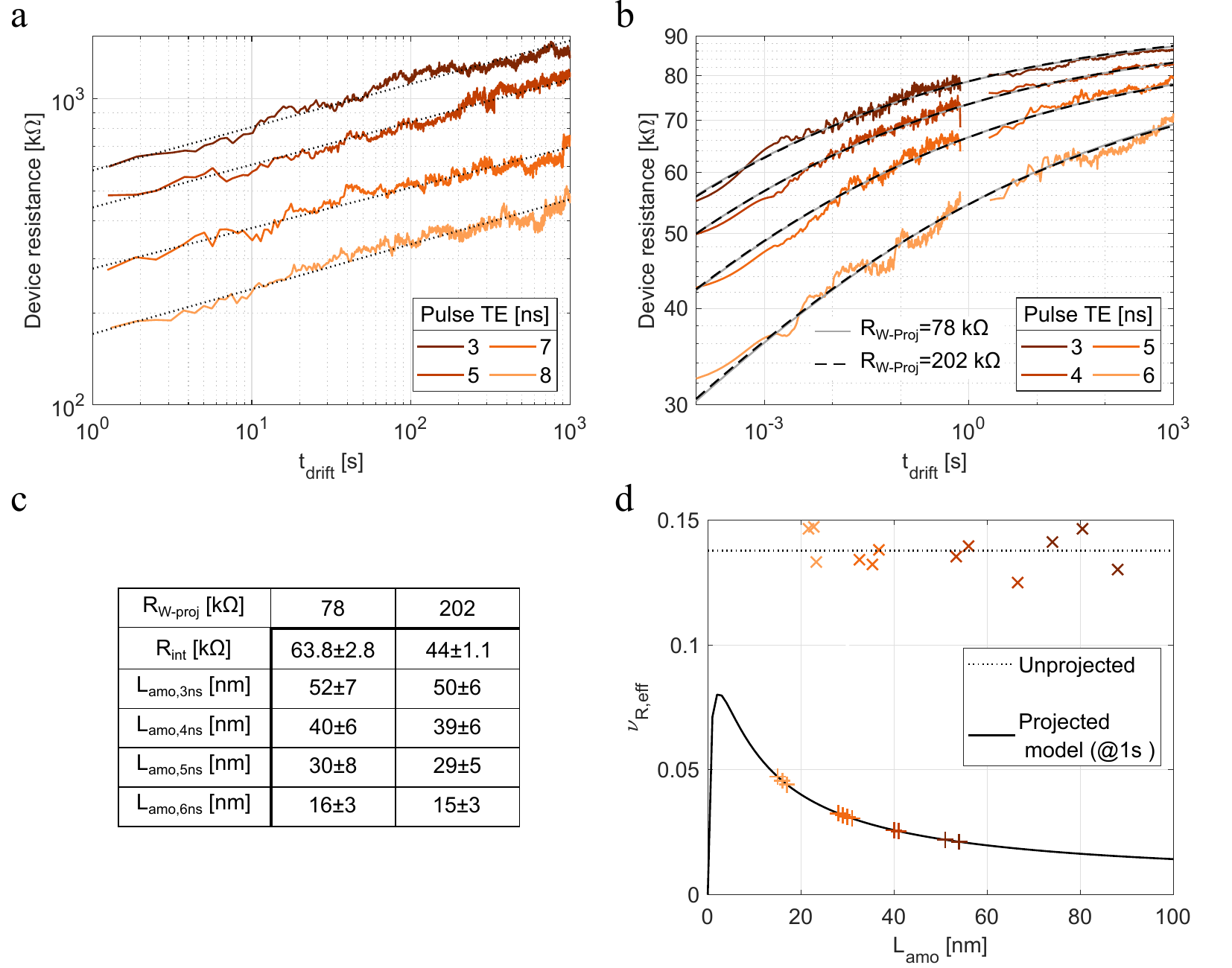}
		\caption[Drift measurements of unprojected and projected Sb bridge cells]{\textbf{Drift measurements of unprojected and projected Sb bridge cells.} Different device states were obtained by varying the RESET pulse trailing edge (TE). \textbf{a}, Unprojected cell drift. \textbf{b}, Projected cell drift. Resistance measurements up to \unit[700]{ms} were obtained with an oscilloscope (\Cref{Sect:FastResMeas}). Measurements from two seconds onward were obtained with a source meter unit. The experimental data are fitted to the device model presented in \Cref{fig:PROJ_DeviceModel}. For both the upper and lower bound of R\textsubscript{W-proj}, the model captures the experimental data. \textbf{c}, Fitted parameters: Interface resistance and an individual amorphous length for each RESET state. The error margin is the propagated error of R\textsubscript{s,amo}~-~Sb (\Cref{tab:PROJ_ModelParams}). \textbf{d}, Effective drift coefficient as a function of the amorphous length. The amorphous length of the unprojected cell is calculated for an amorphous sheet resistance of \unit[410]{k$\Omega$/sq}. The drift coefficients measured on unprojected cells (times symbol) are state-independent, whereas the projected cell shows a notable state (plus symbols) and time dependence (\text{b}) of the resistance drift.} 
		\label{fig:PROJ_Experiment_Drift}
	\end{figure}

	The unprojected bridge cell is programmed to different resistance states with a constant programming current of \unit[610]{$\mu$A} by varying the RESET pulse trailing edge between 3, 5, 7, and \unit[8]{ns}. With increasing pulse trailing edge (duration of the melt-quenching process), the RESET resistance decreases, because a larger amount of molten material crystallizes. After programming, the evolution of the resistance as a function of time is measured over \unit[1000]{s} and fitted to the standard drift equation $R=R_0 \cdot (t/t_0)^{\nu_R}$ (\Cref{fig:PROJ_Experiment_Drift}a). Four different intermediate states have a drift coefficient of 0.14 $\pm$ 0.01 with no apparent dependence of the drift coefficient on the pulse trailing edge. Hence, the drift coefficient is independent of the amorphous length. 
	
	Compared to the measurements at \unit[100]{K} ambient temperature (\Cref{fig:Nano_Drift_Confinement}) the drift coefficient increases by around 0.04 at \unit[200]{K}. Measurements on a single device at both temperatures confirm that this increase from 0.1 to 0.14 is not an artifact of inter-device variability or different cladding materials in the two studies (ZnS:SiO\textsubscript{2} vs. SiO\textsubscript{2}). In general, the collective relaxation model predicts a temperature-independent resistance drift coefficient (assuming constant temperature during the drift measurement). This has been observed experimentally for ambient temperatures in the range from \unit[160]{K} to \unit[420]{K} \cite{LeGallo2018}. However, the statement is only true if the coupling between structural relaxation and electrical conduction does not change. It could happen for example, if the transport mode changes in a certain temperature range (\Cref{Sect:VthDrift_Temperature}), and that might be the case here. The amorphous Sb exhibits an Arrhenius temperature dependence between \unit[225]{K} and \unit[175]{K} and a continuous  deviation from this extrapolation with decreasing temperatures for amorphous Sb (\Cref{fig:Nano_PCMcharacteristics}). This change in temperature dependence indicates a transition to another conduction mode.  

	We now return to the experimental characterization of the projected device. The model predicts a deviation of the resistance drift from the standard relation. The time-dependent resistance of four different RESET states was measured experimentally (\Cref{fig:PROJ_Experiment_Drift}b). In accordance with the model, the evolution of resistance with time exhibits a clear curvature in the double logarithmic plot. Additionally, the separation between different resistance states decreases with time, indicating a non-zero \Rint{} between the Sb and projection layer.
	
	In order to verify the device model, it is fitted to the measured R(t) data. Having predefined all experimentally directly accessible input parameters (\Cref{tab:PROJ_ModelParams,fig:PROJ_AntimonyDevice}b) and the drift coefficient of antimony, two fit parameters are left to describe the R(t) data of an individual RESET state. Those are the \Rint{} between antimony and metal nitride and the L\textsubscript{amo} of the RESET state (\Cref{fig:PROJ_DeviceModel}a). The R(t) traces measured for four different RESET states are fitted collectively. The model can describe the experiment using a single value of \Rint{} and four different L\textsubscript{amo}. 
	
	Of the experimentally determined model input parameters, the contact resistance R\textsubscript{W-proj} was estimated with the greatest uncertainty. For both the lower and upper bounds of R\textsubscript{W-proj}, the fit captures the experimental data well. The amorphous length obtained from the fit is approximately independent of R\textsubscript{W-proj} (\Cref{fig:PROJ_Experiment_Drift}c). Hence, in the model, the two parameters R\textsubscript{W-proj} and R\textsubscript{int} compensate each other. They determine the fraction of read current that is injected into the projection layer and thus not changed by drift. 
	
	With all model parameters at hand, the state-dependent effective drift coefficient of the projected bridge cell, one second after RESET, is calculated (\Cref{fig:PROJ_Experiment_Drift}d). Due to the notable \Rint{}, which is comparable to the sheet resistance of the projection layer, the drift exhibits a pronounced state dependence. Depending on the length of the amorphous volume, the drift coefficient is suppressed by a factor of two (for L\textsubscript{amo}~=~\unit[2]{nm}) up to a factor of 10 (for L\textsubscript{amo}~=~\unit[100]{nm}) compared to the unprojected device.
	
	A critical question to answer is if the magnitude of the obtained interface resistance is plausible. To this end, an estimate of the contact resistivity is required. The interface resistance can be described as 
	
	\begin{equation}
		\Rint = \frac{\rho_c}{L_t \cdot w} \cdot coth(L/L_t)
		\label{equ:PROJ_TransferLength}
	\end{equation} 
	
	where $\rho_c$ is the specific contact resistivity, L\textsubscript{t} the transfer length, w the width of the contact area (perpendicular to the direction of current flow), and L the length of the contact area (in the direction of current flow) \cite{Schroder1991}. In the bridge cell, L is the length of the contact between the crystalline phase change material and the projection layer. For the largest amorphous volume (\unit[50]{nm}), the contact between metal nitride and crystalline Sb is \unit[25]{nm} long. Assuming a transfer length of \unit[10]{nm} (i.e., L \textgreater 1.5 L\textsubscript{t}), the interface resistivity in the device can be approximated as $\rho_c = \Rint\cdot w\cdot L_t \approx \unit[50]{k\Omega} \cdot \unit[45]{nm} \cdot \unit[10]{nm} =$ $\unit[2.25\cdot10^{-7}]{\Omega\,cm^2}$. Specific contact resistivities in the range of \unit[10\textsuperscript{-5} to 10\textsuperscript{-8}]{$\Omega$\,cm\textsuperscript{2}} to metals and metal nitrides have been reported for crystalline \GST{} and GeTe \cite{Cooley2020}. The contact resistance derived in the model fit is well within this range. 
	
	Dedicated contact resistance measurements also show that surface treatment prior to deposition of the phase change material and annealing influence the contact resistance \cite{Cooley2020}. During the device fabrication, the metal nitride, the Sb, and the cladding were deposited without a vacuum break. Subsequent XRR measurements of blanket films indicate a low-density interfacial layer between metal nitride and Sb, which might be attributed to poor adhesion between the two materials. An optimized deposition process could result in better electrical contact and thus improved device characteristics. 
		
	\subsection{Resistance scaling with amorphous volume}
	
	As a second consequence of \Rint, the resistance of the projected device does not scale linearly with the amorphous length. This model prediction is examined experimentally based on the correlation of R\textsubscript{reset} and \Vth{}~\footnote{The threshold switching IV characteristics of the projected devices do not exhibit a sharp snap-back point which typically is used as a threshold criterion. The same was observed for the unprojected device (\Cref{fig:Nano_PCMcharacteristics}b). For this reason threshold switching is defined as the point where the device current exceeds \unit[12.5]{$\mu$A}. }. In an unprojected bridge cell, the threshold voltage depends linearly on the amorphous length \cite{Krebs2009a}.

	\begin{equation}
		V_{th} = E_{th}\cdot L_{amo} + V_{offset}
		\label{equ:VthGeneral}
	\end{equation}
	
\begin{figure}[bth!]
		\centering
		\includegraphics[width=0.6\linewidth]{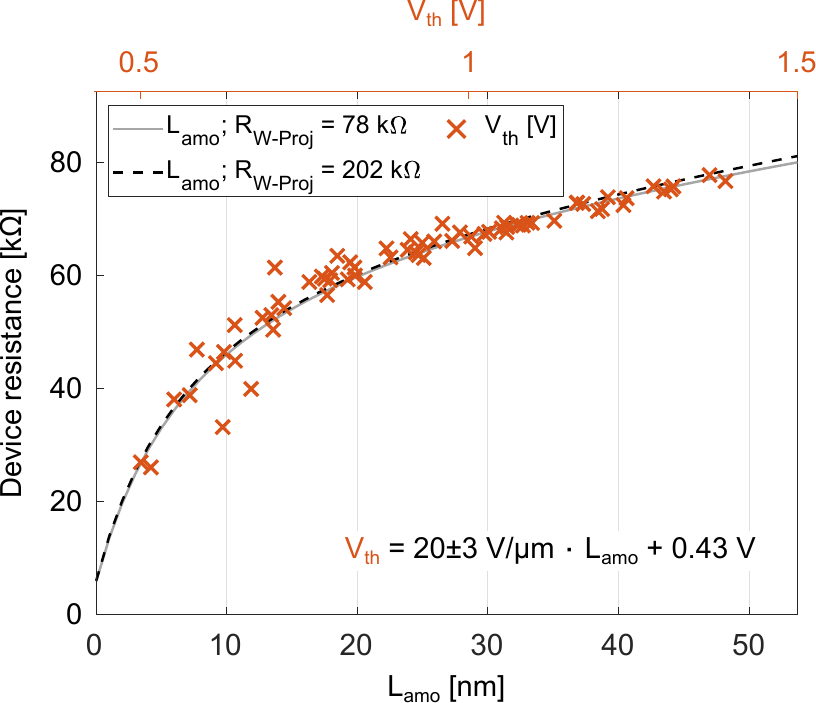}
		\caption[State dependence of RESET resistance in projected Sb bridge cells]{\textbf{State dependence of RESET resistance in projected Sb bridge cells.} State dependence refers to the dependence of the device resistance on the phase configuration, i.e., the amorphous length. Black lines depict the scaling of RESET resistance with the amorphous length predicted by the device model (bottom x-axis). Crosses mark the experimentally obtained data R\textsubscript{RESET} vs V\textsubscript{th} (top x-axis). The linear relation between the top and bottom x-axis is noted in the graph.}
		\label{fig:PROJ_Experiment_ResScaling}
	\end{figure}

	Here E\textsubscript{th} denotes the threshold field and V\textsubscript{offset} an offset voltage. It is expected that this relation is also valid for the projected bridge cell. Switching still occurs at a critical field strength inherent to the phase change material and the field across the amorphous volume is not changed by the electrically parallel projection layer. Accordingly, it should be possible to map R(L\textsubscript{amo}) calculated with the model to the experimental data R(\Vth) using \Cref{equ:VthGeneral}. The best match of the two data sets is obtained for V\textsubscript{th}~=~ \unit[20$\pm$3]{V/$\mu$m} $\cdot$ L\textsubscript{amo} + 0.43 V (\Cref{fig:PROJ_Experiment_ResScaling}). The state dependence predicted with the model is neatly reproduced by the experiment. 
	
	The error margin of the threshold field results from the uncertainty of \Rsa{}, and is independent of R\textsubscript{W-Proj}. Notably, the same threshold field was measured on unprojected Sb bridge cells. Krebs et al., however, report a threshold field of \unit[94$\pm$9]{V/$\mu$m} for as-deposited amorphous Sb (\unit[4]{nm} thick). This discrepancy probably cannot be explained by a more relaxed glass state in the as-deposited amorphous sample. If one assumes that threshold switching is induced by a thermal feedback loop \cite{LeGallo2016}, the different temperatures at which the experiments were performed (\unit[200]{K} vs. room temperature) and different slopes of the voltage increase (\unit[1.5]{V}/\unit[60]{ns} vs. "slow to avoid transient effects" \cite{Krebs2010}) could be two hypothetical reasons. De facto a clear answer to the source of the difference cannot be given at this point.  
		
\section{Guidelines for device optimization}
	
	After validating the proposed device model experimentally, the implications for device applications and the selection of materials for a projected memory device will be discussed in this section. The state dependence of device drift introduced by a large interface resistance is unfavorable even though the projection layer may reduce the absolute device drift. It cannot be corrected with a global scaling factor. Also, a non-linear scaling of RESET resistance with amorphous length may introduce challenges in a device application. The resistance of RESET states with small amorphous lengths changes significantly with the size of the amorphous volume. Thus, it becomes more challenging to program the device to these states. Consequently, only a fraction of the possible device states may be exploited. This could decrease the dynamic resistance range and the number of resistance states that can be distinguished reliably.
	
	Even though finite interface resistances pose a challenge to the concept of the projected PCM, within certain bounds they can be tolerated without detrimental effects on the device characteristics. Three constraints for the device metrics are defined to identify those bounds. First, the maximum drift coefficient should be smaller than 0.01 for amorphous fractions larger than \unit[5]{\%}. For very small amorphous fractions the drift coefficients increase steeply (\Cref{fig:PROJ_Model_FEM_study}d). Most likely these are only hypothetical device states in the model and crystalline percolation paths, reducing drift and resistance, would exist in a real device. The \unit[5]{\%} limit is supposed to exclude these states. Second, to restrict drift variability, the resistance separation between the most and least drifting device states must not change by more than \unit[5]{\%} over 4 orders of magnitude in time (\unit[1]{s} to \unit[10\textsuperscript{4}]{s}). This criterion is expected to be more tangible than a range within which the drift coefficients may vary. Third, the device resistance should deviate for no amorphous fraction by more than \unit[20]{\%} from the linear interpolation between R(amorphous fraction = 0) and R(amorphous fraction = 1). This constraint enforces an approximately linear scaling of the device resistance with L\textsubscript{amo}. 
	
\begin{figure}[bth!]
		\centering
		\includegraphics[width=1\linewidth]{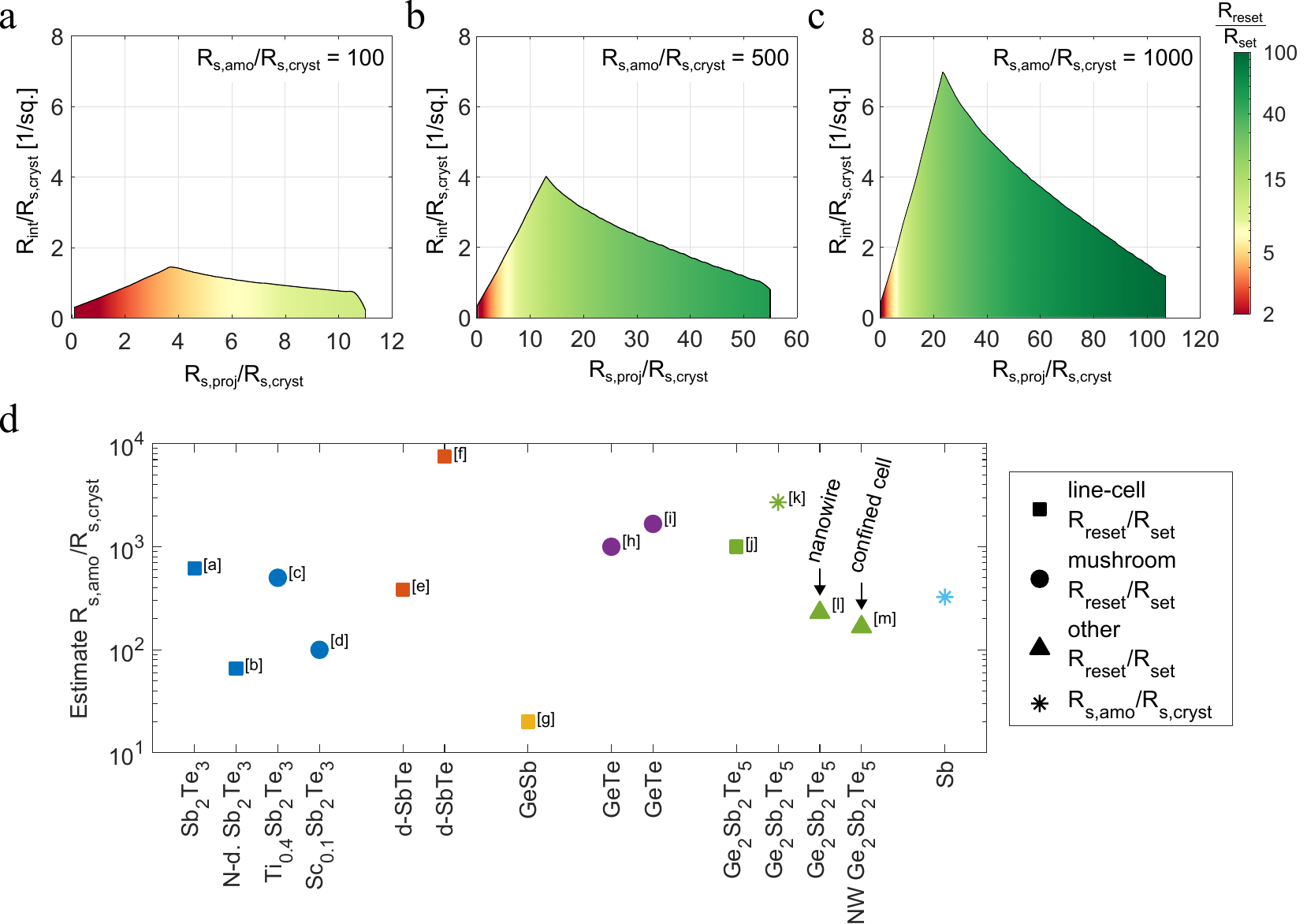}
		\caption[Guidelines for device optimization]{\textbf{Guidelines for device optimization.} Target specifications for future device generations are a drift coefficient smaller than 0.01, a minor state dependence of the drift coefficient, and a close to linear scaling of the resistance with the amorphous length. \textbf{a–c}, The colored area marks feasible projection layer sheet resistances and interface resistances for different ratios \Rsa/\Rsc. The color gradient encodes the dynamic resistance range. \textbf{d}, Estimate of the melt-quenched amorphous to crystalline resistance ratio of different phase change materials. The estimate is obtained from device programming curves. Marker shapes encode the device type. Compounds of comparable compositions are plotted with identical colors. (ref. [a,b] - \cite{Yin2007}; [c] - \cite{Zhu2014a}; [d] - \cite{Rao2017}; [e] - \cite{Lankhorst2005}; [f] - \cite{Oosthoek2011}; [g] - \cite{Chen2006}; [h] - \cite{Perniola2010}; [i] - \cite{Bruns2009}; [j] - \cite{Meister2011}; [k] - \cite{Jeyasingh2011}; [l] - \cite{Kim2010a}; [m] - \cite{Jung2006})}
		\label{fig:PROJ_Model_Extrapolations}
	\end{figure}

	The maximum interface resistance that can be tolerated to fulfill the aforementioned constraints changes depending on the ratio of the sheet resistances of the projection material, crystalline, and amorphous phase change material (\Cref{fig:PROJ_Model_Extrapolations}a-c). The bridge has an aspect ratio (length/width) of two, the drift coefficient is 0.1 and the contact resistances to the metal electrodes are assumed to be negligible, identical to the initial model assessment (\Cref{fig:PROJ_Model_ExtremeRinterface}c). The curves of the highest interface resistance that allows all three criteria to be fulfilled can be split into two regimes to the left and right of the maximum. In regime one, at small projection layer resistances, the maximum interface resistance that can be tolerated increases steeply. The projection layer sheet resistance is much smaller than the amorphous sheet resistance. In this case, the constraint enforcing a linear scaling of the device resistance with L\textsubscript{amo} determines the interface resistance. A drift coefficient smaller than 0.01 could also be achieved with large interface resistances. In the second regime, the maximum interface resistance gradually decreases. The limiting constraint is the requirement of a maximum drift coefficient smaller than 0.01. As the projection layer sheet resistance increases and gets closer to the amorphous sheet resistance, the projection must become “better”, to suppress the drift sufficiently. A better projection is enabled by a smaller interface resistance. Eventually, the projection layer resistance becomes too large to suppress the drift to 0.01. With an increasing ratio of R\textsubscript{s,amo}/R\textsubscript{s,cryst} (\Cref{fig:PROJ_Model_Extrapolations}a–c), the maximum shifts to larger values of R\textsubscript{s,proj}/R\textsubscript{s,cryst} and R\textsubscript{int}/R\textsubscript{s,cryst}. Furthermore, the largest projection layer resistance for which the drift coefficient is smaller than 0.01 increases. Consequently, a larger dynamic range (R\textsubscript{reset}/R\textsubscript{set}) can be realized (color-coded). A high dynamic range is desirable since it allows the quasi-analog device states to be written and stored with higher precision. Thus, phase change materials with large R\textsubscript{s,amo}/R\textsubscript{s,cryst} are preferable. A projection material and phase change material combination that simply maximizes the tolerable interface resistance is not advantageous. Instead, a compromise between dynamic range and acceptable interface resistance must be made.
	
	To assess the suitability of a phase change material for a projected device, knowledge of the resistance ratio of the melt-quenched amorphous and the crystalline state is essential. Whilst numerous papers report the resistivity of as-deposited amorphous samples, only limited data have been reported for the melt-quenched amorphous phase, the material state of interest in device applications. The RESET to SET resistance ratio reported for device programming curves is used to estimate the resistance ratio of the melt-quenched amorphous and the crystalline states (\Cref{fig:PROJ_Model_Extrapolations}d). The maximum RESET state in the analyzed programming curves appears to be a saturation level, suggesting the size of the amorphous volume has reached a maximum. Saturation of the programming curve does not necessarily imply a fully amorphous device. Thus, the analysis may underestimate the materials' resistance contrast. 
	
	The studies of doped-SbTe,\cite{Oosthoek2011} and \GST{} \cite{Meister2011,Jeyasingh2011} show TEM figures of an
	amorphized device, suggesting the complete device volume has been amorphized. These two materials and GeTe, all exhibiting resistance ratios larger than 1000, are the most suitable candidates for a projected memory device. However, device studies of highly confined \GST{} \cite{Kim2010a,Jung2006} report a significantly reduced resistance ratio. The programming data reported for N-doped Sb\textsubscript{2}Te\textsubscript{3}, Sc\textsubscript{0.1}Sb\textsubscript{2}Te\textsubscript{3}, and GeSb suggests these materials are not well suited for a projected device. Due to their small resistance ratios (\textless 100) it may be challenging to apply them for drift and noise resilient multi-level memory devices. For materials with resistance ratios \textless 500 a careful optimization of the projection material, especially the interface resistance, will be required. 
	Since the tolerable \Rint{} scales with \Rsc{}, phase change materials with higher crystalline resistivity would be preferable for a projected device. They would relax the requirements to optimize the contact resistivity between phase change and projection material. Considering there is a certain limit to which the contact resistivity can be reduced, this might prove to be another solution. 

\section{Summary \& Outlook}

In a projected memory cell, resistance drift is suppressed because only a fraction of the read current flows through the amorphous phase change material. Simulations with the equivalent circuit model and measurements over 7 orders of magnitude in time reveal that drift becomes time and state-dependent in a projected device. The increasingly stronger drift suppression with time is inherent to the projected memory and can be considered a desirable effect. Whereas the state dependence causes a gradual loss of information with time, that cannot be recovered. Since resistance drift exhibits variability anyway (inter but also intra device), the overall drift reduction by projection is still expected to be beneficial. Furthermore, the drift coefficient only becomes state-dependent if the current flow from the phase change material into the projection layer is hindered. Thus, it is critical to minimize the contact resistivity between the two materials. 

Based on the device model, a few lessons on how to select projection and phase change material can be learned. First, a drift reduction by 10x implies a reduction of the dynamic range by $\sim$10x, compared to the unprojected device. Additionally, the tolerable interface resistivity can be increased by enlarging the resistance window between the amorphous phase and projection layer (i.e., decreasing \Rsp/\Rsc{}), which would again reduce the dynamic range. Thus, the phase change material should have a large dynamic range (\Rsc/\Rsa). Second, with increasing resistivity of the phase change material a larger interface resistance (interface resistivity) can be tolerated. This makes phase change materials with higher resistivity preferable. 

As a next step to refine the model and improve its predictive character a more elaborate characterization of \Rint{} is required. In particular, separate measurements of the contact resistivity and transfer length could allow \Rint{} to be determined, assuming \Cref{equ:PROJ_TransferLength} can be applied. Potentially, another phenomenological description is required. For example, it could be due to the device geometry or the metal electrodes which are connected to both materials. Furthermore, contact between projection and crystalline phase change material is assumed to be ohmic. If a Schottky barrier forms, \Rint{} to the left and right of the amorphous volume is different. Like in the previous chapter, it becomes apparent, once again, that a good understanding of interfacial and confinement effects is essential to developing future devices. Besides the implications for phase transition dynamics, also those on electronic contacts are critical in a projected memory device. 

Last but not least, the transferability of the model to other projected PCM types will be discussed. Besides the bridge cell, confined cells and mushroom cells with projection have also been studied \cite{Kim2013,Bruce2021,GhaziSarwat2021}. Since the confined cell is basically a vertically flipped bridge cell, the model is applicable to this device type as well. In the mushroom cell (\Cref{fig:Intro_DeviceConcepts}), on the other hand, the geometry of the amorphous volume differs, and, importantly, the phase change material is not in direct contact with the bottom electrode. Projection and phase change material are deposited one after another, directly on top of the heater. All read current must flow across the interface of the two materials. Furthermore, compared to the crystalline state, the interface resistivity to the amorphous phase change material is typically at least three orders of magnitude larger \cite{Cooley2020}. For this reason, a small interface resistance between projection and crystalline phase change material might be less critical in this device type. A different equivalent circuit is required to describe this device because of the different arrangement of the resistor elements \cite{GhaziSarwat2021}.

\textbf{Key findings:}
\begin{itemize}
	\item	A tractable device model to describe the state and time dependence of drift and how the device resistance scales with the size of the amorphous volume was developed. It captures quantitatively the experiments on projected Sb bridge cells. 
	\item	The time-invariant conduction path in parallel with the drifting amorphous phase change material results in a time-dependent drift coefficient in projected memory cells, irrespective of the device type (mushroom or bridge). The more time elapsed after amorphization, the less the device drifts.  
	\item	The drift coefficient also depends on the size of the amorphous volume because of the interface resistance between the Sb and projection material. With increasing interface resistance compared to the resistance of the amorphous volume (i.e., decreasing size), less read current bypasses the amorphous phase. The drift coefficient increases for smaller amorphous lengths. By minimizing the contact resistivity between the projection layer and phase change material, a state-dependent drift coefficient can be avoided. 
\end{itemize}
		
		\chapter{Conclusion}

To close this thesis, we review again how drift can be mitigated in PCM devices. Based on the presented experiments and models, it is discussed which strategies (see \Cref{Sect:MitigateDrift}) have the best prospect of solving the drift problem and two new potential concepts are proposed.  

In \Cref{Chapt:Relax} the onset of relaxation was studied. Since relaxation processes are thermally activated, the onset differs by orders of magnitude in time, dependent on the ambient temperature. By annealing a device relaxation is accelerated. In the same time-interval relaxation processes with higher activation energy take place. Thus, when cooled back to the base temperature, relaxation is put on hold for a while \cite{Oosthoek2012}. Annealing shifts the onset of relaxation to longer timescales and can eliminate drift at least temporarily. Conceptually, it is possible to anneal the device in-situ by applying electrical pulses of the right amplitude. The bias must be sufficiently large to overcome \Vth{}, but the temperature raise must be small enough to avoid crystallization. In the mushroom cell with a series resistor, it is difficult to meet these conditions. The SET window indicates (\Cref{fig:Intro_Programming}a) that the pulses are either subthreshold or induce crystallization of the device. If a PCM device is operated in series with a transistor, that allows controlling the device current after threshold switching, it may be possible to apply such in-situ annealing and shift the onset of drift. However, annealing would increase the energy required to program the device. Furthermore, it is not clear how long drift could be arrested and the realization of such a scheme at scale may be extremely challenging, since the temperature reached for a given annealing power may differ significantly from device to device.      

Also single element glasses may be suitable for building future generations of PCM devices. However, the amorphous Sb drifts as well. Furthermore, the experiments do not indicate that confinement to thicknesses as thin as \unit[3]{nm} or the cladding material has an impact on the drift coefficient. In the devices characterized in \Cref{Chapt:Confine,Chapt:Proj}, Sb was sandwiched between SiO\textsubscript{2}-SiO\textsubscript{2}, SiO\textsubscript{2}-ZnS:SiO\textsubscript{2}, and metal nitride-SiO\textsubscript{2}~\footnote{At \unit[100]{K} ambient temperature the devices with Sb sandwiched between SiO\textsubscript{2}-SiO\textsubscript{2} and SiO\textsubscript{2}-ZnS:SiO\textsubscript{2} exhibit a drift coefficient of 0.1. The drift of the projected device (metal nitride-SiO\textsubscript{2}) is captured with the drift coefficient measured for the unprojected device (SiO\textsubscript{2}-SiO\textsubscript{2}); both were characterized at \unit[200]{K}.}. These observations indicate that material confinement and the selection of an ideal cladding material seem not to offer a solution to drift. In opposition, a study of confined devices with Sb-rich Ge\textsubscript{x}Sb\textsubscript{y}Te\textsubscript{z} reported, in a single measurement, a remarkably small resistance drift coefficient of only 0.011 \cite{Kim2010a}. Given the criticality of drift for PCM technology, it is surprising that no extended follow-up works have been published.  

The drift of \GST{} and GeTe has been linked to defective coordination of Ge atoms in the amorphous state \cite{Zipoli2016}. Additionally, an increase of the Peierls distortion, which might also explain drift in \SbTe{} and pure Sb, has been identified as a cause of drift \cite{Raty2015b}. Both the removal of defective coordinated Ge and an increase of the Peierls distortion define which structural motifs these materials evolve towards. For other, new phase change materials, it might be different motifs. But structural relaxation is observed in all glasses. In principle any structural change must be expected to have some impact on the electronic band structure. Thus, the ambitious goal to find a material that does not drift means designing a material in which the changes of the electronic band structure upon relaxation do not reflect on the electronic transport. 

It is important to keep in mind that not only drift but also absolute resistivity, the resistivity contrast between the crystalline and amorphous state, the threshold field, and crystallization dynamics are critical for a PCM device. Considering how elegantly drift can be addressed in a projected device, it seems most appropriate to concentrate material optimization on those other properties. Measurements on the projected Sb bridge cells and extrapolations with the developed device model show that good electrical contact is essential to building an ideal device. Besides a low interface resistivity to the projection material, the phase change material should also have a large resistance ratio between the crystalline and amorphous phases. Other phase change material properties will be prioritized for a projected device and the choice of a projection material offers another degree of freedom for device optimization. 

\begin{figure}[bth!]
	\centering
	\includegraphics[width=0.6\linewidth]{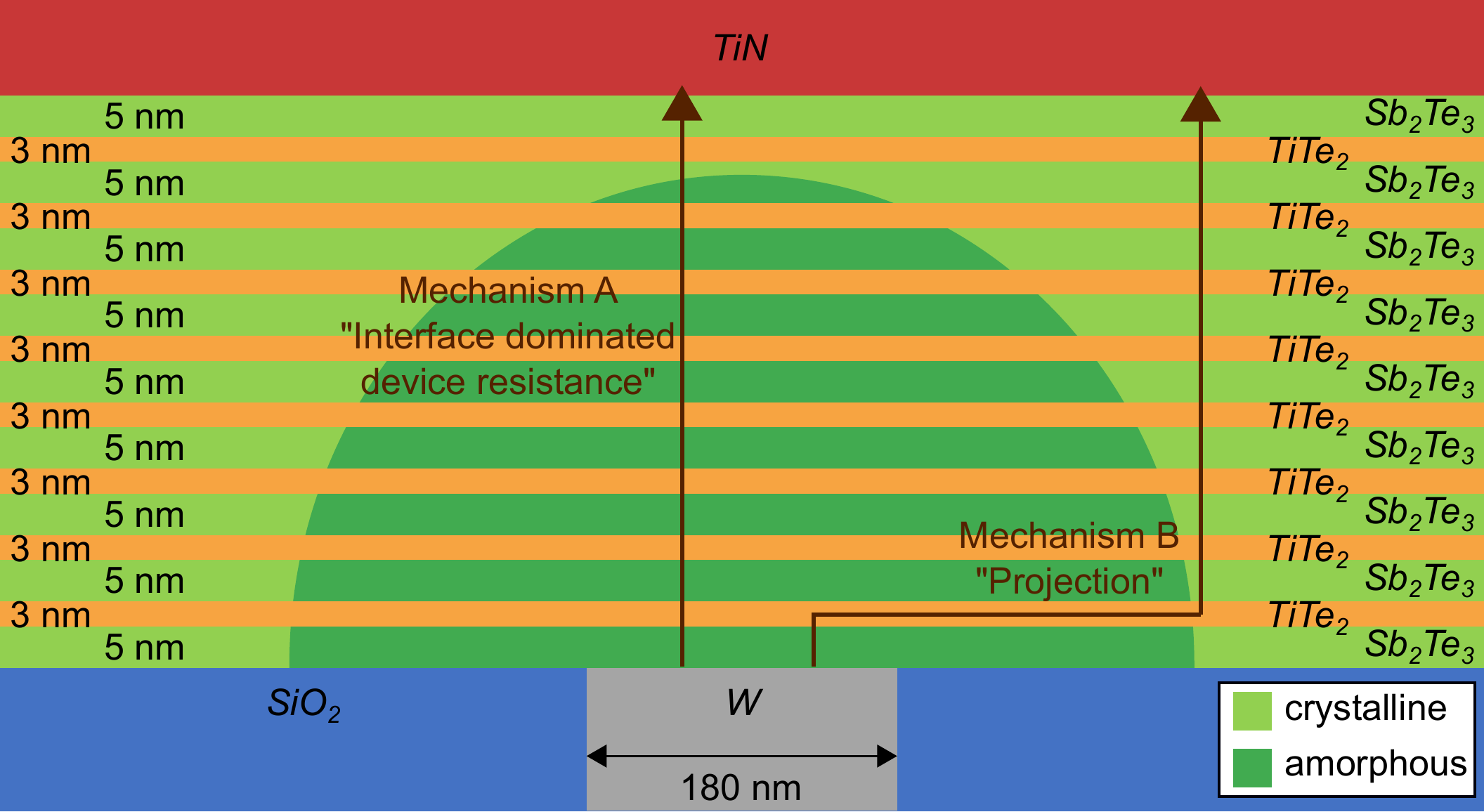}
	\caption[Sketch - Phase change heterostructure]{\textbf{Sketch - Phase change heterostructure.} The device resembles a mushroom cell. The W bottom electrode is patterned to a cylinder with a \unit[180]{nm} diameter. The phase change material (\SbTe) is confined by TiTe\textsubscript{2} layers. The sketch corresponds to the device presented in \cite{Ding2019}. Black arrows indicate two hypotheses that might explain the RESET resistance and drift reduction in this device. A: The device resistance is defined by the number of TiTe\textsubscript{2}-\SbTe interfaces across which the read current must flow. B: The read current bypasses the amorphous material above the first TiTe\textsubscript{2} layer. }
	\label{fig:CONCL_PCHsketch}
\end{figure}

Ideas for new device concepts that allow decoupling even more properties of the phase change material from the device characteristics may lead to the next breakthrough. One example of such a new concept is the phase change hetero structure (PCH). It is a modified mushroom cell device. For a prototype device, thin films of phase change material (\unit[5]{nm} \SbTe) and an inert confinement material (\unit[3]{nm} TiTe\textsubscript{2}) were deposited alternating on a W bottom electrode with a contact area of \unit[28252]{nm\textsuperscript{2}} (\Cref{fig:CONCL_PCHsketch}). The confinement material has a low thermal conductivity, a higher T\textsubscript{m} than the phase change material, and functions as a diffusion barrier. Compared to the conventional reference device (\GST), cycling endurance, power efficiency, programming speed, and importantly drift ($\nu_R$\textless 0.005) and read-noise are remarkably improved. These benefits are attributed to limited diffusion, better heat-confinement, and suppression of Peierls distortion, i.e., drift suppression by confinement \cite{Ding2019}. Unquestionably, the device exhibits significantly improved characteristics. However, the results of this thesis give reason to reconsider the explanation of the RESET resistance and the drift reduction in this device. 

The device comprises 8 layers of TiTe\textsubscript{2}, which is estimated to have a resistivity of $\sim$\unit[1.3$\cdot$10\textsuperscript{-6}]{$\Omega$\,m} and 9 layers of \SbTe, for which the authors measured a resistivity of $\sim$\unit[5$\cdot$10\textsuperscript{-3}]{$\Omega$\,m} in the as-deposited amorphous state \cite{Ding2019}. The device can be approximated as multiple discs the size of the bottom electrode. The maximum RESET resistance of such a device is  [\unit[1.3$\cdot$10\textsuperscript{-6}]{$\Omega$\,m} $\frac{\unit[3]{nm}}{\unit[28252]{nm\textsuperscript{2}}}$]$\cdot$8 + [\unit[5$\cdot$10\textsuperscript{-3}]{$\Omega$\,m} $\frac{\unit[5]{nm}}{\unit[28252]{nm\textsuperscript{2}}}$]$\cdot$9 = \unit[7965]{$\Omega$}. Note that this overestimates the RESET resistance since the top electrode is much larger. But experimentally a 125x higher RESET resistance was observed. Two hypotheses might explain the notably higher RESET resistance. First, the as-deposited amorphous state typically exhibits a notably higher resistivity than the less relaxed, melt-quenched state. But, the resistivity of the as-deposited thin film sample may not correspond to an amorphous resistivity if crystalline percolation paths exist. Another study reported a considerably higher resistivity ($\sim$\unit[0.1]{$\Omega$\,m}) for as-deposited amorphous \SbTe{} \cite{DamodaraDas1987}. Second, the device current must flow across one W-\SbTe(amorphous) and up to 16 TiTe\textsubscript{2}-\SbTe(amorphous) interfaces. Potentially, the RESET resistance could be mainly defined by the contact resistance of these interfaces. In this scenario, the device resistance would be decoupled from the amorphous resistivity in the PCH.

Drift suppression in \SbTe{} upon confinement was deduced from the resistance measurements of an as-deposited amorphous thin film sample (\unit[5]{nm}). As mentioned before, other studies measured a 500 times higher as-deposited amorphous resistivity \cite{DamodaraDas1987}. If the lower resistivity is due to crystalline percolation paths, those would also explain the apparently negligible drift. Even more critically, the sample is as-deposited amorphous and must be expected to be in a much more relaxed state than a melt-quenched material. Thus, the onset of relaxation is shifted to longer timescales. Wimmer et al. showed nicely that even heated to \unit[323]{K}, as-deposited amorphous GeTe only begins to exhibit drift after more than \unit[2000]{s} \cite{Wimmer2014b}. The resistance measurement of the as-deposited \SbTe{} sample for $\sim$\unit[3000]{s} at room temperature cannot serve as conclusive evidence that the confined \SbTe{} does not drift. Additionally, the studies show that, at least for Sb, confinement has no impact on the drift coefficient of the amorphous material. 

Another mechanism similar to projection could explain the drift reduction observed in the melt-quenched PCH. Besides a smaller resistor in parallel (projection), a larger resistor in series with the resistor defined by the amorphous phase change material also results in a reduced device drift. Consider a device comprising a time-invariant \unit[1]{M$\Omega$} resistor in series with a \unit[0.1]{M$\Omega$} phase change material resistor, which has a drift coefficient of 0.066. As one order of magnitude in time elapses the device resistance increases from \unit[1.1]{M$\Omega$} to \unit[1.12]{M$\Omega$}. The device drift coefficient of 0.0078 is comparable to those reported for the PCH. The contact resistance at the material interfaces may correspond to this comparably large, time-invariant series resistor (\Cref{fig:CONCL_PCHsketch}, mechanism A). Alternatively, if the current flow from TiTe\textsubscript{2} into \SbTe{} is hindered, the read current could flow horizontally in the first TiTe\textsubscript{2} plane above the bottom electrode and bypass, like in a projected device, the second and all subsequent amorphous \SbTe{} layers (\Cref{fig:CONCL_PCHsketch}, mechanism B). The RESET resistance would scale with the diameter of the amorphous volume in the second \SbTe{} layer above the bottom electrode, like in a projected device. The in-plane resistance of the \unit[3]{nm} thin TiTe\textsubscript{2} film might be notably larger than the bulk material resistivity of \unit[1.3$\cdot$10\textsuperscript{-6}]{$\Omega$\,m} \cite{Koike1983} and could thus also explain the high RESET resistance. In both scenarios, the RESET resistance is decoupled from the amorphous resistivity. It would be defined by either the interface resistance (mechanism A) or the in-plane resistivity of the confinement material (mechanism B).

Admittedly, these alternative explanations for the drift mitigation in the PCH are speculative. Confinement might indeed suppress drift in \SbTe. Potentially, a combination of confinement and a huge time-invariant series resistor causes the drift reduction. But even if the ideas proposed here do not apply to the \SbTe-TiTe\textsubscript{2} PCH device, they are still noteworthy concepts. In a multi-layer device structure, like the PCH, a huge contact resistance to the amorphous phase change material can reduce the drift coefficient and help to decouple the device RESET resistance from the amorphous resistivity. To realize such a device concept, the contrast of the contact resistance between the amorphous and crystalline phase to the confinement material should be huge. Additionally, structural relaxation of the amorphous phase should not alter the contact resistivity. To the author's best knowledge, it has not yet been studied if or how much the interface resistivity drifts. It is certainly not trivial to build a device that allows melt-quenching of the phase change material and measuring the interface resistance directly, without compromising the integrity of the interface during device fabrication. 

In the last decades, the focus of PCM research has been on the phase change material itself. As devices are scaled to ever-smaller dimensions, confinement and interfacial effects become increasingly important. Not only will they determine the crystallization kinetics and accordingly write speed as well as retention time, but they will also define how well a projected device works and may enable new memory mechanisms (switching of the interface resistance) or drift suppression methods (time-invariant series resistor). The research and development of the next generation phase change memory devices rely crucially on quantitative knowledge of effects related to nano-confinement. Instead of studying only a phase change material, it will be essential to characterize a combination of phase change material and interface under well-controlled conditions.

		\chapter*{Acknowledgment}
\markboth{Acknowledgement}{Acknowledgment}
\addcontentsline{toc}{chapter}{Acknowledgment}

It has been a long way from my first day as a Ph.D. student to this moment of writing the last lines of my thesis. Within these years I have been granted the time, freedom, and support I needed to grow as a researcher and scientist. Privately I experienced the brightest but also most challenging moments of my life. I am grateful for the mentors, colleagues, and friends who joined and supported me on this journey.

I would like to thank Martin Salinga for his commitment as my mentor and thesis advisor. Our regular intense and inspiring discussions taught me to rigorously assess experiments and hypotheses. To appropriately describe the importance and value of his role, I must use a German expression. He is my Doktorvater. 

My second, equally important mentor, and manager at IBM Research is Abu Sebastian. Besides the frequent exchange of ideas with an exceptional researcher, I found his expertise in abstracting systems while describing them accurately particularly valuable. Furthermore, I learned to approach tasks more result-oriented and how to prioritize problems accordingly.  

Next, I want to thank Gerhard Wilde and Nikos Doltsinis for reading and refereeing this thesis. 

Without the hard work of many experienced cleanroom engineers, this thesis would have been impossible. I acknowledge the support of the cleanroom operations team of the Binnig and Rohrer Nanotechnology Center (BRNC). Especially, I thank Vara Prasad Jonnalagadda, who fabricated multiple generations of bridge cell devices and continuously optimized the process. Furthermore, I thank our colleagues from IBM Research Albany for providing the mushroom cell devices. 

For a relatively short, yet memorable time, I have been sharing an office with Sebastian Walfort and Xuan Thang Vu at RWTH Aachen University before our ways separated to M{\"u}nster and Zurich. It was nice having both of you as office mates. 
 
At IBM Research Zurich I had the opportunity to work with a group of excellent scientists and engineers. I enjoyed the projects and moments when we managed to combine our expertise and strengths to achieve results and a degree of productivity, the individual would not have been capable of. Thanks to Irem Boybat, Syed Ghazi Sarwat, Manuel Le Gallo, Urs Egger, Vara Prasad Jonnalagadda, and Abu Sebastian. Working with you and getting to know you as a person has been a pleasure. I would also like to thank the master's students and interns I got to supervise and work with, Michele Martemucci, Vladimir Ovuka, and Francesco Marrone. I am pleased each of your projects led to a publication. 

Last but not least I thank my family, my wife Lena Kersting, our daughter Frieda, my parents Doris and Norbert Kersting, and my brother Thomas. I feel immensely fortunate and grateful to have you in my life. You are my peaceful haven. Without your support and encouragement, I could never have accomplished this work.


%






\appendix
\renewcommand{\thechapter}{A}
\renewcommand{\thefigure}{A.\arabic{figure}}
\setcounter{figure}{0}

\chapter*{Appendix}
\addcontentsline{toc}{chapter}{Appendix}
\markboth{Appendix}{Appendix}

\section{Cryogenic probe station}
\label{Sect:ProbeStation}

\begin{figure}[htb]
	\centering
	\includegraphics[width=0.8\linewidth]{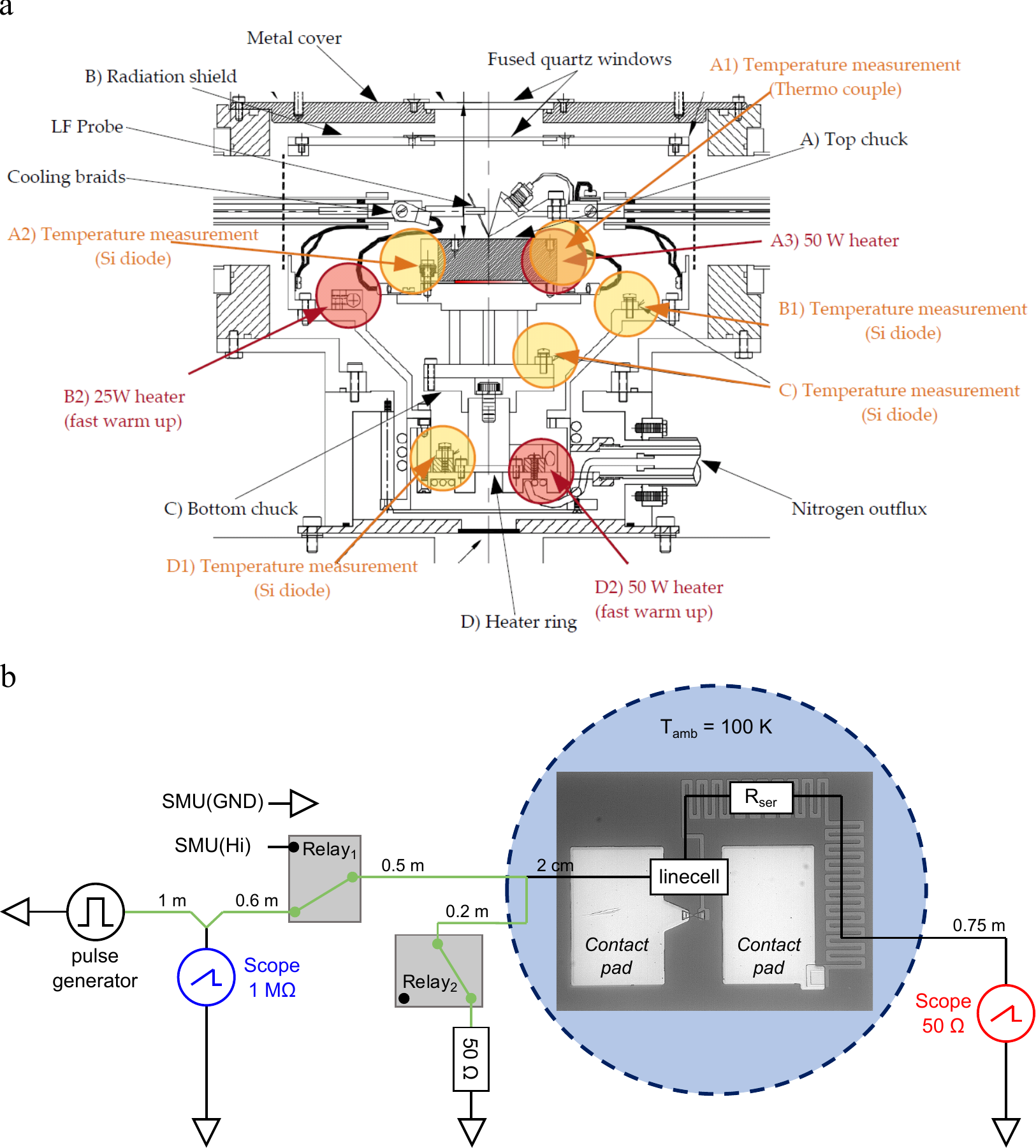}
	\caption[Cryogenic probe station]{\textbf{Cryogenic probe station. a}, Technical drawing of the probe chamber. Functional elements are annotated with arrows. Reproduced with permission from \cite{LeGallo2017d} \textbf{b}, Sketch of the electronic circuit in AC configuration. The \unit[50]{$\Omega$} terminated transmission path is marked green. In DC configuration both relays are switched; relay two is open.}
	\label{fig:APPEND_ProbeStation}
\end{figure}

The experiments were performed in a liquid-nitrogen-cooled cryogenic probe station (JANIS ST-500-2- UHT, \Cref{fig:APPEND_ProbeStation}). The ambient temperature in the vacuum chamber is probed at four positions (A: Chuck top, B: Radiation shield housing, C: Chuck bottom, D: Heater ring). A Lakeshore 336 Automatic temperature controller, connected to two resistive heaters is used to stabilize the temperature with an accuracy of \textless \unit[0.5]{K}. Heat transfer into the system through radiation and convection are minimized by a radiation shield and evacuating the probe chamber to \unit[10\textsuperscript{-5}]{mbar}, respectively. Heating the samples with the high-frequency probe (Cascade Microtech Dual-Z) is avoided by connecting it with cooling braids to the copper chuck, holding the sample.

The electronic test setup combines a DC measurement path (high current resolution; nA) with an AC path (high time resolution; 200 ps). For DC measurements relay 1 connects the device to the SMU, relay 2 is connected to open and the scope channel (\unit[50]{$\Omega$} terminated) and SMU(GND) are connected to common ground. The AC pulse is sent through a \unit[50]{$\Omega$} terminated transmission path (marked green). By making the signal lines at the impedance miss-matches (\unit[1]{M$\Omega$} terminated oscilloscope channel and \textgreater \unit[2000]{k$\Omega$}) as short as possible, signal reflections are superimposed on the timescale of several ps and cancel out. For this reason, the voltage reference trace measured in the oscilloscope and the signal applied to the device are practically undistorted. The shape of signals with a transition time of only \unit[3]{ns} is preserved (RESET - \Cref{fig:RELA_Experiment_VthDrift}a). Mechanical relays (OMRON G6Z-1F-A) are used to switch between the DC and AC measurement paths. 

\newpage
\section{Resistance measurement on $\mu$s to ms timescales}
\label{Sect:FastResMeas}

\begin{figure}[htb]
	\centering
	\includegraphics[width=1\linewidth]{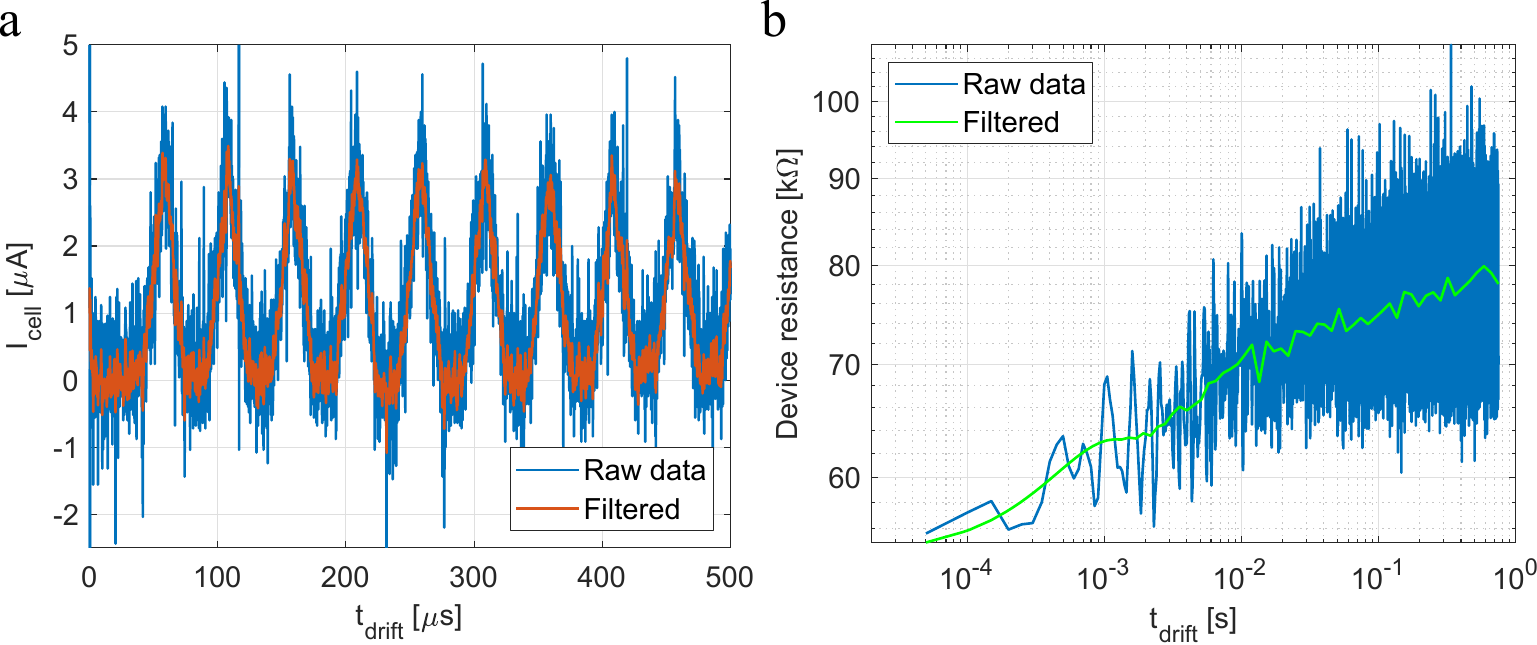}
	\caption[Fast resistance measurements - raw data]{\textbf{Fast resistance measurements - raw data. a}, Clipping of the current trace measured with the oscilloscope to obtain the time-resolved device resistance. The spike at \unit[0]{$\mu$s} results from the RESET pulse. However, the sampling rate is too small to resolve it properly. The triangular read pulses (\unit[13]{$\mu$s} leading edge and trailing edge) are applied with a \unit[50]{$\mu$s} period. \textbf{b}, Time-resolved reset resistance. The data is smoothened for further analysis and model fitting.}
	\label{fig:APPEND_FastResistanceRead}
\end{figure}

The measurement of the crystallization time (\Cref{fig:Nano_CrystallizationTime}) and drift in the projected bridge cell (\Cref{fig:PROJ_Experiment_Drift}), rely on resistance measurement beginning microseconds after RESET. An Agilent 81150 A Pulse Function Arbitrary Generator that allows to combine the signal of two independent internal pulse generators was used for these experiments. One generator sends the RESET pulse and the second a burst of triangular read pulses (0.2 V). The transient voltage signal and device current are measured with an oscilloscope (TDS3054B/DPO5104B). Each pulse of the burst gives the device's IV characteristic. By fitting those, the time-resolved device resistance is obtained. The raw scope traces are passed through a low-pass filter with a cutoff frequency of 1 MHz. Additionally, the time-dependent resistance data is processed with a Savitzky-Golay finite impulse response smoothening filter to obtain a steady resistance measurement (\Cref{fig:APPEND_FastResistanceRead}).  

\newpage

\section{Projected bridge cell - preliminary experiments}
\label{Sect:ProjBridge_preliminary}

\textit{Metal nitride - sheet resistance}

\begin{figure}[htb]
	\centering
	\includegraphics[width=1\linewidth]{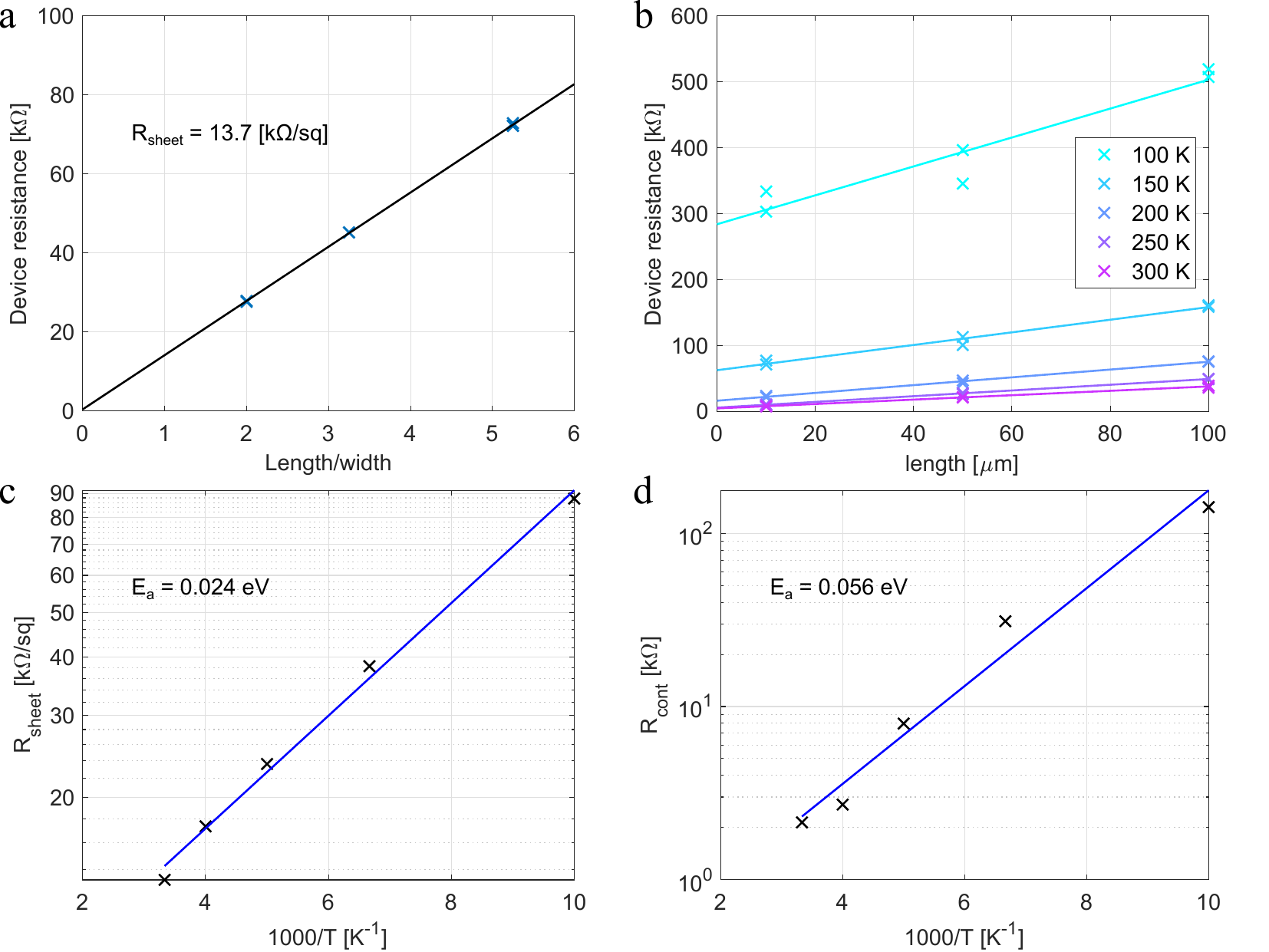}
	\caption[Metal nitride sheet resistance]{\textbf{Metal nitride sheet resistance. a}, 4 Probe resistance measured on bar structures at room temperature. \textbf{b}, Temperature-dependent resistance measured on two probe structures of different lengths. The data is fitted linearly to obtain the sheet resistance and contact resistance as a function of temperature \textbf{c\&d}, Both temperature dependencies can be described by an Arrhenius behavior.}\textbf{\textit{\textbf{}}}	
	\label{fig:APPEND_Rsheet_MN}
\end{figure}

The metal nitride sheet resistance was measured on bar structures in four-probe configuration at room temperature (\Cref{fig:APPEND_Rsheet_MN}a). To account for its temperature dependence, the resistance two-probe reference structures of different lengths were measured in a temperature range from \unit[100]{K} to \unit[300]{K} (\Cref{fig:APPEND_Rsheet_MN}b). From these measurements, the temperature dependence of the sheet resistance and contact resistance are obtained (\Cref{fig:APPEND_Rsheet_MN}c\&d). Both follow an Arrhenius behavior with an activation energy of \unit[0.024]{eV} and \unit[0.056]{eV}, respectively. At \unit[200]{K}, the metal nitride has a sheet resistance of \unit[21.8]{k$\Omega$/sq}.

\newpage
\textit{Characterization of contact resistances to the metal electrodes}

\textit{Metal electrode to antimony}

\begin{figure}[htb]
	\centering
	\includegraphics[width=0.6\linewidth]{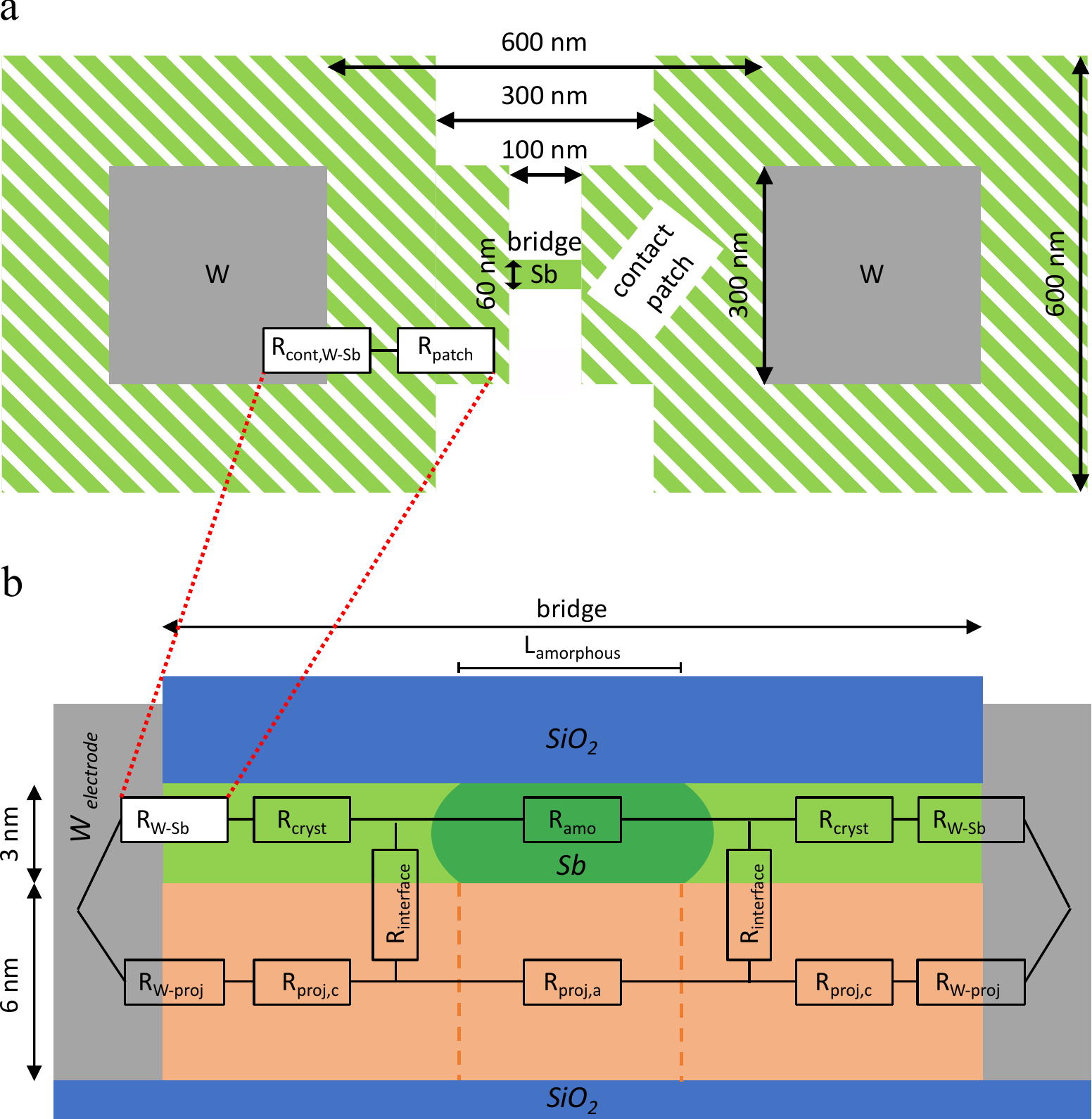}
	\caption[Device geometry]{\textbf{Device geometry. a}, Top view sketch of the bridge cell geometry. Antimony is marked in green and tungsten electrodes in grey. Only the central confined region of the device is the active volume in which material is melt-quenched and recrystallized. The resistance of the green striped area labeled contact patch is in the device model \textbf{b} part of the resistor R\textsubscript{W-Sb}. In the model, the PCM is reduced to the active region (bridge) of the device.}
	
	\label{fig:APPEND_Rcont_sketch}
\end{figure}

In the device model, the device geometry is simplified. In the real device, the phase change and projection material are patterned to a dog bone shaped geometry, whereas the model is abstracted to a confined bridge in direct contact with the metal electrodes (\Cref{fig:APPEND_Rcont_sketch}). Since the extended area of the dog bone does not participate in the switching process, its resistance is added to the resistor element that describes the contact to the metal electrode. The contact resistance to the electrode in the model is the sum of the material interface contact resistance between W and Sb, and the resistance of the extended area in the dog bone structure (e.g. R\textsubscript{W-Sb} = R\textsubscript{cont(W-Sb)} + R\textsubscript{patch(Sb)}). 

R\textsubscript{W-Sb} as defined here corresponds to the R\textsubscript{’cont’} that was defined to obtain the sheet resistance of crystalline antimony (\Cref{fig:PROJ_AntimonySheetResistance}b). The total contact resistance is \unit[1622]{$\Omega$}. To verify this value, R\textsubscript{cont(W-Sb)} and R\textsubscript{patch(Sb)} are obtained separately. Four probe contact resistance structures on the bridge cell chip enable us to measure R\textsubscript{cont(W-Sb)} directly for a contact area identical to the bridge cell. The resistance is \unit[120$\pm$10]{$\Omega$}. To obtain the R\textsubscript{patch(Sb)} resistance, we perform a COMSOL-FEM simulation of an unprojected device with the geometry depicted in \Cref{fig:APPEND_Rcont_sketch}. In the COMSOL simulation, R\textsubscript{cont(W-Sb)} = \unit[0]{$\Omega$}. The resistance of the patch is calculated as R\textsubscript{patch(Sb)} = (R\textsubscript{sim} - R\textsubscript{s,cryst} $\cdot$ length/width)/2 = \unit[1410]{$\Omega$}, where R\textsubscript{sim} is the total device resistance obtained in the COMSOL simulation, R\textsubscript{s,cryst} the sheet resistance of crystalline Sb and length and width refer to the geometry of the bridge. The sum of the separately obtained R\textsubscript{cont(W-Sb)} and R\textsubscript{patch(Sb)} matches well with the directly measured value for R\textsubscript{W-Sb}. In the device model, a contact resistance of \unit[1622]{$\Omega$} is assumed.

\textit{Metal electrode to metal-nitride}

To define the resistor element R\textsubscript{W-proj} in the equivalent circuit, R\textsubscript{cont(W-proj)} and R\textsubscript{patch(proj)} are measured separately. The contact resistance of W to metal nitride, R\textsubscript{cont(W-proj)}, was measured on micrometer scale reference structures in four-probe configuration. The structures have a square contact area of varying sizes (\unit[75x75]{$\mu$m\textsuperscript{2}}; \unit[50x50]{$\mu$m\textsuperscript{2}} and \unit[25x25]{$\mu$m\textsuperscript{2}}). The scaling of contact resistance with contact area can be described by the following formula

\begin{equation}
	R_c = \frac{\rho_c}{L_t \cdot w} coth(L/L_t).
\end{equation}

\begin{figure}[htb!]
	\centering
	\includegraphics[width=1\linewidth]{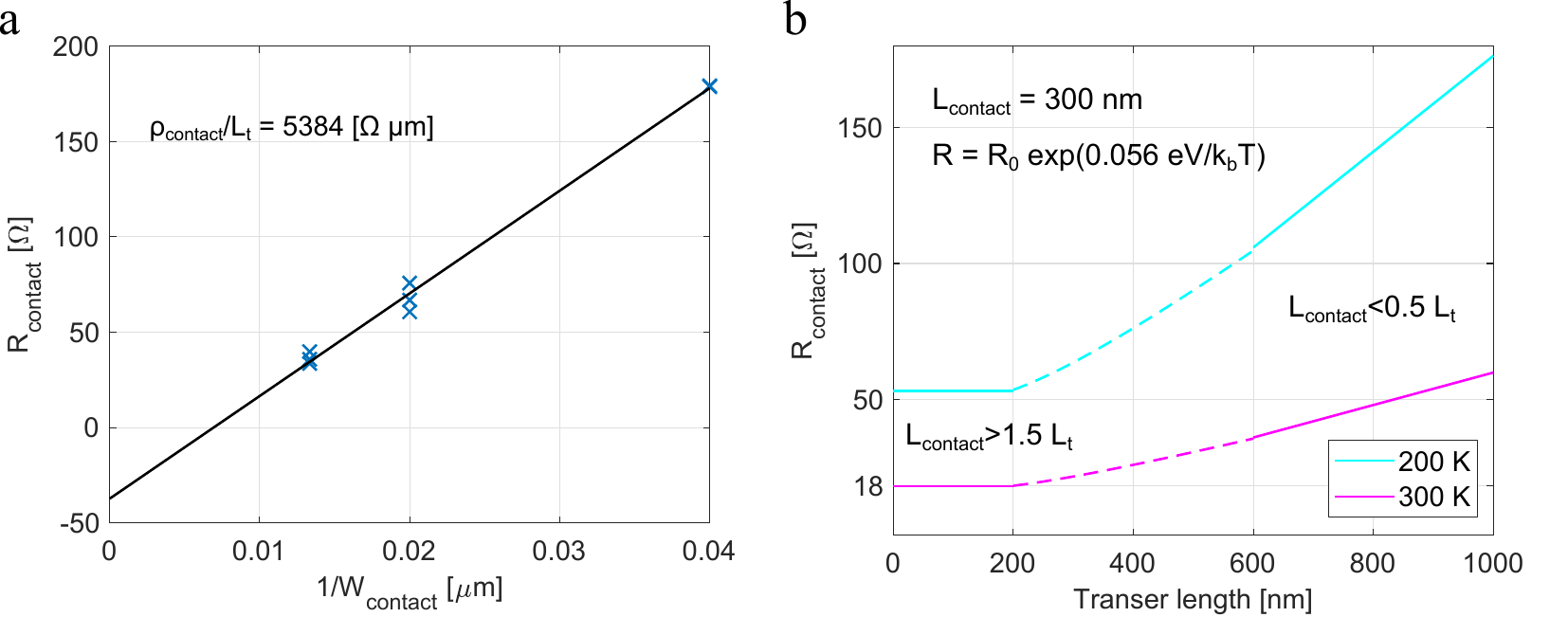}
	\caption[Contact resistance W-metal nitride]{\textbf{Contact resistance W-metal nitride. a}, Four-Probe contact resistance measured on geometries with a square contact area of varying size. The slope of the plot corresponds to the contact resistivity over transfer length [R\textsubscript{c} $\sim$ $\rho$\textsubscript{c} /(w$\cdot$L\textsubscript{t}) for L\textsubscript{contact} \textgreater 1.5L\textsubscript{t}]. \textbf{b}, Extrapolation of the contact resistance measured on reference structures to the \unit[300x300]{nm\textsuperscript{2}} contact area in the bridge cell. The contact resistance is a function of the transfer length.}
	
	\label{fig:APPEND_Rcont_MN}
\end{figure}

Here $\rho_c$ is the specific contact resistivity, L\textsubscript{t} the transfer length, w the width of the contact area (perpendicular to the direction of current flow), and L the length of the contact area (in direction of current flow)\cite{Schroder1991}. For L \textgreater 1.5$\cdot$L\textsubscript{t}, the coth converges to unity and R\textsubscript{c} should scale with 1/w. If L \textless 0.5$\cdot$L\textsubscript{t}, the coth(L/L\textsubscript{t}) can be approximated as L\textsubscript{t}/L. Accordingly, R\textsubscript{c} would be proportional to the contact area (L$\cdot$w). R\textsubscript{c} of the reference structures scales linearly with 1/w (\Cref{fig:APPEND_Rcont_MN}a). The slope tallies with $\rho_c$/L\textsubscript{t}. Since the transfer length is still an unknown variable, one can only estimate the contact resistance in the bridge cells dependent on the transfer length (\Cref{fig:APPEND_Rcont_MN}b). The contact area is \unit[300x300]{nm\textsuperscript{2}}; i.e. L \& w are \unit[300]{nm}. It is assumed that the transfer length is in the range of \unit[100]{nm} to \unit[1]{$\mu$ m}, which corresponds to a contact resistance R\textsubscript{cont(W-proj)} in the range of \unit[18]{k$\Omega$} (L\textsubscript{t}$\cdot$1.5 \textless \unit[300]{nm}) to \unit[60]{k$\Omega$} (L\textsubscript{t} = \unit[1]{$\mu$m}) at \unit[300]{K}. With the activation energy of \unit[0.056]{eV} (\Cref{fig:APPEND_Rsheet_MN}d) the measurement is extrapolated to a resistance R\textsubscript{cont(W-proj)} of \unit[53]{k$\Omega$} to \unit[177]{k$\Omega$} at \unit[200]{K}.

Like for the phase change material, the input parameter to the model R\textsubscript{W-proj} is the sum R\textsubscript{cont(W-proj)} + R\textsubscript{patch(proj)}. The value R\textsubscript{patch(proj)} = \unit[25]{k$\Omega$} is derived from the COMSOL simulation, with the same procedure described above for the metal to PCM interface. The model input parameter R\textsubscript{W-proj} is estimated in the range of \unit[78]{k$\Omega$} to \unit[202]{k$\Omega$}.

\newpage
\section{Threshold voltage drift - definition \Vth}
\label{Sect:VthAlgorithm}

\begin{figure}[htb]
	\centering
	\includegraphics[width=0.6\linewidth]{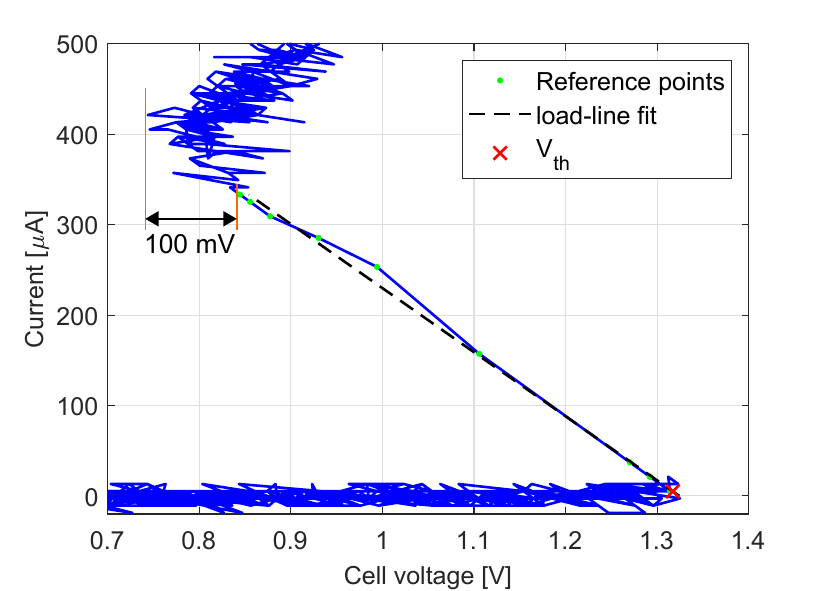}
	\caption[Threshold switching IV characteristic]{\textbf{Threshold switching IV characteristic.} The snap-back of the IV curve is fitted linearly to obtain \Vth. It is defined at a load-line current of 5µA. Green dots mark the measurement points used to fit the load-line. The blue line is the current and voltage trace measured with the oscilloscope. To obtain the voltage applied to the PCM device, the voltage drop over the series resistor ($\mathrm{V_{Rser} = Rser * Current}$)is subtracted.}
	
	\label{fig:APPEND_Vth_Algorithm}
\end{figure}

\Vth{} is extracted from the threshold switching IV characteristic. To this end, the load-line of the voltage snap-back is fitted. The mushroom cell is fabricated with an on-chip series resistor. In the moment of switching, the cell resistance drops to values similar to the series resistor (R\textsubscript{ser}), and thus the voltage drop over the cell decreases. By fitting the load-line, instead of choosing the largest voltage drop prior to switching as threshold voltage value, the analysis scheme becomes more resilient to noise in the transient voltage and current trace. The reference points to fit the load-line range from the last point where the current is smaller than \unit[20]{$\mu$A} to \unit[100]{mV} above the minimum of the load-line (\Cref{fig:APPEND_Vth_Algorithm}). \Vth{} is defined at a load-line current of \unit[5]{$\mu$A}.

\newpage
\section{RESET programming at different ambient temperatures}
\label{Sect:SimilarRESETstates}
To create comparable RESET states at different ambient temperatures, the programming power was scaled such that the initially molten volume remained approximately constant. An easily accessible metric to compare the molten volume at different ambient temperatures is the hot-spot temperature ($\mathrm{T_{hs}}$) inside the device. If $\mathrm{T_{hs} = R_{th} P_{inp} +T_{amb}}$ remains constant, so does the molten volume. $\mathrm{R_{th}}$ is the average thermal resistance of the device, $\mathrm{P_{inp}}$ is the input power associated with the voltage pulse and $\mathrm{T_{amb}}$ the ambient temperature \cite{Boniardi2012, Sebastian2014a}.

\begin{figure}[htb]
	\centering
	\includegraphics[width=1\linewidth]{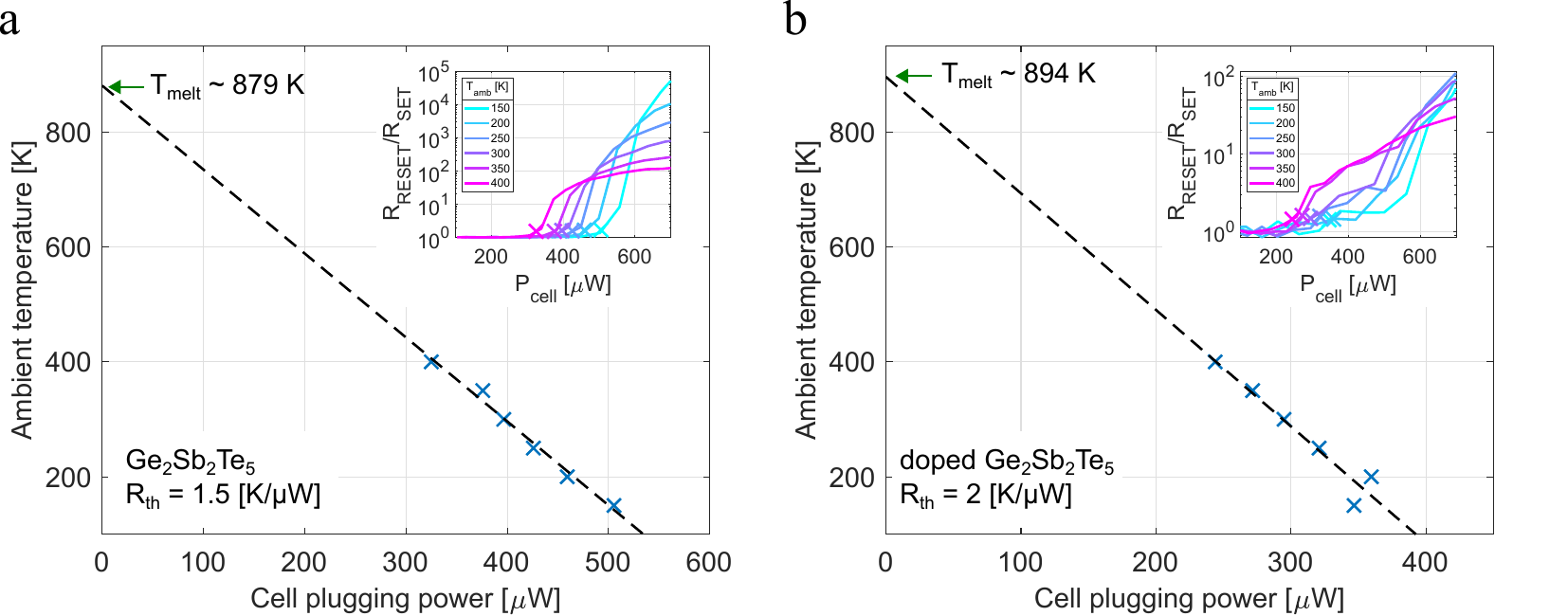}
	\caption[Thermal resistance]{\textbf{Thermal resistance.} Programming curves of \textbf{a} \GST{} and \textbf{b} doped \GST{} cells are measured at ambient temperatures ranging from \unit[150]{K} to \unit[400]{K} (insets). The cell plugging power, defined as the programming power required to induce the first increase of cell resistance, increases linearly with ambient temperature. The slope of the linear fit is the device's thermal resistance.}
	\label{fig:APPEND_Rth}
\end{figure}

The thermal resistance of \GST{} and doped \GST{} mushroom cells is obtained from the programming curves measured at different ambient temperatures (\Cref{fig:APPEND_Rth}). The programming power at which the device resistance begins to increase marks the point at which $\mathrm{T_{hs}}$ rises to T\textsubscript{m} and an amorphous volume begins to cover the device's bottom electrode. Extrapolated to \unit[0]{$\mu$W}, the temperature-dependent programming power coincides fairly well with the T\textsubscript{m} reported in the literature (\GST:\unit[858]{K}; comparable type of doped \GST: \unit[877]{K} \cite{Sebastian2014a}). 

\begin{figure}[htb]
	\centering
	\includegraphics[width=0.6\linewidth]{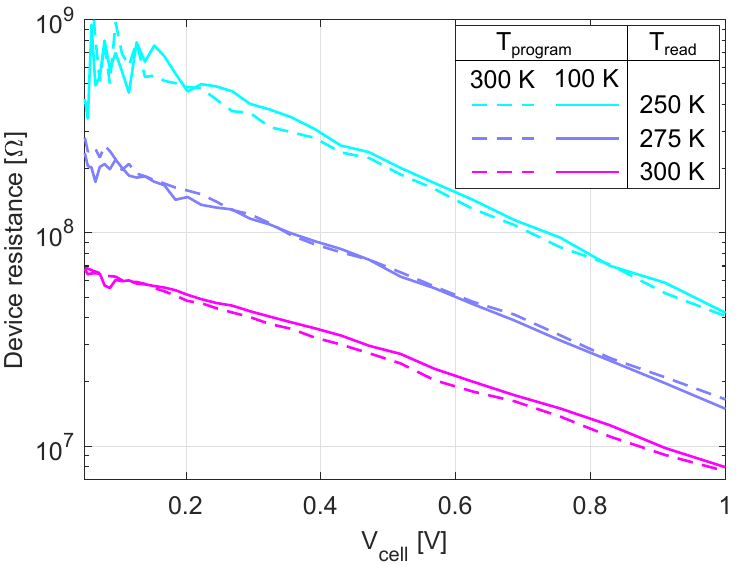}
	\caption[Programming comparable device states]{\textbf{Programming comparable device states.} Two devices were programmed with the same $\mathrm{T_{hs}}$ an ambient temperature of \unit[100]{K} and \unit[300]{K} and annealed at \unit[320]{K} to create akin relaxed glass states. The matching field and temperature-dependent resistance indicate amorphous volumes of similar size were programmed. }
	\label{fig:APPEND_SameRESET}
\end{figure}

A comparison of two devices, which were programmed at \unit[100]{K} and \unit[300]{K}, shows its fidelity (\Cref{fig:APPEND_SameRESET}). The two cell states have a different thermal history and thus represent differently relaxed glass states. An annealing step to \unit[320]{K} for \unit[15]{minutes} was applied to erase the differing thermal history. After annealing the devices show an identical field and temperature-dependent transport. This remarkable match demonstrates two things. First, that RESET states of comparable size were created. Second, upon annealing the glass state in both devices exhibits identical transport characteristics, which implies that the annealing step created similarly relaxed glass states. 

\newpage



\listoffigures

\listoftables

\bibliographystyle{apalike}
\bibliography{bib/library}

\chapter*{Curriculum Vitae}
\addcontentsline{toc}{chapter}{Curriculum Vitae}
\markboth{Curriculum Vitae}{Curriculum Vitae}

Benedikt Johannes Kersting \\
born on 06.01.1990 \\
citizen of Germany \\

	\begin{table}[h!]
	\centering
	\begin{tabular}{ l p{9 cm}}
	
		07.2018 - present 	& IBM RESEARCH - EUROPE (in Zurich) \\ [0.5ex]
							& Doctoral student \\[0.5ex]
							& \textit{Thesis: "Quest for a solution to drift in phase change memory devices"} \\
							& \\
		09.2017 - 06.2018 	& RWTH AACHEN UNIVERSITY - 1. INSTITUTE OF PHYSICS (IA) \\[0.5ex]
							& Research assistant \\
							& \\
		10.2016 - 08.2017	& IBM RESEARCH - ZURICH \\[0.5ex]
							& Intern \\
							& \\
		10.2013 - 08.2016	& RWTH AACHEN UNIVERSITY \\[0.5ex]
							& M.Sc. student in Material Science \\[0.5ex]
							& \textit{Thesis: "Fabrication and characterization of a micro-heater for the electrical analysis of nanoscale phase change devices"} \\
							& \\
		02.2014 - 06.2014	& ISTANBUL TECHNICAL UNIVERSITY \\[0.5ex]
							& Erasmus program \\
							& \\
		10.2010 - 09.2013	& RWTH AACHEN UNIVERSITY \\[0.5ex]
							& B.Sc. student in Material Science \\[0.5ex]
							& \textit{Thesis: "Influence of the Al\textsubscript{2}O\textsubscript{3}-content on the tensile strength of glass filaments"} \\
							& \\
		09.2009 - 05.2010   & CHILDREN'S HOME ST. AGNES \\[0.5ex]
							& Civil service \\
							& \\
		08.2000 - 06.2009 	& GYMNASIUM CANISIANUM \\ [0.5ex]
							& Abitur \\

	\end{tabular}
	\end{table}

\chapter*{Publications}
\addcontentsline{toc}{chapter}{Publications}
\markboth{Publications}{Publications}

\begin{table}[h!]
	\centering
	\begin{tabular}{ l p{14 cm}}

	1. 	& Martin Salinga, Benedikt Kersting, Ider Ronneberger, Vara Prasad Jonnalagadda,
	Xuan Thang Vu, Manuel Le Gallo, Iason Giannopoulos, Oana Cojocaru-Mirédin, Riccardo Mazzarello and Abu Sebastian \\ 
		& \textbf{Monatomic phase change memory} \\ 
		& \textit{Nature Materials}, 2018 \\
		& \\
		
	2.	& Benedikt Kersting and Martin Salinga \\ 
		& \textbf{Exploiting nanoscale effects in phase change memories} \\ 
		& \textit{Faraday Discussions}, 2019 \\ 
		& \\
		
	3.	& Vara Prasad Jonnalagadda, Benedikt Kersting, Wabe W. Koelmans, Martin Salinga and Abu Sebastian \\
		& \textbf{Resistive memory device} \\ 
		& \textit{US Patent}, 2019 \\
		& \\
		
	4.	& Benedikt Kersting, Vladimir Ovuka, Vara Prasad Jonnalagadda, Marilyne Sousa,
	Valeria Bragaglia, Syed Ghazi Sarwat, Manuel Le Gallo, Martin Salinga and
	Abu Sebastian \\ 
		& \textbf{State dependence and temporal evolution of resistance in projected phase change memory} \\
		& \textit{Scientific Reports}, 2020 \\
		& \\
		
	5.	& Benedikt Kersting, Syed Ghazi Sarwat, Manuel Le Gallo, Kevin Brew, Sebastian Walfort, Nicole Saulnier, Martin Salinga and Abu Sebastian \\ 
		& \textbf{Measurement of Onset of Structural Relaxation in Melt-Quenched Phase Change Materials} \\
		& \textit{Advanced Functional Materials}, 2021 \\
		& \\
		
	6.	& Michele Martemucci, Benedikt Kersting, Riduan Khaddam-Aljameh, Irem Boybat, SR Nandakumar, Urs Egger, Matthew Brightsky, Robert L Bruce, Manuel Le Gallo and Abu Sebastian \\ 
		& \textbf{Accurate weight mapping in a multi-memristive synaptic unit} \\
		& \textit{IEEE International Symposium on Circuits and Systems}, 2021 \\
		& \\
		
		\end{tabular}
\end{table}  
\newpage		
\begin{table}[t!]
	\centering
	\begin{tabular}{ l p{14 cm}}

	7.	& Riduan Khaddam-Aljameh, Michele Martemucci, Benedikt Kersting, Manuel Le Gallo, Robert L Bruce, Matthew BrightSky and Abu Sebastian \\
		&\textbf{ A Multi-Memristive Unit-Cell Array With Diagonal Interconnects for In-Memory Computing} \\
		& \textit{IEEE Transactions on Circuits and Systems II: Express Briefs}, 2021 \\
		& \\

	8.	& Irem Boybat*, Benedikt Kersting*, Syed Ghazi Sarwat, Xavier Timoneda, Robert L Bruce, Matthew BrightSky, Manuel Le Gallo and Abu Sebastian; *equal contribution \\ 
		&\textbf{Temperature sensitivity of analog in-memory computing using phase-change memory} \\
		& \textit{IEEE International Electron Devices Meeting}, 2021 \\
		& \\
		
	9.	& Francesco Marrone, Jacopo Secco, Benedikt Kersting, Manuel Le Gallo, Fernando Corinto, Abu Sebastian and Leon O Chua \\
		& \textbf{Experimental validation of state equations and dynamic route maps for phase change memristive devices} \\
		& \textit{Scientific Reports}, 2022 \\ 
		
	\end{tabular}
\end{table}  
\null
\vfill

%
%
%
%
%
%
%
%
%

\end{document}